%% file: MainV12.tex
\documentclass[twocolumn,aps,prx,groupedaddress,superscriptaddress,amsmath,amssymb,noeprint,nofootinbib,floatfix,10pt]{revtex4-2}
\input{AuxiliaryFiles/preamble_packages}

\input{AuxiliaryFiles/preamble_commands}

\makeatletter 

\renewcommand\onecolumngrid{% <<<<<<
\do@columngrid{one}{\@ne}%
\def\set@footnotewidth{\onecolumngrid}% <<<<<<<<<<<<<<<<
\def\footnoterule{\kern-6pt\hrule width 1.5in\kern6pt}%
}

\renewcommand\twocolumngrid{% <<<<<<
\def\footnoterule{% restore rule
\dimen@\skip\footins\divide\dimen@\thr@@
\kern-\dimen@\hrule width.5in\kern\dimen@}
\do@columngrid{mlt}{\tw@}
}%

\makeatother    
\begin{document}

\title{A Flash-Based Relativistic Collapse Model for Quantum Fields}

\author{Nicol\`{o} Piccione}
\email{nicolo.piccione@units.it}
\affiliation{Department of Physics, University of Trieste, Strada Costiera 11, 34151 Trieste, Italy}
\affiliation{Istituto Nazionale di Fisica Nucleare, Trieste Section, Via Valerio 2, 34127 Trieste, Italy}

\begin{abstract}
Spontaneous collapse models provide a possible, testable solution to the quantum measurement problem, but constructing a relativistic version applicable to quantum field theory remains an important open challenge.
Challenges in constructing relativistic collapse models include avoiding superluminal signaling and foliation dependence, controlling energy production, and ensuring vacuum stability.
Here, we present a relativistic collapse framework that addresses these problems at the cost of introducing a new space of physical states, whose complete characterization is left open.
A modified, history-dependent Aharonov--Albert prescription describes instantaneous nonlocal state updates on arbitrary spacelike foliations while preserving foliation independence.
To ensure vacuum stability, we modify the standard unitary evolution through a prescription that we name dynamical vacuum subtraction.
The resulting dynamical map is linear and trace-preserving but not positive on the full set of standard density operators.
We establish no-signaling by proving microscopic parameter independence, and identify conditions under which standard nonrelativistic collapse dynamics are recovered.
Finally, we apply our framework to a free massive scalar field, obtaining finite particle-number and energy-increase rates at leading order in the long-time regime, together with further promising evidence for the consistency of the framework.
\end{abstract}

\maketitle

\section{Introduction\label{Sec:Introduction}}

Quantum mechanics is an extraordinarily successful theory, yet its standard formulation is afflicted by the so-called measurement problem: the linear and unitary evolution prescribed by the Schrödinger equation does not, by itself, explain the emergence of definite outcomes in individual experiments~\cite{Book_Bell2004Speakable,Book_Norsen2017foundations,Book_Durr2020understanding,Book_Tumulka2022Foundations}.
Spontaneous collapse models address this issue by supplementing standard quantum dynamics with stochastic jumps or continuous stochastic modifications~\cite{Review_Bassi2003Dynamical,Review_Bassi2013Models,Book_Tumulka2022Foundations,Review_Bassi2023CollapseModels,Review_Carlesso2025CollapseModelsShortReview}.
These mechanisms, negligible for microscopic systems but dominant for macroscopic ones, suppress macroscopic superpositions while approximately reproducing ordinary quantum predictions in the microscopic regime~\cite{Review_Bassi2003Dynamical,Review_Bassi2013Models,Book_Tumulka2022Foundations,Review_Bassi2023CollapseModels,Review_Carlesso2025CollapseModelsShortReview}.

Spontaneous collapse models also provide a route toward hybrid classical--quantum (CQ) theories of gravity, in which quantum matter consistently interacts with classical gravitational degrees of freedom~\cite{Kafri2014LOCCGravity,Kafri2015BoundsNewtonianGravity,Tilloy2016CSLGravity,Tilloy2017LeastDecoherence,GaonaReyes2021GravitationalFeedback,Oppenheim2023PostQuantum,Piccione2025NewtonianPSL,Piccione2026EnergyIncreaseGPSL,Oppenheim2026CovariantPathIntegrals}.
In Newtonian constructions, spontaneous monitoring of quantum matter produces a classical stochastic record that sources the gravitational field, which in turn acts back on the quantum dynamics~\cite{Tilloy2016CSLGravity,Tilloy2017LeastDecoherence}.
More generally, continuous Markovian hybrid dynamics can be formulated in terms of monitoring and feedback~\cite{Tilloy2024HybridDynamics}.
In fact, Refs.~\cite{Diosi2024HybridRelativisticDynamicsDifficulties,Tilloy2024HybridDynamics} point out that relativistic collapse models seem to be necessary ingredients of this approach to relativistic CQ gravity.

As argued in Ref.~\cite{Mucino2026DemystifyingRelativisticCollapse}, conceptual objections to relativistic collapse do not establish an incompatibility with relativity.
Bell himself noted that the original GRW theory ``takes away the ground of my fear that any exact formulation of quantum mechanics must conflict with fundamental Lorentz invariance''~\cite[p.~209]{Bell1987JumpsV2}.
Since 1990~\cite{Pearle1990RelativisticStatevectorReduction,Ghirardi1990RelativisticCSL}, proposals have explored stochastic state-vector and field-theoretic dynamics~\cite{Dove1996Local,Breuer1998Relativistic,Pearle1999Relativistic,Adler2001RelativisticCSL,Nicrosini2003Relativistic,Bedingham2011Relativistic,Pearle2015RelativisticCSL,Bedingham2019csl}, discrete spacetime~\cite{Dowker2004SpontaneousLattice,Bedingham2021Collapse}, and flash or matter-density ontologies~\cite{Tumulka2006relativistic,Tumulka2009rGRWf,Bedingham2014Matter}.
Tumulka's flash models accommodate both non-interacting and interacting distinguishable particles, but retain a fixed particle number~\cite{Tumulka2006relativistic,Tumulka2021relativistic}.
New proposals continue to appear~\cite{Bedingham2025RelativisticQuantizedTime,Wang2026RelativisticColoredNoise,Gundhi2026RelativisticCollapseStatisticalDynamics}.
Recurring challenges include vacuum instability and divergent energy production~\cite{Pearle1999Relativistic,Myrvold2017RelativisticMarkovian,Gundhi2026RelativisticCollapseStatisticalDynamics}, alongside the requirements of foliation independence~\cite{Ghirardi1990RelativisticCSL} and no-signaling~\cite{Jones2021MassCoupled}.
Combining these requirements with a consistent stochastic ontology remains an open challenge for collapse theories applicable to standard interacting quantum field theory (QFT) on continuous Minkowski spacetime.

One of the most challenging aspects of building a relativistic collapse theory compatible with quantum fields comes from the requirement of vacuum stability.
Generally speaking, a relativistic theory is not required to preserve the number of particles. However, a non-vanishing Poincaré-invariant rate of particle production from the vacuum is necessarily divergent~\cite{Pearle1999Relativistic,Pearle2015RelativisticCSL,Myrvold2017RelativisticMarkovian}.
Intuitively, the vacuum selects no preferred inertial frame, so an invariant production law must assign equal statistical weight to momentum regions related by Lorentz boosts.
However, the corresponding mass shells have infinite invariant volume, preventing a nonzero Lorentz-invariant production law from yielding a finite particle-production rate per unit spacetime volume~\cite{Myrvold2017RelativisticMarkovian}.

Within a Markovian framework with completely positive dynamics, Myrvold showed that vacuum stability and commutation of the evolution operators with spacelike-separated observables force deterministic evolution when all fields satisfy the standard spectrum condition~\cite{Myrvold2017RelativisticMarkovian}.
Related obstructions arise for particular mass-coupled models~\cite{Jones2021MassCoupled} and proposed relativistic GKLS dynamics~\cite{Diosi2022RelativisticGKSLNope}.
These restrictions motivate exploring non-Markovian dynamics, already extensively developed in nonrelativistic collapse theory~\cite{Adler2007NonWhiteNoises,Adler2008NonWhiteNoisesII,Bassi2009NonMarkovianTrajectories,Diosi2014NonMarkovianGaussianDynamics,Carlesso2018ColoredCSLExperimentalTests,Piccione2026NonMarkovianPSL}, although non-Markovianity alone guarantees neither vacuum stability nor relativistic consistency.
Recently, Gundhi et al.~\cite{Gundhi2026RelativisticCollapseStatisticalDynamics} proposed a relativistic statistical evolution combining local field operators with Lorentz-invariant colored noise and an additional nonstandard normal-ordering prescription.
Their construction addresses several of these difficulties at the statistical level, but leaves open the specification of a physical collapse process and its conditional state evolution.

We take as our starting point the Collapse Points Quantum Mechanics (CPQM) framework underlying the Poissonian Spontaneous Localization (PSL) models~\cite{Piccione2024PoissonianCollapse,Piccione2025ExploringMassDependence,Piccione2026NonMarkovianPSL}.
In this framework, we assume that spacetime is randomly filled with special points distributed according to a homogeneous Poisson process and we call these points Collapse Points (CPs).
We then associate auxiliary qubits with the CPs and couple them to local quantum fields in the causal past of the corresponding CPs.
The qubits are initialized in $\ket{0}$ and, following standard terminology in the foundations of physics~\cite{Book_Tumulka2022Foundations}, we identify their spontaneous outcomes in the state $\ket{1}$ with so-called \emph{flashes}.
The Poisson sprinkling is statistically Poincar\'{e} invariant~\cite{Bombelli2009DiscretenessWithoutSymmetryBreaking,Review_Surya2019CausalSetApproach}, while the flashes provide a local spacetime ontology~\cite{Bell1987JumpsV2,Tumulka2006relativistic}.

Our proposal makes use of the Tomonaga--Schwinger formalism~\cite{Book_Tumulka2022Foundations} for the quantum state evolution and of the Aharonov--Albert prescription~\cite{Aharonov1980States,Aharonov1981RelativisticMeasurementProcess,Aharonov1984TimeRelativistic} for the collapse.
In standard quantum mechanics, the former allows the interaction-picture quantum state to evolve from any Cauchy surface to any other in a foliation independent manner~\cite{Book_Tumulka2022Foundations}, while the latter allows for a covariant prescription of measurements and induced wavefunction collapse~\cite{Book_Tumulka2022Foundations}.

The above construction based on CPQM formally works but cannot preserve the vacuum.
In particular, for this matter--qubit coupling, the local vacuum structure of standard QFT leads to vacuum excitation, as explained using the Reeh--Schlieder theorem and its corollaries in Sec.~\ref{Sec:ModifyingQM}.
Therefore, we introduce the dynamical vacuum subtraction (DVS) prescription, a modification of the usual unitary evolution.
This modified evolution retains linearity, trace preservation, and the composition law but is not positive on the full conventional state space.
Physical states are instead required to assign non-negative probabilities to every finite flash/no-flash history; the existence and characterization of a sufficiently rich non-vacuum physical state space remain open.
We also provide a modified, history-dependent collapse prescription formulated relative to a fixed initial Cauchy surface.
Moreover, given these modifications to unitary dynamics and collapse, we establish no-signaling by proving microscopic parameter independence~\cite[page 378]{Book_Tumulka2022Foundations}.

In the long-time regime and in the limit of weak qubit coupling and infinite CP density, the unmodified statistical map formally coincides with the statistical dynamics investigated in Ref.~\cite{Gundhi2026RelativisticCollapseStatisticalDynamics}.
Nevertheless, our dynamics differ from those of Ref.~\cite{Gundhi2026RelativisticCollapseStatisticalDynamics} due to our DVS modification, despite agreeing on the evolution of the expectation values of certain Hermitian observables, such as (normal-ordered) energy and particle number.
In this regime, we apply our framework to a free massive scalar field with a quadratic field coupling, obtaining finite leading-order particle-number and energy-increase rates for suitable smearing functions and initial states diagonal in particle number.
For standard positive states in this class, the leading one-point flash intensity is also non-negative, providing preliminary evidence for the viability of the framework.
We further identify conditions under which standard nonrelativistic collapse dynamics are recovered.

The paper is organized as follows.
Section~\ref{Sec:AharonovAlbertCollapse} reviews the Aharonov--Albert prescription, and Sec.~\ref{Sec:CPQM_ModelBuilding} introduces the relativistic CPQM construction.
Section~\ref{Sec:ModifyingQM} presents DVS, restricted positivity, the modified collapse rule, and the no-signaling argument.
Sections~\ref{Sec:CPQM_ModelPresentation}--\ref{Sec:FlashIntensities} derive the reduced matter dynamics, its PSL limit, and the flash intensities.
Section~\ref{Sec:KleinGordonApplication} applies the framework to a real scalar field.
Sections~\ref{Sec:Discussion} and~\ref{Sec:Conclusions} contain the discussion and conclusions.
Detailed derivations are collected in the appendices.

\section{The Aharonov-Albert relativistic measurement-induced collapse\label{Sec:AharonovAlbertCollapse}}

To properly understand the construction presented in this paper, we need to invoke the Aharonov-Albert prescription~\cite{Aharonov1980States,Aharonov1981RelativisticMeasurementProcess,Aharonov1984TimeRelativistic} for relativistic collapse of the wavefunction. We will mainly follow the presentation given in Chapter 7 of Ref.~\cite{Book_Tumulka2022Foundations}.
In the following, we set $c=\hbar=1$ and use the mostly minus sign convention, i.e., $\eta^{00} = 1$ and $\eta^{jj}=-1$.

Let us consider a Minkowski spacetime $\mcM$ and Cauchy surfaces\footnote{\textbf{Def. 7.3} of Ref.~\cite{Book_Tumulka2022Foundations}: Given a spacetime $\mcM$, a Cauchy surface $\Sigma$ is a subset of $\mcM$ which is intersected by every inextendible timelike-or-lightlike curve exactly once. In Minkowski spacetime, every Cauchy surface is spacelike but not every spacelike surface is a Cauchy surface. E.g., a hyperboloid is not a Cauchy surface.} $\Sigma$ over this spacetime. We assume that to each such surface we can associate a quantum state $\ket{\psi_\Sigma}$ and that, in the absence of measurements, these states evolve from one surface to another by means of the Tomonaga-Schwinger (TS) equation
\begin{equation}\label{eq:TomonagaSchwingerEquation}
\frac{\delta}{\delta \Sigma(x)}\ket{\Psi_\Sigma}
=
-i \hat\mcH(x)\ket{\Psi_\Sigma},
\end{equation}
where $\hat\mcH(x)$ is the interaction-picture Hamiltonian density and $\frac{\delta}{\delta \Sigma(x)}\ket{\Psi_\Sigma}$ denotes the change induced on $\ket{\Psi_\Sigma}$ by a normal (with respect to $\Sigma$) deformation of $\Sigma$. In order for the dynamics to be foliation-independent, one needs the TS evolution to respect the so-called integrability condition~\cite{Diosi2026CommentRelativisticCovarianceNonLinear}:
\begin{equation}\label{eq:IntegrabilityCondition}
\comm{\frac{\delta}{\delta \Sigma(x)}}{\frac{\delta}{\delta \Sigma(y)}}\ket{\Psi_\Sigma}=0,
\qq{for}
(x-y)^2 < 0,
\end{equation}
which is equivalent~\cite[Sec.~4.1]{Petrat2014MultiTimeQFT} to requiring that
\begin{equation}
\comm{\hat\mcH(x)}{\hat\mcH(y)}=0,
\qq{for}
(x-y)^2<0.
\end{equation}
The evolution operator between Cauchy surfaces satisfies the TS equation
\begin{equation}\label{eq:TomonagaSchwingerDifferentialEquation_UnitaryVersion}
\frac{\delta}{\delta \Sigma(x)} \hU_{\Sigma_0}^{\Sigma} = -i \hat\mcH(x) \hU_{\Sigma_0}^{\Sigma},
\end{equation}
with initial condition $\hU_{\Sigma_0}^{\Sigma_0}=\Id$.
The formal solution of the TS equation between two Cauchy surfaces $\Sigma_0$ and $\Sigma_1$ is given by
\begin{equation}\label{eq:FormalSolutionTS}
\hU_{\Sigma_0}^{\Sigma_1} = \mcT_{\hat\mcH} \exp{-i \int_{\Sigma_0}^{\Sigma_1} \dd[4]{x} \hat\mcH(x)},
\end{equation}
where $\mcT_{\hat\mcH}$ denotes the time-ordering of the interaction-picture Hamiltonian density $\hat\mcH(x)$.
By integrating over constant-time slices and recalling that the full interaction-picture Hamiltonian $\hH_I (t)$ is given by $\hH_I (t) =\int \dd[3]{\bx} \hat\mcH (t,\bx)$, one also gets the standard (interaction-picture) \Schr equation:
\begin{equation}\label{eq:StandardInteractionPictureDynamics}
\dv{t} \hU_{t_0}^{t} = - i \hH_I (t) \hU_{t_0}^{t}.
\end{equation}

Let us now add measurements and collapses to the dynamics. We consider $N$ idealized measurements $\hA_j$ (with $j=1,\dots,N$) being performed at spacetime points $X_1,\dots,X_N$. Let us consider a Cauchy surface $\Sigma$ passing through $X_j$ such that we know how to describe the action of the measurement of $\hA_j (\Sigma)$. This will usually be a constant-time hypersurface in a given Lorentz frame. In particular, the measurement $\hA_j$ will have possible results $\alpha_j$ and projection operators $\hP_{j,\alpha_j} (\Sigma)$ so that
\begin{equation}
\hA_j (\Sigma) = \sum_{\alpha_j} \alpha_j \hP_{j,\alpha_j} (\Sigma).
\end{equation}
The action of this measurement on any other surface $\Sigma'$ can be obtained by writing
\begin{equation}
\hA_j (\Sigma') = \hU_{\Sigma}^{\Sigma'} \hA_j (\Sigma) \hU_{\Sigma'}^{\Sigma}.
\end{equation}
Given the measurement outcomes $\alpha_1,\dots,\alpha_N$, and an initial wavefunction $\ket{\psi_{\Sigma_0}}$\footnote{It makes sense to consider $\Sigma_0$ to be in the past of all $X_j$ even though it is not necessary at the mathematical level.}, the wavefunction on any other surface can be computed as
\begin{equation}
\ket{\psi_\Sigma} = \mcN \hU_{\Sigma_0}^{\Sigma} 
\mcT_{\hP} \prtq{\prod_{j: X_j\in J^{-}(\Sigma)} \hP_{j,\alpha_j} (\Sigma_0)} \ket{\psi_{\Sigma_0}},
\end{equation}
where $\mcN$ is the normalization constant, $J^{-}(\Sigma)$ denotes the causal past\footnote{Notice that $\Sigma \subset J^{-} (\Sigma)$.} of $\Sigma$ and $\mcT$ is a time-ordering operator.
The product includes precisely those measurement events in the past of $\Sigma$. All projectors are represented on the same reference surface $\Sigma_0$ and $\mcT_{\hP}$ makes so that, in any given frame, they are time-ordered.
Then, the evolution operator $\hU_{\Sigma_0}^{\Sigma}$ transports the resulting state to $\Sigma$.
The above formula is foliation-independent because one assumes that $\hP_{\alpha_j} (\Sigma)$ and $\hP_{\alpha_k} (\Sigma)$ commute whenever $X_j$ and $X_k$ are spacelike.
The probability of obtaining a certain set of outcomes is instead given by
\begin{equation}\label{eq:AharonovAlbertTotalProbabilities}
\mbP (\alpha_1,\dots,\alpha_N|\psi_{\Sigma_0}) = \norm{\mcT \prtq{\prod_{j} \hP_{j,\alpha_j} (\Sigma_0)} \ket{\psi_{\Sigma_0}}}^2.
\end{equation}
The conditional probability of certain outcomes (say for $j=n+1,\dots,N$) in the future of $\Sigma$ given the outcomes in the past of it ($j=1,\dots,n$) is instead given by
\begin{equation}\label{eq:AharonovAlbertConditionalProbabilities}
\mbP (\alpha_{n+1},\dots,\alpha_N|\psi_{\Sigma})
=
\norm{\mcT \prtq{\prod_{j=n+1}^N \hP_{j,\alpha_j} (\Sigma)}\ket{\psi_\Sigma}}^2,
\end{equation}
where we recall that $\ket{\psi_\Sigma}$ also encodes the outcomes of measurements between $\Sigma_0$ and $\Sigma$, and  $\hP_{j,\alpha_j} (\Sigma) = \hU_{\Sigma_0}^{\Sigma} \hP_{j,\alpha_j} (\Sigma_0) \hU_{\Sigma}^{\Sigma_0}$.

A potential problem with the Aharonov--Albert proposal is that their construction relies on the seemingly unrealistic premise that measurements occur at specific spacetime points even though measurement apparatuses are macroscopic and realistic measurement procedures are not instantaneous. This is a problem that, as we will show, our proposal solves by taking this premise at face value and assuming that spontaneous collapses really do happen at precise points in spacetime.

\section{Construction of the model\label{Sec:CPQM_ModelBuilding}}

As anticipated in the introduction, the model presented here is based on the same premises as the framework proposed in Ref.~\cite{Piccione2024PoissonianCollapse}.
For the reader's convenience, however, this section will be self-contained so that the reader does not need to read Ref.~\cite{Piccione2024PoissonianCollapse}. 

We start by fixing some terminology and notation and presenting the ideas behind the model.
We will refer to this general model as the CPQM (Collapse Points Quantum Mechanics) model.
Later, we will focus on the PSL (see Sec.~\ref{Sec:EvolutionPSL_Limit}) limit of CPQM.

We assume that spacetime is randomly filled with special points that we call Collapse Points (CPs). These points are distributed according to a Poisson distribution with intensity proportional to the infinitesimal spacetime volume. Therefore, in every infinitesimal spacetime volume $\dd[4]{x}$ there is a probability $\mu_C \dd[4]{x}$ of finding one of these special points, where $\mu_C$ is the spacetime density of the CPs or, mathematically, the intensity of the homogeneous Poisson point process. In fact, this constitutes a so-called Poisson sprinkling~\cite{Review_Surya2019CausalSetApproach} and its constant density assures us that the distribution of such points is invariant under Poincaré transformations~\cite{Review_Surya2019CausalSetApproach,Bedingham2021Collapse}\footnote{Moreover, in a general Lorentzian manifold, the Poisson sprinkling is coordinate-invariant and diffeomorphism-invariant, a fact which has also been used to propose prototypical versions of spontaneous collapse models in curved spacetimes~\cite{Bedingham2016RelativisticCollapseBlackHole,Juarez2018CollapseModelsHadamardCondition}. Furthermore, Ref.~\cite{Bombelli2009DiscretenessWithoutSymmetryBreaking} argues that such a Poisson sprinkling is, in some sense, relativistically invariant for every realization of the process.}

To each of the CPs we associate an internal Hilbert space that we take to be that of a two-level system, i.e., a qubit. We will refer to the quantum system associated with a CP as a CP-qubit.
The initial state on a given Cauchy surface is, generally, an entangled state of standard quantum matter and CP-qubits.
Given a CP at spacetime point $x$, with $x=(t,\bx)$ in a given frame, the corresponding CP-qubit interacts with quantum matter by means of an interaction Hamiltonian $\sqrt{\Upsilon}\hH_{x} (t)$, where $\sqrt{\Upsilon}$ quantifies the coupling\footnote{The square root has been introduced for later convenience.}.
Denoting the spacetime location of the $n$-th CP by $x_n$, the full Hamiltonian of the model is then given by
\begin{equation}\label{eq:TotalHamiltonian}
\hH (t) = \hH_S + \sqrt{\Upsilon} \sum_n \hH_{x_n} (t),
\end{equation}
where $\hH_S$ is the standard Hamiltonian of system $S$, and we assume the following structure:
\begin{equation}\label{eq:InteractionHamiltonianStructure}
\hH_{x_n} (t) = \hL_{x_n} (t) \otimes \hX_{n}.
\end{equation}
In the above equation, $\hX_{n}$ is the standard Pauli operator $\hX = \dyad{0}{1} + \dyad{1}{0}$ acting on the $n$-th CP-qubit, and $\hL_{x_n} (t)$ is a time-dependent operator acting on the quantum matter Hilbert space. In particular, we assume the following form:
\begin{equation}\label{eq:RelativisticCollapseOperator}
\hL_x (t) = \int \dd[3]{\by} g (t-x^0;\by-\bx) \hQ (\by),
\end{equation}
where $\hQ (\by)$ is a Hermitian scalar local field operator, and $g(x)$ is a Lorentz-invariant real function such that\footnote{We recall that $\abs{x}:=\sign(x^2)\sqrt{\abs{x^\mu x_\mu}}$.}
\begin{equation}\label{eq:CondotionsOnSmearingFunction}
g(x)=
\begin{cases}
0, \qquad & x^0 > 0,\\
0, \qquad & x^2 = x^\mu x_\mu < 0,\\
f(\abs{x}), \qquad &\textrm{elsewhere}.
\end{cases}
\end{equation}
In other words, the interaction with a CP-qubit only takes place in the past light cone of the CP and with a coupling that only depends on the spacetime interval between that point and the CP. Therefore, moving to the interaction picture with respect to $\hH_S$ we can use the TS formalism of Eq.~\eqref{eq:TomonagaSchwingerEquation}, where $\ket{\Psi_\Sigma}$ is now the (interaction-picture) total state of quantum matter and CP-qubits associated with the Cauchy surface $\Sigma$. In this case, $\hat\mcH (x)$ is the interaction-picture Hamiltonian density given by
\begin{equation}\label{eq:HamiltonianDensity_CPQM}
\hat\mcH(x) = \sqrt{\Upsilon} \sum_n g(x-x_n) \hQ_I (x) \otimes \hX_n,
\end{equation}
where $\hQ_I (x):= e^{i \hH_S t}\hQ (\bx) e^{-i \hH_S t}$ and we assume it is a Lorentz-scalar operator.
Since the operator $\hQ(\bx)$ is a standard local Quantum Field Theory (QFT) operator, one has that 
\begin{equation}
\comm{\hat Q_I (x)}{\hat Q_I (y)}=0,
\qq{for}
(x-y)^2<0.
\end{equation}
Moreover, the $\hX_n$ commute with each other.
Hence, $[\hat\mcH(x),\hat\mcH(y)]=0$ for spacelike-separated $x$ and $y$, ensuring the integrability of the TS evolution and its foliation independence.

\begin{figure}[t]
\centering
\includegraphics[width=\linewidth]{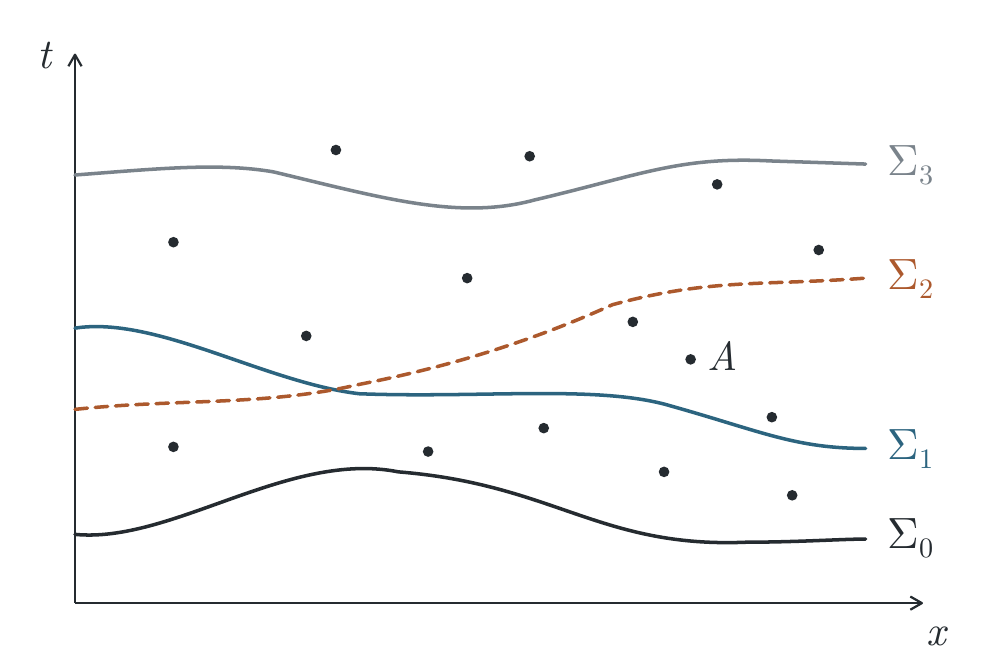}
\caption{State assignment on Cauchy surfaces. Dots denote Collapse Points (CPs), without specifying their flash/no-flash outcomes. Starting from the state on $\Sigma_0$, the state assigned to a later Cauchy surface is conditioned on the CP-qubit outcomes between the two surfaces. For example, $A$ lies to the future of $\Sigma_1$ and to the past of $\Sigma_2$: its outcome is included in the state assignment on $\Sigma_2$, but not on $\Sigma_1$.}
\label{fig:FlashDynamicsCauchySurfaces}
\end{figure}

To finalize the model, we have to add the spontaneous collapse element to it. To do this, we assume that each CP-qubit spontaneously collapses into $\ket{0}$ or $\ket{1}$ according to the Born rule whenever we consider a surface $\Sigma$ that is in the future of the related CP (see Fig.~\ref{fig:FlashDynamicsCauchySurfaces}). Thus, the collapse happens instantaneously on every Cauchy surface intersecting a given CP, following the relativistic collapse construction due to Aharonov and Albert~(see Sec.~\ref{Sec:AharonovAlbertCollapse}). If a CP-qubit located at $x_C$ spontaneously collapses to $\ket{1}$, we say, following standard terminology in the Foundations of Physics~\cite{Book_Tumulka2022Foundations}, that a \emph{flash} occurred at spacetime location $x_C$.
In the absence of matter, we intuitively expect absence of flashes. For this reason, we now also assume that, in any frame, as $t\to -\infty$ every CP-qubit is in the state $\ket{0}$.

Under the assumption that the underlying QFT is Poincaré invariant, the model we have presented is statistically Poincaré invariant by construction: initial states and Cauchy surfaces having the same description in their respective Lorentz frames yield the same laws for state evolution and flash histories when expressed in those frames.
The reason is that only two random processes are employed to build it. The first consists of the Poisson sprinkling which is known to be statistically Poincaré invariant~\cite{Bombelli2009DiscretenessWithoutSymmetryBreaking,Review_Surya2019CausalSetApproach}. The second stochastic process consists of the spontaneous measurement of the CP-qubits in the $\hZ$ basis\footnote{The operator $\hZ= \dyad{1}-\dyad{0}$ is one of the three standard Pauli operators for a two-level system.}. This spontaneous measurement process realizes without approximations the covariant measurement scheme proposed by Aharonov and Albert for relativistic measurements. Since the interaction of every CP-qubit with quantum matter is built to have the same form in every frame and the distribution of CPs is statistically Poincaré invariant, the distribution of flashes is also statistically Poincaré invariant.
This invariance concerns the stochastic law and does not require the initial state itself to possess Poincaré symmetry.

\section{Modifying quantum mechanics to obtain vacuum preservation\label{Sec:ModifyingQM}}

The relativistic CPQM model that we have presented in Sec.~\ref{Sec:CPQM_ModelBuilding} formally works but, as it stands, will inevitably produce particles from the vacuum when applied to QFT.
The reason for this is the following. Let us consider a bounded open region $\mcR$ of the spacetime $\mcM$. According to the Reeh--Schlieder theorem and its corollaries~\cite{Reeh1961ReehSchliederTheorem,Myrvold2017RelativisticMarkovian,Review_Witten2018EntanglementQFT,Fewster2020AQFTIntroduction}, any local operator\footnote{In this context, by a local operator $\hO(\mcR)$ we mean an operator localized in the spacetime region $\mcR$, e.g., an operator constructed from quantum fields smeared with functions supported in $\mcR$. Such operators satisfy microcausality by assumption: namely, they commute with operators localized in spacelike-separated regions.} $\hO(\mcR)$ that acts trivially on the vacuum is trivial, i.e., if a local operator $\hO (\mcR)$ is such that $\hO (\mcR) \ket{\Omega} = c \ket{\Omega}$, where $c$ is a complex number, then $\hO (\mcR) = c \Id$, where $\Id$ is the identity operator.
When $c=0$, we have that $\hO (\mcR) \ket{\Omega}=0$ implies that $\hO (\mcR) =0$.
In the limit in which $\mcR$ tends to a single point, we can heuristically replace $\hO (\mcR)$ with $\hO (x)$.
Thus, to have non-trivial dynamics, the action of the Hamiltonian density of Eq.~\eqref{eq:HamiltonianDensity_CPQM} cannot leave the vacuum invariant.
This is quite problematic because the spacetime density of particles that come out of the vacuum in a relativistically invariant theory is either zero or infinite~\cite{Pearle1999Relativistic,Pearle2015RelativisticCSL,Myrvold2017RelativisticMarkovian}.
Thus, a nontrivial local matter operator $\hQ_I(x)$ cannot be required to annihilate the QFT vacuum.\footnote{Ordinary local QFT fields have the same property, but this does not imply that ordinary time evolution creates particles from the vacuum: the physical vacuum is an eigenstate of the global Hamiltonian $\hH_S$.}

Vacuum instability is one of the sources of the divergences usually encountered in relativistic collapse theories, but it is not the only one.
In particular, Ref.~\cite{Gundhi2026RelativisticCollapseStatisticalDynamics} analyzed a Poincaré invariant statistical dynamics where the energy increase rate divergence due to vacuum instability was cured through a nonstandard normal ordering prescription.
Still, they also showed that a purely white noise in spacetime would have made the energy increase rate divergent anyway.
In the following, we will propose a solution to the vacuum stability problem, and we will then consider some consequences of our solution.

\subsection{Dynamical Vacuum Subtraction\label{SubSec:DVS}}

We propose the following solution to the above problem: to modify the basic dynamical law of quantum mechanics itself.
In particular, we propose to subtract the vacuum contribution at the level of the evolution map, hence the name Dynamical Vacuum Subtraction (DVS).
This may be considered analogous in spirit to vacuum subtraction (via normal ordering) in the definition of QFT observables, but here it modifies the basic dynamical law of quantum mechanics itself, not just the observable at the initial time.

In super-operator form, the TS dynamics of Eq.~\eqref{eq:FormalSolutionTS} applied to quantum matter plus CP-qubits is given by
\begin{equation}
\mcU_{\Sigma_0}^{\Sigma_1} 
=
\mcT_{\mcL} \exp{\int_{\Sigma_0}^{\Sigma_1} \dd[4]{x} \mcL_x},
\end{equation}
where $\mcL_x := -i [\hat \mcH (x),\bullet]$, and $\hat \mcH (x)$ is given in Eq.~\eqref{eq:HamiltonianDensity_CPQM}. However, we assume that the actual physical evolution of standard quantum matter plus CPs is not given by the above unitary super-operator but by a new super-operator $\tl{\mcU}_{\Sigma_0}^{\Sigma_1}$. To define this operator, we first introduce the states $\Omega_S$ and $\Omega_{\text{CP}}$.
The former, $\Omega_S$, represents the vacuum of the standard QFT we are working with, which we assume to be non-degenerate.
The latter, $\Omega_{\text{CP}}:= \bigotimes_n \dyad{0}_n$, is the state of all CP-qubits in their ground state. Thus, we define the vacuum of the CPQM model as $\Omega = \Omega_S \otimes \Omega_{\text{CP}}$. For a generic operator $\hA$, we define
\begin{equation}\label{eq:DVS_Definition}
\tl{\mcU}_{\Sigma_0}^{\Sigma_1} \hA =
\mcU_{\Sigma_0}^{\Sigma_1}\hA
+
\Tr{\hA}\prt{\Omega-\mcU_{\Sigma_0}^{\Sigma_1}\Omega}.
\end{equation}
In particular, given a physical state $\sigma_{\Sigma_0}$, its standard evolution would lead to $\sigma_{\Sigma} = \mcU_{\Sigma_0}^{\Sigma} \sigma_{\Sigma_0}$ while its physical evolution is given by $\tl{\sigma}_{\Sigma} = \tl{\mcU}_{\Sigma_0}^{\Sigma} \tl{\sigma}_{\Sigma_0}$, with $\tl{\sigma}_{\Sigma_0} = \sigma_{\Sigma_0}$. For an operator with unit trace, the evolution of Eq.~\eqref{eq:DVS_Definition} can be thought of as stemming from the following modified TS equation (see Appendix~\ref{APPSec:ModifiedUnitaryDynamics_MathProperties} for the derivation):
\begin{equation}\label{eq:DVS_DifferentialEquationEvolution}
\frac{\delta}{\delta \Sigma(x)} \tl{\sigma}_{\Sigma}
=
\mcL_x \prt{\tl{\sigma}_{\Sigma}-\Omega}
=
- i \comm{\hat\mcH(x)}{\tl{\sigma}_{\Sigma}-\Omega}.
\end{equation}

We investigate mathematical properties of the modified dynamics of Eq.~\eqref{eq:DVS_Definition} in Appendix~\ref{APPSec:ModifiedUnitaryDynamics_MathProperties}.
In the following, we list some relevant ones:
\begin{itemize}
\item 
$\tl{\mcU}_{\Sigma_0}^{\Sigma_1}$ preserves the vacuum by construction:
\begin{equation}
\tl{\mcU}_{\Sigma_0}^{\Sigma_1} \Omega =
\mcU_{\Sigma_0}^{\Sigma_1}\Omega
+
\Tr{\Omega}\prt{\Omega-\mcU_{\Sigma_0}^{\Sigma_1}\Omega}
=\Omega,
\end{equation}
and is equal to the identity when $\Sigma_1=\Sigma_0$.
\item 
$\tl{\mcU}_{\Sigma_0}^{\Sigma_1}$ is linear, trace-preserving, and (as one can see from direct computation) also respects the composition law
\begin{equation}
\tl{\mcU}_{\Sigma_1}^{\Sigma_2}\tl{\mcU}_{\Sigma_0}^{\Sigma_1}
=
\tl{\mcU}_{\Sigma_0}^{\Sigma_2},
\end{equation}
as the standard unitary super-operator $\mcU_{\Sigma_0}^{\Sigma_1}$ does. Moreover, this implies that 
\begin{equation}
\prt{\tl{\mcU}_{\Sigma_0}^{\Sigma_1}}^{-1} = \tl{\mcU}_{\Sigma_1}^{\Sigma_0},
\end{equation}
as for the standard unitary operator. If the unitary evolution already preserves the vacuum, i.e., $\mcU_{\Sigma_0}^{\Sigma_1} \Omega = \Omega$, then we have that $\tl{\mcU}_{\Sigma_0}^{\Sigma_1}=\mcU_{\Sigma_0}^{\Sigma_1}$, which is the case in standard QFT.
\item
The evolution given by $\tl{\mcU}_{\Sigma_0}^{\Sigma_1}$ can be seen as being generated by the super-operator $\tl{\mcL}_x := \mcL_x - \Tr{\bullet}\mcL_x\Omega$, i.e., the TS equation becomes
\begin{equation}
\frac{\delta}{\delta \Sigma(x)} \tl{\sigma}_{\Sigma}
=
\tl{\mcL}_x (\tl{\sigma}_{\Sigma}),
\end{equation}
and its solution is
\begin{equation}
\tl{\mcU}_{\Sigma_0}^{\Sigma_1} 
=
\mcT_{\tl{\mcL}} \exp{\int_{\Sigma_0}^{\Sigma_1} \dd[4]{x} \tl{\mcL}_x}.
\end{equation}
The super-operator $\tl{\mcL}_x$ is not local in the standard QFT sense. For example, it need not commute with $\mcL_y$ even when $(x-y)^2<0$. However, these super-operators commute among themselves at spacelike separation, so that they satisfy the super-operator version of the integrability condition of Eq.~\eqref{eq:IntegrabilityCondition}. Moreover, they do still satisfy some sort of weak locality condition in the sense that, considering a standard local QFT operator $\hB (y)$, one has that
\begin{equation}
(x-y)^2<0 \implies \tl{\mcL}_x^\dg \hB (y) = 0.
\end{equation}
\item The DVS evolution preserves norm-induced distances between states whenever the norm $\norm{\bullet}$ is unitarily invariant, i.e.,
\begin{equation}
\norm{\tl{\mcU}_{\Sigma_0}^{\Sigma}\prt{\tl{\sigma}_{\Sigma_0}-\tl{\sigma}'_{\Sigma_0}}}
=
\norm{\tl{\sigma}_{\Sigma_0}-\tl{\sigma}'_{\Sigma_0}}.
\end{equation}
This implies that the vacuum is not attractive in the DVS evolution.

\end{itemize}

\subsection{Positivity and restricted positivity\label{SubSec:RestrictedPositivity}}

One may worry that $\tl{\mcU}_{\Sigma_0}^{\Sigma_1}$ may not be completely positive. In fact, formally speaking, it is not even positive: if one takes the initial state $\sigma_{\Sigma_0} = \mcU_{\Sigma_1}^{\Sigma_0}\Omega$, then one has that
\begin{equation}
\tl{\mcU}_{\Sigma_0}^{\Sigma_1} \sigma_{\Sigma_0}
=
2\Omega - \mcU_{\Sigma_0}^{\Sigma_1} \Omega,
\end{equation}
which is not positive\footnote{For example, consider a state $\psi$ which is orthogonal to $\Omega$ but not to $\mcU_{\Sigma_0}^{\Sigma_1} \Omega$. Then one has that $\Tr \{\psi\ \tl{\mcU}_{\Sigma_0}^{\Sigma_1} \sigma_{\Sigma_0} \} < 0$.} when $\mcU_{\Sigma_0}^{\Sigma_1} \Omega \neq \Omega$. However, we have to consider why positivity is demanded in standard quantum theory and how this translates to the CPQM model. In standard quantum theory, every Hermitian observable corresponds to an observable that is, in principle (set aside superselection rules), measurable. In particular, considering all rank-one projectors (i.e., all possible states, and yes-or-no measurements), one requires the probability of finding the system in any given state to be between zero and one. In formulas, given that the state of the system is $\rho$ and we measure an arbitrary pure state $\ket{\psi}$, we have that
\begin{equation}
\Tr{\dyad{\psi}\rho}
=
\ev{\rho}{\psi}\geq 0.
\end{equation}
However, in the CPQM model there are no measurements.
It is, after all, a quantum theory without observers~\cite{Bell2004AgainstMeasurement}.
At the fundamental level, the only probabilities we are interested in and that have to be well-defined are all the possible joint probabilities for the CP-qubits to end up in $\ket{0}$ or $\ket{1}$ once the corresponding CP is in the past.
In other words, we are not interested in the positivity property but in what we could call a \emph{restricted positivity} condition.

Let us consider the simplest possible example of the above prescription: a 
two-level system in which the only possible measurement is the binary 
projective measurement $\{\dyad{0},\dyad{1}\}$ and whose Hamiltonian is 
$\hH=\omega(t)\hX$, where $\omega(t)$ is nonzero only between $t=0$ and $t=T$. 
The analogue of the CP measurement occurs at $t=T$. Clearly, $\ket{0}$ 
is not an eigenstate of $\hH$ whenever $\omega(t)\neq 0$ and is therefore not 
preserved by the standard dynamics, so that our prescription will 
genuinely modify the dynamics of such a system. A general state 
$\tl{\sigma}_0$ at the initial time $t=0$ can be written as
\begin{equation}
\tl{\sigma}_0
=
\frac{1}{2}\prt{
a_0 \Id + a_1 \hX + a_2 \hY + a_3 \hZ
},
\end{equation}
where all the $a_j$ are complex numbers. Notice that we are not assuming that 
$\tl{\sigma}_0$ is a standard density operator. If it were, we would have 
$a_0=1$, $a_1,a_2,a_3\in\mathbb{R}$, and
\begin{equation}
\abs{a_1}^2+\abs{a_2}^2+\abs{a_3}^2\leq 1.
\end{equation}
If the evolution were the standard one, we would get
\begin{multline}
\mcU_{0}^{T}\tl{\sigma}_0
=\\
\frac{1}{2}\prt{
a_0 \Id
+
a_1 \hX
+
(a_2 c_T-a_3 s_T)\hY
+
(a_2 s_T+a_3 c_T)\hZ
},
\end{multline}
where
\begin{equation}
c_T:=\cos(\Lambda_T),
\quad
s_T:=\sin(\Lambda_T),
\quad
\Lambda_T:=2\int_0^T\omega(t)\dd{t}.
\end{equation}
The modified dynamics instead gives
\begin{multline}
\tl{\mcU}_{0}^{T}\tl{\sigma}_0
=
\frac{1}{2}\prtq{
a_0 \Id
+
a_1 \hX
+
\prt{a_2 c_T-(a_3+a_0)s_T}\hY
}
\\
+
\frac{1}{2}
\prt{a_2 s_T+(a_3+a_0)c_T-a_0}\hZ.
\end{multline}
At time $t=T$, the quantity associated with the outcome $\ket{1}$ is
\begin{equation}
\Tr{\dyad{1}\tl{\mcU}_{0}^{T}\tl{\sigma}_0}
=
\frac{a_0 c_T+a_2 s_T+a_3 c_T}{2},
\end{equation}
while that associated with the outcome $\ket{0}$ is
\begin{equation}
\Tr{\dyad{0}\tl{\mcU}_{0}^{T}\tl{\sigma}_0}
=
a_0
-
\frac{a_0 c_T+a_2 s_T+a_3 c_T}{2}.
\end{equation}
For these quantities to define probabilities, they must be real, 
non-negative, and sum to one. Since their sum is equal to $a_0$, normalization 
immediately implies that $a_0=1$. We must also have that $\Im{a_2 s_T + (a_3+1)c_T}=0$. Finally, non-negativity requires
\begin{equation}
0
\leq
\Re\left\{
a_2 s_T+(a_3+1)c_T
\right\}
\leq
2.
\end{equation}
Notice that, in this example, the probabilities are completely independent of $a_1$, which can therefore be regarded as an operationally redundant degree of 
freedom with respect to the only allowed measurement, somewhat analogous to a 
gauge degree of freedom. Thus, the set of physical states is very different 
from the set of standard density operators.

Let us now consider the totality of CP-qubits, a sub-ensemble $S_N$ with $N$ CPs and an initial Cauchy surface $\Sigma_0$ in the past of this sub-ensemble. Let us denote by $\Pi_{\bF_{S_N}}$ a generic rank-one projector on their states (in the $\prtg{\ket{0},\ket{1}}$ basis).
Here, $\bF_{S_N}$ denotes the set of flashes within the $N$ CP-qubits.
For example, if $N=5$ with CP locations $(x_1,x_2,x_3,x_4,x_5)$ one could have $\bF_{S_N}=\prtg{x_2,x_4}$ and the projector would be $\Pi_{\bF_{S_N}} = \dyad{0,1,0,1,0}$.
The no-flash outcome is obtained by setting $\bF_{S_N} = \emptyset$.
Clearly, we have that $\sum_{\bF_{S_N}} \Pi_{\bF_{S_N}} = \Id$, which, due to the fact that $\tl{\mcU}$ is trace-preserving, automatically implies that all the possible probabilities sum up to one if and only if $\tl{\sigma}_{\Sigma_0}$ has trace one. We conclude that $\Tr \{\tl{\sigma}_{\Sigma_0}\}=1$ is a necessary condition for $\tl{\sigma}_{\Sigma_0}$ to be a physical state.
We therefore require that
\begin{equation}\label{eq:PhysicalStateCondition_DVS}
\forall\ \Sigma \succ \Sigma_0, S_N\subset J^{-}(\Sigma), \bF_{S_N},
\quad
\Tr{\Pi_{\bF_{S_N}}
\tl{\mcU}_{\Sigma_0}^{\Sigma}
\tl{\sigma}_{\Sigma_0}}
\geq 0,
\end{equation}
where $\Sigma' \succ \Sigma$ means that every point of $\Sigma'$ lies either on $\Sigma$ or in its future. In words, we require that every finite flash/no-flash history has a well defined probability to occur.

The set of physical states is not empty.
Indeed, $\Tr\{\Omega\}=1$ and $\tl{\mcU}_{\Sigma_0}^{\Sigma}\Omega=\Omega$, so that $\Tr\{\Pi_{\bF_{S_N}}\tl{\mcU}_{\Sigma_0}^{\Sigma}\Omega\}=\Tr\{\Pi_{\bF_{S_N}}\Omega\}$ equals one for the all-no-flash history and zero for every history containing at least one flash.
The more important question is whether this set contains a sufficiently rich family of non-vacuum states satisfying all these conditions at all orders; the leading-order results in Sec.~\ref{SubSec:KleinGordon_FirstOrderIntensities} do not by themselves establish this but make our approach promising.

A proper characterization of the set of physical states seems to be a formidable mathematical task that we leave for future work. One thing we can say, however, is the following. For a fixed initial Cauchy surface, CP configuration, interaction operator, smearing function, and underlying QFT, the resulting space of physical states is convex.
Indeed, if $\tl{\sigma}_{\Sigma_0}$ and $\tl{\sigma}'_{\Sigma_0}$ are physical states, then $p\tl{\sigma}_{\Sigma_0}+(1-p)\tl{\sigma}'_{\Sigma_0}$ is also a physical state for every $0\leq p\leq1$, since both trace normalization and all the restricted-positivity conditions in Eq.~\eqref{eq:PhysicalStateCondition_DVS} are preserved under convex combinations.

\subsection{Dynamical evolution with flashes\label{SubSec:CollapseUpdateRules_DVS}}

The DVS prescription allows to evolve a physical state from a Cauchy surface $\Sigma$ to a later Cauchy surface $\Sigma'$ while keeping the vacuum stable.
For a fixed CP configuration, let us denote by $\bF_{\Sigma}^{\Sigma'}$ the set of flashes between $\Sigma$ and $\Sigma'$.
Then, given an initial state $\tl{\sigma}_{\Sigma_0}$, one would like to also associate (as in Sec.~\ref{Sec:CPQM_ModelBuilding}) a collapsed state $\tl{\sigma}_{\Sigma_1}^{\bF}$ with the outcome $\bF_{\Sigma_0}^{\Sigma_1}$ through the Aharonov--Albert prescription.
However, having replaced the standard unitary evolution by the DVS evolution, we also need to examine the consistency of the collapse rule.

To investigate this problem, we now consider three Cauchy surfaces $\Sigma_0$, $\Sigma_1$, and $\Sigma_2$ with $\Sigma_2 \succ \Sigma_1 \succ \Sigma_0$. We can then give two different (but equally intuitive) formulas for the state $\tl{\sigma}_{\Sigma_2}^{\bF}$ given a certain flash history $\bF$ between $\Sigma_0$ and $\Sigma_2$. Unfortunately, as we will see, these two formulas need not give the same result.\footnote{For mathematical definiteness, the argument that follows may be understood for a finite set of CPs, or after restricting the spacetime region between the surfaces to a bounded region; the obstruction discussed below is already present in this finite setting.}
The first, obvious state update formula would be
\begin{equation}\label{eq:StateUpdateRuleV1_DVS}
\tl{\sigma}_{\Sigma_2}^{\bF}
=
\frac{\mcF_{\Sigma_0}^{\Sigma_2} \tl{\mcU}_{\Sigma_0}^{\Sigma_2} \tl{\sigma}_{\Sigma_0}}{\Tr{\mcF_{\Sigma_0}^{\Sigma_2} \tl{\mcU}_{\Sigma_0}^{\Sigma_2} \tl{\sigma}_{\Sigma_0}}},
\qquad
\mcF_{\Sigma}^{\Sigma'}:= \hat \Pi_{\bF_{\Sigma}^{\Sigma'}} \bullet \hat \Pi_{\bF_{\Sigma}^{\Sigma'}},
\end{equation}
while the second would be
\begin{equation}\label{eq:StateUpdateRuleV2_DVS}
\tl{\sigma}_{\Sigma_2}^{\bF}
=
\frac{\mcF_{\Sigma_1}^{\Sigma_2} \tl{\mcU}_{\Sigma_1}^{\Sigma_2} \tl{\sigma}^{\bF}_{\Sigma_1}}{\Tr{\mcF_{\Sigma_1}^{\Sigma_2} \tl{\mcU}_{\Sigma_1}^{\Sigma_2} \tl{\sigma}^{\bF}_{\Sigma_1}}}
=
\frac{\mcF_{\Sigma_1}^{\Sigma_2} \tl{\mcU}_{\Sigma_1}^{\Sigma_2}\mcF_{\Sigma_0}^{\Sigma_1} \tl{\mcU}_{\Sigma_0}^{\Sigma_1} \tl{\sigma}_{\Sigma_0}}{\Tr{\mcF_{\Sigma_1}^{\Sigma_2} \tl{\mcU}_{\Sigma_1}^{\Sigma_2}\mcF_{\Sigma_0}^{\Sigma_1} \tl{\mcU}_{\Sigma_0}^{\Sigma_1} \tl{\sigma}_{\Sigma_0}}},
\end{equation}
where we have defined $\tl{\sigma}_{\Sigma_1}^{\bF}$ similarly to $\tl{\sigma}_{\Sigma_2}^{\bF}$ in Eq.~\eqref{eq:StateUpdateRuleV1_DVS}.
Normally, the two equations above would be equal because $[\mcU_{\Sigma_1}^{\Sigma_2},\mcF_{\Sigma_0}^{\Sigma_1}]=0$ and $\mcF_{\Sigma_1}^{\Sigma_2}\mcF_{\Sigma_0}^{\Sigma_1}=\mcF_{\Sigma_0}^{\Sigma_2}$. However, because of the DVS modification, using Eq.~\eqref{eq:DVS_Definition} one gets that
\begin{multline}
\comm{\tl{\mcU}_{\Sigma_1}^{\Sigma_2}}{\mcF_{\Sigma_0}^{\Sigma_1}}\hA
=
\Tr{\mcF_{\Sigma_0}^{\Sigma_1}\hA}
\prt{\Omega-\mcU_{\Sigma_1}^{\Sigma_2}\Omega}
\\
-
\Tr{\hA}
\mcF_{\Sigma_0}^{\Sigma_1}
\prt{\Omega-\mcU_{\Sigma_1}^{\Sigma_2}\Omega},
\end{multline}
which is generally nonvanishing.

To solve the above problem, we propose the following modification: \emph{every projection and probability computation must be done with respect to the initial surface $\Sigma_0$.}
Indeed, such a rule gives a prominent role to $\Sigma_0$ and we will discuss some of the conceptual implications of this in Sec.~\ref{SubSec:Discussion_Collapse}.
Under the assumption that there is at least one flash between $\Sigma_0$ and $\Sigma_1$, we show (see Appendix~\ref{APPSec:CollapseRuleModification}) that the following holds:
\begin{equation}\label{eq:CompleteHistoryComposition_DVS}
\mcF_{\Sigma_0}^{\Sigma_2}
\tl{\mcU}_{\Sigma_1}^{\Sigma_2}
\mcF_{\Sigma_0}^{\Sigma_1}
\tl{\mcU}_{\Sigma_0}^{\Sigma_1}
=
\mcF_{\Sigma_0}^{\Sigma_2}
\tl{\mcU}_{\Sigma_0}^{\Sigma_2}.
\end{equation}
Thus, we now recover the equality between Eq.~\eqref{eq:StateUpdateRuleV1_DVS} and a modified version of Eq.~\eqref{eq:StateUpdateRuleV2_DVS} in which $\mcF_{\Sigma_1}^{\Sigma_2}$ is replaced by $\mcF_{\Sigma_0}^{\Sigma_2}$. The probability of the partial flash history between any two surfaces in the future of $\Sigma_0$ is then given by
\begin{equations}\label{eq:ModifiedProbabilityRule_DVS}
p \prt{\bF_{\Sigma}^{\Sigma'} \mid \bF_{\Sigma_0}^{\Sigma}}
&=
\Tr{\mcF_{\Sigma_0}^{\Sigma'} \tl{\mcU}_{\Sigma}^{\Sigma'} \tl{\sigma}_{\Sigma}^{\bF}},
\\
\tl{\sigma}_{\Sigma}^{\bF}
&=
\frac{\mcF_{\Sigma_0}^{\Sigma} \tl{\mcU}_{\Sigma_0}^{\Sigma} \tl{\sigma}_{\Sigma_0}}{\Tr{\mcF_{\Sigma_0}^{\Sigma} \tl{\mcU}_{\Sigma_0}^{\Sigma} \tl{\sigma}_{\Sigma_0}}},
\end{equations}
again under the assumption that there is at least one flash between $\Sigma_0$ and $\Sigma$. Moreover, for a realized flash history $\bF_{\Sigma_0}^{\Sigma'}$, the collapsed state is
\begin{equation}\label{eq:ModifiedCollapseRule_DVS}
\tl{\sigma}_{\Sigma'}^{\bF}
=
\frac{\mcF_{\Sigma_0}^{\Sigma'} \tl{\mcU}_{\Sigma}^{\Sigma'} \tl{\sigma}_{\Sigma}^{\bF}}{\Tr{\mcF_{\Sigma_0}^{\Sigma'} \tl{\mcU}_{\Sigma}^{\Sigma'} \tl{\sigma}_{\Sigma}^{\bF}}}
=
\frac{\mcF_{\Sigma_0}^{\Sigma'} \tl{\mcU}_{\Sigma_0}^{\Sigma'} \tl{\sigma}_{\Sigma_0}}{\Tr{\mcF_{\Sigma_0}^{\Sigma'} \tl{\mcU}_{\Sigma_0}^{\Sigma'} \tl{\sigma}_{\Sigma_0}}}.
\end{equation}
Of course, the above definitions make sense when the requirement of Eq.~\eqref{eq:PhysicalStateCondition_DVS} is satisfied.
Importantly, the distinguished role of $\Sigma_0$ does not introduce a preferred foliation: for fixed boundary surfaces, both the DVS evolution and the flash-history projector are independent of the interpolating foliation.

The assumption behind Eq.~\eqref{eq:ModifiedProbabilityRule_DVS} that at least one flash has occurred between $\Sigma_0$ and $\Sigma$ is only needed to define $p (\bF_{\Sigma}^{\Sigma'} \mid \bF_{\Sigma_0}^{\Sigma})$ in that precise way. The general formula is instead given by
\begin{equation}\label{eq:GeneralConditionalProbability_DVS}
p\left(
\bF_{\Sigma}^{\Sigma'}
\mid
\bF_{\Sigma_0}^{\Sigma}
\right)
=
\frac{p\left(
\bF_{\Sigma}^{\Sigma'},
\bF_{\Sigma_0}^{\Sigma}
\right)
}{p\left(
\bF_{\Sigma_0}^{\Sigma}
\right)}
=
\frac{
\Tr{
\mcF_{\Sigma_0}^{\Sigma'}
\tl{\mcU}_{\Sigma_0}^{\Sigma'}
\tl{\sigma}_{\Sigma_0}
}
}{
\Tr{
\mcF_{\Sigma_0}^{\Sigma}
\tl{\mcU}_{\Sigma_0}^{\Sigma}
\tl{\sigma}_{\Sigma_0}
}
},
\end{equation}
provided the denominator is nonzero. Moreover, even if the history between $\Sigma_0$ and $\Sigma$ contains only no-flash outcomes, the state $\tl{\sigma}_{\Sigma}^{\bF}$ can still be defined as in Eq.~\eqref{eq:ModifiedProbabilityRule_DVS} but it cannot be used to compute $p (\bF_{\Sigma}^{\Sigma'} \mid \bF_{\Sigma_0}^{\Sigma})$.
Thus, pure no-flash histories are probabilistically well defined.
What is lost before the first flash is only the possibility of reproducing the same conditional dynamics by restarting the DVS evolution autonomously from the intermediate state.
Once a flash has occurred, the projector $\mcF$ annihilates the vacuum, and the recursive representation of Eq.~\eqref{eq:ModifiedProbabilityRule_DVS} becomes available on all later Cauchy surfaces.

We point out that when $\tl{\mcU}=\mcU$, i.e., when the DVS correction vanishes, the collapse-rule modification also vanishes, and the collapse rule reduces to the standard one employed in Sec.~\ref{Sec:CPQM_ModelBuilding}. This is particularly relevant when considering the non-relativistic limit of our framework, in which it is expected that the DVS modification vanishes.

\subsection{No-signaling under external potentials
\label{SubSec:NoSignalingExternalPotentials}}

An important question is whether our proposal allows for superluminal signaling.
Even with the DVS modification, the dynamics generated by $\tl{\mcU}$ remains linear.
While linearity is usually considered a necessary condition for no-signaling~\cite{Gisin1989StochasticDynamics,Bassi2015LinearityNoSignaling},\footnote{There may be, however, loopholes~\cite{Kent2005NonlinearityWithoutSuperluminality,Kent2019SummoningNoSignalling,Kent2021QuantumStateReadout}.} it alone does not imply the absence of superluminal signaling, since linearity does not prevent a dynamical map from transmitting information between spacelike-separated regions~\cite{Beckman2001CausalLocalizableOperations,Borsten2021ImpossibleMeasurementsRevisited}.\footnote{A simple example is the following. Consider two qubits. At random times, a spontaneous projective measurement of the nonlocal observable $\hL=\dyad{0}_A\otimes\Id_B+\dyad{1}_A\otimes\dyad{1}_B$ occurs. Alice can modify Bob's qubit dynamics by choosing to prepare her qubit in $\dyad{0}_A$ or $\dyad{1}_A$.}

A way to argue that our framework does not allow for superluminal signaling consists of assuming that an agent can turn on an external potential in a bounded spacetime region $\mcB$. If doing this does not change the marginal distribution of flashes in any bounded region $\mcA$ which is spacelike separated from $\mcB$, we can say that no superluminal signaling is possible~\cite[p.~408]{Book_Tumulka2022Foundations}. This specific property is usually called ``microscopic parameter independence''~\cite[page 378]{Book_Tumulka2022Foundations} and is considered stronger than no-signaling, since no-signaling can hold without it.\footnote{For example, Bohmian mechanics violates it but still does not allow for superluminal signaling~\cite{Book_Tumulka2022Foundations}.}

Consider two bounded spacetime regions $\mcA$ and $\mcB$ at spacelike separation and lying in the future of $\Sigma_0$.
We compare the dynamics with and without a local external perturbation $\delta\hat{\mcH}_S(x)=v(x)\hR(x)$ supported in $\mcB$, where $v$ is real and $\hR$ is a local Hermitian operator of system $S$.
The external potential is included in the standard QFT evolution defining the interaction picture, while the reference state $\Omega$ is kept fixed.
We also assume an initial state $\tl{\sigma}_{\Sigma_0}$ that satisfies the restricted-positivity requirement for both situations. We denote the evolution super-operators in the presence of the external potential by $\mcU[v]$ and $\tl{\mcU}[v]$.

Let us now fix a realization of the CP configuration.
Let $\Pi_{\bF_\mcA}$ be the projector associated with a specific complete flash/no-flash pattern for the CPs in $\mcA$.
The marginal probability $p_v (\bF_\mcA)$ of the flash realization within $\mcA$ is given by
\begin{align}
p_v(\bF_\mcA)
&=
\Tr{\Pi_{\bF_\mcA}
\tl{\mcU}_{\Sigma_0}^{\Sigma}[v]
\tl{\sigma}_{\Sigma_0}}
\notag\\
&=
\Tr{\Pi_{\bF_\mcA}\Omega}
+
\Tr{\Pi_{\bF_\mcA}
\mcU_{\Sigma_0}^{\Sigma}[v]
(\tl{\sigma}_{\Sigma_0}-\Omega)},
\label{eq:FlashMarginalExternalPotential}
\end{align}
where $\Sigma$ lies to the future of both $\mcA$ and $\mcB$.
Because $\mcA$ and $\mcB$ are spacelike separated, we can choose an intermediate Cauchy surface $\Sigma_*$ lying after $\mcA$ and before $\mcB$.
The evolution from $\Sigma_0$ to $\Sigma_*$ is independent of $v$.
Moreover, for every CP $x_n\in \mcA$, the causal-support condition on $g$ implies
$g(x-x_n)=0$ throughout the region between $\Sigma_*$ and $\Sigma$.
Thus, the subsequent interaction Hamiltonian contains no operators acting on the CP-qubits in $\mcA$, and
\begin{equation}\label{eq:ExternalPotentialIdentities}
\mcU_{\Sigma_0}^{\Sigma_*}[v]
=
\mcU_{\Sigma_0}^{\Sigma_*}[0],
\qquad
\left(\mcU_{\Sigma_*}^{\Sigma}[v]\right)^\dg
\Pi_{\bF_\mcA}
=
\Pi_{\bF_\mcA}.
\end{equation}
Using the first identity and the composition law of the standard unitary evolution, we obtain
\begin{equation}\label{eq:UnitaryCompositionExternalPotential}
\mcU_{\Sigma_0}^{\Sigma}[v]=\mcU_{\Sigma_*}^{\Sigma}[v]\mcU_{\Sigma_0}^{\Sigma_*}[0].
\end{equation}
Substituting Eq.~\eqref{eq:UnitaryCompositionExternalPotential} into Eq.~\eqref{eq:FlashMarginalExternalPotential}, we move $\mcU_{\Sigma_*}^{\Sigma}[v]$ onto $\Pi_{\bF_\mcA}$ using its adjoint. The second identity in Eq.~\eqref{eq:ExternalPotentialIdentities} then removes this factor, yielding
\begin{align}\label{eq:MicroscopicParameterIndependence}
p_v(\bF_\mcA)
&=
\Tr{\Pi_{\bF_\mcA}\Omega}
+
\Tr{\Pi_{\bF_\mcA}
\mcU_{\Sigma_0}^{\Sigma_*}[0]
(\tl{\sigma}_{\Sigma_0}-\Omega)}
\notag\\
&=
p_0(\bF_\mcA).
\end{align}
This argument shows that flash probabilities before any Cauchy surface are independent of local external potentials supported after it, thus establishing microscopic parameter independence.

The equality in Eq.~\eqref{eq:MicroscopicParameterIndependence} holds before averaging over the CP configuration.
Since the Poisson sprinkling is independent of the external potential, averaging over the CP configurations preserves the equality. 
Consequently, the entire marginal flash distribution in $\mcA$ is independent of the potential in $\mcB$.
Therefore, microscopic parameter independence survives the averaging and our framework does not allow superluminal signaling.

\subsection{Evolution of expectation values\label{SubSec:ExpectationValuesEvolution_DVS}}

Let us consider an \enquote{observable}\footnote{The quotation marks have been inserted because associating Hermitian operators with observables is an operation that makes sense in standard quantum theory. In the approach proposed in this paper, standard quantum mechanical observations emerge from the patterns of the flashes.} $\hO$ of system $S$ with vanishing vacuum expectation, $\Tr \{ \hO\Omega_S \}=0$. Moreover, let us also assume that $[\hH_S,\hO]=0$. Such an observable may be the normal-ordered $\hH_S$ itself or another quantity that is globally conserved in standard QFT. In particular, let us consider an initial state at time $t=0$ and let us compute the expectation value of such an observable at time $t$. We get
\begin{equation}\label{eq:EvolutionExpectationValueNormalOrderedOperators}
\ev{\hO}_t
=
\Tr{\hO \tl{\mcU}_t \sigma_0}
=
\Tr{\hO \mcU_t \sigma_0} - \Tr{\hO\mcU_t\Omega},
\end{equation}
that is, we subtract from the expectation value the number we would obtain by evolving the vacuum.
Thus, whenever normal ordering the evolved observable amounts to subtracting its vacuum expectation value, our modified dynamics reproduces the expectation values obtained with the non-standard \enquote{normal ordering} prescription used in Ref.~\cite{Gundhi2026RelativisticCollapseStatisticalDynamics} to avoid the vacuum divergence.
In contrast, however, the DVS prescription is not imposed separately for each observable; it follows from a single modification of the dynamics and applies to every conserved quantity satisfying the assumptions above.

\subsection{Non-relativistic limit\label{SubSec:NonRelativisticLimit_DVS}}

In order to follow the DVS prescription, we had to abandon the general requirement of positivity taken for granted in standard quantum theory.
Fortunately, we expect reasonable choices of the collapse operators $\hat Q_I (x)$ in Eq.~\eqref{eq:HamiltonianDensity_CPQM} to lead to standard non-relativistic collapse model dynamics that are, instead, free of this problem.
For example, the typical collapse operator of the mass-proportional Continuous Spontaneous Localization (CSL) model is $\hat Q(\bx) = \hat \mu (\bx)$, where $\hat \mu (\bx)$ is the mass-density operator.
If, in the relativistic version, we choose $\hat Q_I (x) = \hat T^{\mu}_\mu (x)$, where $\hat T^{\mu\nu} (x)$ are the components of the energy--momentum tensor, and so $\hat T^{\mu}_\mu (x)$ is its trace, it follows that the non-relativistic limit of this operator should approximately be the mass-density operator.

Importantly, the mass density operator $\hat \mu (\bx)$, despite appearances, is not a local operator in the QFT sense.
For example, this is illustrated in Sec.~\ref{SubSec:KleinGordon_NonRelativisticLimit} for the case of a free scalar theory with \mbox{$\hQ_I(x)=\ :\!\hat\varphi^2(x)\!:/2$}.
Consequently, even though $\hmu(\bx)\ket{\Omega}=0$, this does not contradict the Reeh--Schlieder theorem corollary mentioned right before Sec.~\ref{SubSec:DVS}.
When considering $\hat Q(\bx) \to \hat \mu (\bx)$, the interaction preserves the vacuum and the DVS correction vanishes, so that $\tl{\mcU}=\mcU$.
One then recovers the non-Markovian dynamics analyzed in Ref.~\cite{Piccione2026NonMarkovianPSL} using Eq.~\eqref{eq:RelativisticCollapseOperator} with $\hat Q(\by)\to \hmu (\by)$.
The conditions for neglecting particle production and heating out of the low-energy sector are examined explicitly for the Klein--Gordon field example in Sec.~\ref{SubSec:KleinGordon_NonRelativisticLimit}.

We point out that the above considerations suggest that any standard quantum state of matter that is nonrelativistic should also belong (in some approximate sense) to the set of physical states. 
This is because, in the effective non-relativistic theory, the joint matter and CP-qubits evolution is unitary and the collapse rule reduces to the standard projection rule.
Therefore, when the initial state is a standard density operator, all flash probabilities will be (approximately) well defined.

\section{Quantum matter dynamics in the CPQM model\label{Sec:CPQM_ModelPresentation}}

We now start investigating the dynamics entailed by the CPQM model, following similar steps to those of Ref.~\cite{Piccione2026NonMarkovianPSL} for the non-relativistic non-Markovian CPQM model. In particular, we are interested in characterizing the average evolution of the state of standard quantum matter.
To do so, we will initially consider the unmodified dynamics described in Sec.~\ref{Sec:CPQM_ModelBuilding}.

\begin{figure}[t]
\centering
\includegraphics[width=\linewidth]{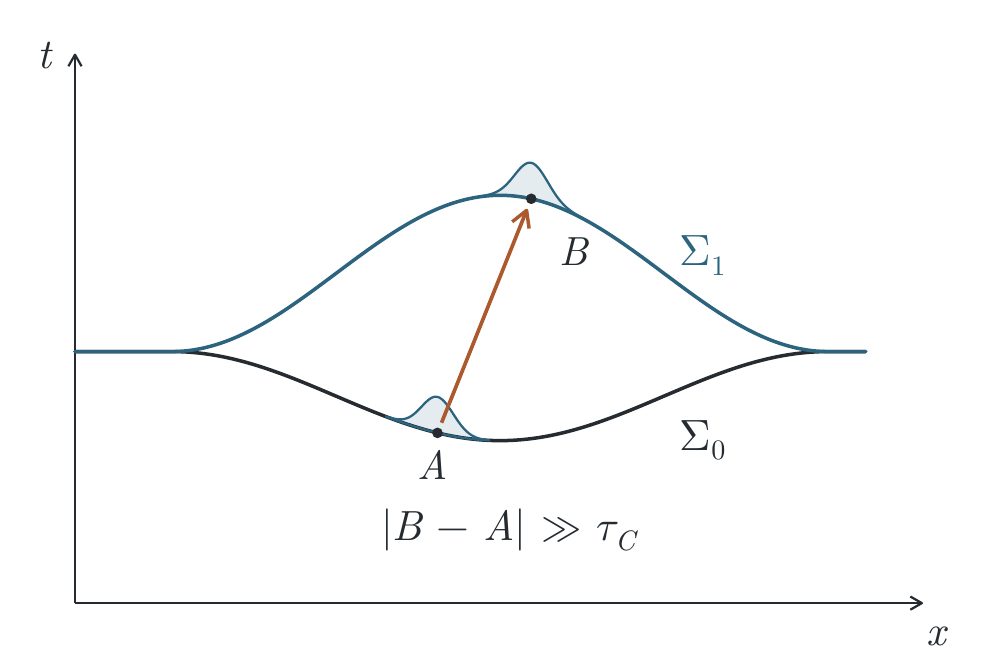}
\caption{Heuristic proper-time criterion for the late-time approximation. A massive one-particle wavepacket with mean momentum $\bk_0$ is localized near $A$ on $\Sigma_0$ on the spatial scales relevant to the diagram. The arrow represents its approximate classical trajectory, which intersects $\Sigma_1$ at $B$; the profiles schematically indicate the wavepacket near the two intersections. The proper time along this inertial segment is $\abs{B-A}$. The condition $\abs{B-A}\gg\tau_C$, with $\tau_C$ estimated in the particle's mean rest frame, motivates neglecting the initial matter--CP correlations and the use of $\Phi_{\Sigma_0}^{\Sigma_1}$.}
\label{fig:LongTimeEffectiveDynamicsCondition}
\end{figure}

First, we recall the definition of the state $\Omega_{\text{CP}} := \bigotimes_n \dyad{0}_n$, which is the state of all CP-qubits in their ground state. Any initial state $\sigma_\Sigma$ of matter and CP-qubits can be decomposed as
\begin{equations}
\sigma_\Sigma &= \rho_{\Sigma} \otimes \Omega_{\text{CP}}+\chi_\Sigma,
\\
\rho_{\Sigma} :&= \Tr_{\rm CP}\prtg{\sigma_\Sigma}
\\
\chi_\Sigma :&= \sigma_\Sigma - \rho_{\Sigma} \otimes \Omega_{\text{CP}},
\end{equations}
where $\rho_{\Sigma}$ is the reduced initial state of quantum matter and $\chi_\Sigma$ contains both matter--CP correlations and any departure of the CP marginal from $\Omega_{\text{CP}}$.
Since we do not know the disposition of the CPs in the future of $\Sigma$, nor the actual total initial state $\sigma_\Sigma$, one may wonder if $\rho_\Sigma$ is the initial state of quantum matter that one should use for making predictions.
However, the point is that we \emph{assume} that $\rho_\Sigma$ is the initial state of quantum matter so that whatever the disposition of the CPs and the initial state $\sigma_\Sigma$, one has that $\Tr_{\rm CP}\prtg{\sigma_\Sigma}=\rho_\Sigma$.
Therefore, for each CP configuration and an initial surface $\Sigma_0$, we consider a compatible total initial state $\sigma_{\Sigma_0}$ such that $\Tr_{\rm CP}\prtg{\sigma_{\Sigma_0}}=\rho_{\Sigma_0}$ and the corresponding $\chi_{\Sigma_0}$ generally depends on the CP configuration.
Within these specifications, the average final state of quantum matter $\rho_{\Sigma_1}$ on the surface $\Sigma_1 \succ \Sigma_0$\footnote{We recall that the symbol $\succ$ between two Cauchy surfaces indicates that every point of the first surface does not lie in the past of the second one. Intuitively, $\Sigma_1 \succ \Sigma_0$ means that $\Sigma_1$ comes after $\Sigma_0$ in a given foliation of spacetime.} will be given by
\begin{multline}\label{eq:TrueCPQMDynamicsWithCorrelations}
\rho_{\Sigma_1} = \mbE_{\rm CP}\prtq{\Tr_{\rm CP}\prtg{
\mcU_{\Sigma_0}^{\Sigma_1} (\rho_{\Sigma_0}\otimes \Omega_{\text{CP}})}}
\\+
\mbE_{\rm CP}\prtq{\Tr_{\rm CP}\prtg{\mcU_{\Sigma_0}^{\Sigma_1} \chi_{\Sigma_0} (\text{CP})}},
\end{multline}
where $\mbE_{\rm CP}$ denotes the average over the Poisson sprinkling.
In each realization, $\mcU_{\Sigma_0}^{\Sigma_1}$, $\Omega_{\text{CP}}$, and $\chi_{\Sigma_0}$ refer to that same CP configuration, and the CP trace is taken before averaging.
Equation~\eqref{eq:TrueCPQMDynamicsWithCorrelations} is an exact identity for the specified total initial states, but does not define a closed evolution from $\rho_{\Sigma_0}$ alone.
We will assume below that the contribution of $\chi_{\Sigma_0}$ becomes negligible at late times and focus on the quantum map defined by
\begin{equation}
\Phi_{\Sigma_0}^{\Sigma_1} \rho
:=
\mbE_{\rm CP}\prtq{\Tr_{\rm CP}\prtg{
\mcU_{\Sigma_0}^{\Sigma_1} (\rho \otimes \Omega_{\text{CP}})}}.
\end{equation}
In fact, as discussed in more detail in Ref.~\cite{Piccione2026NonMarkovianPSL}, if the interaction time between a CP-qubit and quantum matter is characterized by a timescale $\tau_C$, and one additionally assumes that the last term of Eq.~\eqref{eq:TrueCPQMDynamicsWithCorrelations} becomes negligible at late times, one obtains $\rho_{t_1} \simeq \Phi_{t_0}^{t_1} \rho_{t_0}$ for $(t_1-t_0)\gg \tau_C$, where here the symbols $t_j$ ($j=0,1$) stand for constant time hyperplanes.
In contrast to the non-relativistic model of Ref.~\cite{Piccione2026NonMarkovianPSL}, however, here we do not expect $\tau_C$ to be a state-independent quantity. In fact, taking as an example the case of a single massive particle whose state is well localized in momentum space, $\tau_C$ should be thought of as the typical interaction time between the particle and a CP-qubit in the frame in which the particle is, on average, at rest. In another frame, where the particle is instead moving at a relativistic speed of magnitude $v$, one expects the interaction time to be given by $\tau_C \gamma_v$, where $\gamma_v$ is the Lorentz factor. In any frame, however, one can estimate the largest relevant interaction time and use the map $\Phi_{\Sigma_0}^{\Sigma_1}$ with the volume enclosed between the two surfaces extending far enough in the relevant directions for the approximation to hold (see Fig.~\ref{fig:LongTimeEffectiveDynamicsCondition} for an illustrative example). Thus, the map $\Phi_{\Sigma_0}^{\Sigma_1}$ is an effective late-time expression. Equation~\eqref{eq:TrueCPQMDynamicsWithCorrelations} remains the exact reduced dynamics, and the contribution of $\chi_{\Sigma_0}$ cannot in general be neglected uniformly at short times~\cite{Piccione2026NonMarkovianPSL}.

To find a nicer expression for the map $\Phi_{\Sigma_0}^{\Sigma_1}$ we consider the following fact: tracing over the CP-qubits is equivalent to measuring them in any basis and averaging over the possible results. Thus, we can choose to \enquote{measure} the observable $\hX$ for each CP-qubit, the advantage being that $\forall x,n,\ [\hat\mcH(x),\hX_n]=0$, so that this measurement can be done before evolving the state. Starting from $\rho_\Sigma \otimes \Omega_{\text{CP}}$, this measurement gives $\pm 1$ for each CP-qubit with $50\%$ probability and all outcomes are independent of each other. Therefore, one discovers that
\begin{equation}\label{eq:CPQM_UnitaryUnraveling}
\Phi_{\Sigma_0}^{\Sigma_1}
=
\mbE_{\rm CP}\mbE_{X} \prtq{ \mcU_{\Sigma_0}^{\Sigma_1} (\textrm{CPs},X)},
\end{equation}
where $X$ represents the total string of outcomes of the measurement of $\hX$ for each CP-qubit and $\mcU_{\Sigma_0}^{\Sigma_1} (\textrm{CPs},X)$ is the unitary map acting on quantum matter according to the outcome $X$ and the CP locations, i.e., it is generated by the interaction Hamiltonian
\begin{equation}\label{eq:CPQM_UnitaryUnravelingHamiltonianDensity}
\hat\mcH^{(\textrm{CPs},X)} (x)  = \sqrt{\Upsilon} \sum_n X_n g(x-x_n) \hQ_I (x),
\end{equation}
where $x_n$ runs over all CP locations, each $X_n = \pm 1$ with $50\%$ probability, and all outcomes $X_n$ are independent of each other. The infinite family $X$ is well-defined; to handle the random sum, one can first consider only the CPs inside a finite spacetime volume and then take the limit as this volume tends to infinity. A more in-depth analysis of this procedure can be found in Ref.~\cite{Piccione2026NonMarkovianPSL}.

Given that the underlying QFT is Poincaré invariant, let us note that not only is the true dynamics of the relativistic CPQM model statistically Poincaré invariant, but the dynamics generated by $\Phi_{\Sigma_0}^{\Sigma_1}$ is also Poincaré invariant. In fact, the formal extended form of $\Phi_{\Sigma_0}^{\Sigma_1}$ is given by
\begin{multline}\label{eq:CQPM_UnitaryUnravelingExtended}
\Phi_{\Sigma_0}^{\Sigma_1}
=
\mbE_{\rm CP}\mbE_{X} \prtqB{
\\
\mcT_{\mcQ_I}\! \exp{-i\sqrt{\Upsilon}\sum_n \int_{\Sigma_0}^{\Sigma_1}\!\!\!\! \dd[4]{x} X_n g(x-x_n)\mcQ_I (x)}},
\end{multline}
where $\mcQ_I (x) := \comm{\hat Q_I (x)}{\bullet}$. As already pointed out, averaging over the CP locations is a statistically Poincaré invariant process, and the $X$ process is frame-independent. Hence, the stochastic Hamiltonian density transforms as a scalar in distribution and one concludes that, in any two different frames, the same statistical evolution would take place, i.e.,
\begin{equation}
\Phi_{\Sigma_0}^{\Sigma_1} = \Phi_{\Lambda\Sigma_0+a}^{\Lambda\Sigma_1+a},
\end{equation}
where $\Lambda$ denotes a proper orthochronous Lorentz transformation and $a$ denotes a spacetime translation.

Using the modified dynamics introduced in Sec.~\ref{Sec:ModifyingQM}, the physical map $\tl{\Phi}_{\Sigma_0}^{\Sigma_1}$ acts on an initial physical state $\tl{\rho}_{\Sigma_0}$ as follows
\begin{equations}
\tl{\Phi}_{\Sigma_0}^{\Sigma_1}\tl{\rho}_{\Sigma_0}
&=
\mbE_{\rm CP}\prtq{\Tr_{\rm CP}\prtg{
\tl{\mcU}_{\Sigma_0}^{\Sigma_1} (\tl{\rho}_{\Sigma_0} \otimes \Omega_{\text{CP}})}},
\\
&=
\Phi_{\Sigma_0}^{\Sigma_1}\tl{\rho}_{\Sigma_0}
+\Omega_S
-\Phi_{\Sigma_0}^{\Sigma_1}\Omega_S.
\end{equations}
Since $\Phi_{\Sigma_0}^{\Sigma_1}$ does not have the semi-group property, neither does its modified version. It is, however, trace preserving and linear as it inherits these properties from $\tl{\mcU}_{\Sigma_0}^{\Sigma_1}$. 
Moreover, assuming that $\Omega_S$ is Poincaré invariant in the underlying QFT,\footnote{For example, Poincaré invariance need not hold in the presence of an external potential.} the DVS-modified map $\tl{\Phi}_{\Sigma_0}^{\Sigma_1}$ inherits the same Poincaré invariance of $\Phi_{\Sigma_0}^{\Sigma_1}$. Applied to a generic operator that may  not have unit trace, the above equation becomes
\begin{equation}
\tl{\Phi}_{\Sigma_0}^{\Sigma_1} \hA
=
\Phi_{\Sigma_0}^{\Sigma_1}\hA +\Tr{\hA}\prt{\Omega_S - \Phi_{\Sigma_0}^{\Sigma_1}\Omega_S}.
\end{equation}

\section{The PSL limit\label{Sec:EvolutionPSL_Limit}}

An important physical regime is achieved by taking what we call the PSL limit:
\begin{equation}\label{eq:PSL_limit}
\Upsilon \to 0,
\quad 
\mu_C \to \infty,
\quad 
\Upsilon \mu_C \to \lambda >0.
\end{equation}
We start by finding the form that $\Phi_{\Sigma_0}^{\Sigma_1}$ assumes in the PSL regime. To do this, we again follow the route used in Ref.~\cite{Piccione2026NonMarkovianPSL} for the non-relativistic non-Markovian model. We divide spacetime into cells of spacetime volume $(\Delta r)^4$ and number the cells.
For the moment, we keep the size $(\Delta r)^4$ of these cells finite, but it will become infinitesimal in the end\footnote{Importantly, the limit $(\Delta r)^4 \to \dd[4]{r}$ is taken \emph{after} the limit $\mu_C \to \infty$.}. Thus, when $\mu_C \to \infty$, there will be a huge number of CPs within each cell, and we can also set $g(x-x_n) = g(x-x_m)$ when $x_n$ and $x_m$ belong to the same cell.
The sum of all outcomes of the $\hX$ measurement within a cell becomes a random walk variable in the limit of infinitely many steps, i.e., a Gaussian variable with zero mean and variance $\mu_C (\Delta r)^4$~\cite{Book_Oksendal2003StochasticDifferentialEquations,Schondorf2019WienerMeasureDonsker,Book_Durrett2019ProbabilityTheoryExamples}. So, our stochastic Hamiltonian density becomes
\begin{equation}\label{eq:StochasticHamiltonian_GaussianPSL}
\hat\mcH^{(\textrm{CPs},X)}(x)
\to
\hat\mcH^{(W)}(x) = \sqrt{\lambda} \sum_n \Delta W_n g(x-x_n)\hat Q_I (x)
\end{equation}
where
\begin{equation}
\Delta W_n = \sqrt{\Upsilon/\lambda}\sum_{m} X_m,
\qquad
\text{for $m\in n$-th cell},
\end{equation}
and the $\Delta W_n$ are uncorrelated Gaussian variables such that
\begin{equation}
\mbE_W [\Delta W_n]=0,
\qquad
\mbE_W [\Delta W_n \Delta W_m]= \delta_{n,m} (\Delta r)^4.
\end{equation}
We can now take the limit $\Delta r \to 0$ to rewrite the stochastic Hamiltonian. First, we write\footnote{$w(r):= \lim_{\Delta r\to 0} \Delta W_n/(\Delta r)^4$ is white noise in spacetime. Thus, one has $\mbE[w(x)w(y)]=\delta(x-y)$.}:
\begin{multline}
\sum_n \Delta W_n g(x-x_n)\hat Q_I (x)
=\\
=
\sum_n (\Delta r)^4 \frac{\Delta W_n}{(\Delta r)^4} g(x-x_n)\hat Q_I (x)
\to
\\
\to
\int \dd[4]{r} w(r) g(x-r)\hat Q_I (x)
=
\xi(x)\hat Q_I (x),
\end{multline}
where we defined the colored Gaussian noise $\xi(x):= \int \dd[4]{r} w(r) g(x-r)$, characterized by
\begin{equation}\label{eq:LorentzInvariantNoiseCharacterization}
\mbE[\xi (x)]=0,
\quad
\mbE[\xi (x) \xi (y)] = G(x-y),
\end{equation}
where
\begin{equations}\label{eq:NoiseCorrelationFunction}
G(x-y)
:&= \int \dd[4]{r} g (x-r) g (y-r),
\\
&= \int \dd[4]{r} g (r) g (r+(x-y)),
\end{equations}
i.e., $G(x-y)$ is the autocorrelation function of $g(x)$. Hence, in the PSL limit, the unitary unraveling of $\Phi_{\Sigma_0}^{\Sigma_1}$ is generated by
\begin{equation}\label{eq:StochasticHamiltonianUnraveling_PSL}
\hat\mcH^{(W)}(x) = \sqrt{\lambda}\xi(x)\hat Q_I (x),
\end{equation}
similarly to the relativistic dynamics studied in Ref.~\cite{Gundhi2026RelativisticCollapseStatisticalDynamics}.
When considering the map $\Phi_{t_0}^{t_1}$ between constant time hyperplanes, we have that~\cite{Diosi2014NonMarkovianGaussianDynamics,Piccione2026NonMarkovianPSL}
\begin{equation}\label{eq:DynamicalMapLongTime}
\Phi_{t_0}^{t_1}
=
\mcT_{\mcQ_I} \exp{-\frac{\lambda}{2}\int_{t_0}^{t_1}\!\! \dd[4]{y}\dd[4]{z} G(y-z) \mcQ_I (y) \mcQ_I (z)},
\end{equation}
where $\mcQ_I (y) := [\hat Q_I (y),\bullet]$, and $\int_{t_0}^{t_1} \dd[4]{y}\dd[4]{z}$ means that both $y^0$ and $z^0$ range between $t_0$ and $t_1$. As shown in Appendix~\ref{APPSec:ProofReducedDynamicsFormulaBetweenCurvedSurfaces}, the above result can be generalized to the evolution between arbitrary Cauchy surfaces:
\begin{equation}\label{eq:GeneralFormalMapTomonagaSchwinger}
\Phi_{\Sigma_0}^{\Sigma_1}
=
\mcT_{\mcQ_I} \exp{-\frac{\lambda}{2}\int_{\Sigma_0}^{\Sigma_1}\!\! \dd[4]{y}\dd[4]{z} G(y-z) \mcQ_I (y) \mcQ_I (z)}.
\end{equation}
Considering now the correction coming from the modifications to the underlying matter-qubits dynamics discussed in Sec.~\ref{Sec:ModifyingQM}, we get the new map
\begin{equation}\label{eq:ModifiedLongTimeStatisticalEvolution}
\tl{\Phi}_{\Sigma_0}^{\Sigma_1}
=
\Phi_{\Sigma_0}^{\Sigma_1} +\Tr{\bullet}\prt{\Omega_S - \Phi_{\Sigma_0}^{\Sigma_1}\Omega_S}, 
\end{equation}
where now $\Phi_{\Sigma_0}^{\Sigma_1}$ is given by Eq.~\eqref{eq:GeneralFormalMapTomonagaSchwinger}. We can also find its adjoint by demanding that $\Tr \{\hO \tl{\Phi}\hA\}=\Tr\{\hA \tl{\Phi}^\dg \hO\}$ for any $\hA$ and $\hO$:
\begin{equation}\label{eq:ModifiedLongTimeStatisticalEvolution_Adjoint}
[\tl{\Phi}_{\Sigma_0}^{\Sigma_1}]^\dg \hO 
=
[\Phi_{\Sigma_0}^{\Sigma_1}]^\dg\hO +\Tr{\Omega_S\prt{1-[\Phi_{\Sigma_0}^{\Sigma_1}]^\dg}\hO}\Id,
\end{equation}
where
\begin{equation}
[\Phi_{\Sigma_0}^{\Sigma_1}]^\dg
\!=\!
\bar{\mcT}_{\mcQ_I} \exp{-\frac{\lambda}{2}\!\int_{\Sigma_0}^{\Sigma_1}\!\! \dd[4]{y}\dd[4]{z} G(y-z) \mcQ_I (y) \mcQ_I (z)\! },
\end{equation}
and $\bar{\mcT}$ denotes the anti-chronological time-ordering.

\section{Flash intensities\label{Sec:FlashIntensities}}

Assuming that the initial state of quantum matter and CP-qubits is factorized and that the CP-qubits are in the initial state $\Omega_{\rm CP}$ allowed us to derive the formal CPTP map $\Phi_{\Sigma_0}^{\Sigma_1}$ on the state of quantum matter and its modified version of Eq.~\eqref{eq:ModifiedLongTimeStatisticalEvolution}. In the non-relativistic model of Ref.~\cite{Piccione2026NonMarkovianPSL}, we derived the long-time flash intensities for flashes located far in the future of the initial constant-time hyperplane. We follow similar steps here for the relativistic model.

Given the initial matter state $\tl{\rho}_{\Sigma_0}$ at the Cauchy surface $\Sigma_0$, and assuming the dynamics is given by Eq.~\eqref{eq:ModifiedLongTimeStatisticalEvolution}, we are interested in finding $\mu_F^{(n)}(x_1,\dots,x_n|\tl{\rho}_{\Sigma_0})$, i.e., the $n$-point joint intensity density of the flash process. Thus,
$\mu_F^{(n)}(x_1,\dots,x_n|\tl{\rho}_{\Sigma_0})\dd[4]{x_1}\cdots\dd[4]{x_n}$
is the probability, to leading order in the infinitesimal spacetime volumes, that
one flash occurs in each of the neighborhoods $\dd[4]{x_i}$ of the points $x_i$, irrespective of possible flashes elsewhere. Unlike a Poisson point process, the flash process is not entirely determined by its
first-order intensity density $\mu_F^{(1)}(x|\tl{\rho}_{\Sigma_0})$.\footnote{For a Poisson
point process, all factorial moment densities factorize into products of the
first-order intensity density. In our case this is generally not true. For
instance, if a single particle is prepared in a superposition of two distant
wave packets, the occurrence of a flash near one packet changes the conditional
state and thereby suppresses the probability of subsequent flashes near the
other packet. Hence, in general,
$\mu_F^{(2)}(x_1,x_2|\tl{\rho}_{\Sigma_0})\neq
\mu_F^{(1)}(x_1|\tl{\rho}_{\Sigma_0})\mu_F^{(1)}(x_2|\tl{\rho}_{\Sigma_0})$.}
A complete characterization requires the full set of joint
intensity densities $\mu_F^{(n)}(x_1,\ldots,x_n|\tl{\rho}_{\Sigma_0})$, for all $n\geq 1$. In Appendix~\ref{APPSec:PointProcessMathematics}, we summarize why knowledge of all the $\mu_F^{(n)}(x_1,\ldots,x_n|\tl{\rho}_{\Sigma_0})$ completely characterizes the stochastic flash process.

In order to derive a formula for $\mu_F^{(n)}(x_1,\dots,x_n|\tl{\rho}_{\Sigma_0})$, we start by assuming a given configuration of CPs and select $n$ of them at locations $x_1,\dots,x_n$. We introduce the notation $\ket{\mathbf{1}_n}:=\ket{1}_1\cdots\ket{1}_n$ and similarly for $\ket{\mathbf{0}_n}$. Let $\Sigma_f$ be any Cauchy surface lying in the future of all the selected CPs. Because the selected CP-qubits are then decoupled, evolution beyond $\Sigma_f$ acts only on the remaining degrees of freedom and cancels from the trace defining their joint outcome probability. We then use $+\infty$ as shorthand for such a final surface; the probabilities are independent of its choice. Then, the conditional probability of flashes occurring at those $n$ points is given by
\begin{multline}
\mbP_{\rm F} (x_1,\dots,x_n|\CP,\tl{\rho}_{\Sigma_0})
=\\
\Tr{\dyad{\mathbf{1}_n} \tl{\mcU}_{\Sigma_0}^{{+\infty}}\prt{\tl{\rho}_{\Sigma_0}\otimes\Omega_{\rm CP}}}.
\end{multline}
The above probability is obtained for a given configuration of CPs, but one should average it over all the CP configurations in which there are CPs at the designated points.
Since the positions of the other CPs in a Poisson sprinkling are completely uncorrelated with those of the fixed ones, we can directly focus on
\begin{multline}
\mbP_{\rm F} (x_1,\dots,x_n|\tl{\rho}_{\Sigma_0})
=\\
\Tr{\dyad{\mathbf{1}_n} \mbE_{\CP}\prtq{\tl{\mcU}_{\Sigma_0}^{{+\infty}}\prt{\tl{\rho}_{\Sigma_0}\otimes\Omega_{\rm CP}}}}.
\end{multline}
Here, $\mbE_\CP$ is the average with respect to all the CPs except those at points $x_1,\dots,x_n$. As in Eqs.~\eqref{eq:CPQM_UnitaryUnraveling} and~\eqref{eq:CPQM_UnitaryUnravelingHamiltonianDensity}, we can unravel the standard dynamics in terms of a new stochastic Hamiltonian density, but this time we must explicitly retain the evolution of the CP-qubits at points $x_1,\dots,x_n$. Thus, we get
\begin{equation}\label{eq:FlashHamiltonianDensityFiniteUpsilon}
\hat\mcW_n (x)
=
\sqrt{\Upsilon} \hat\mcH_n (x) + \hat\mcH^{(\textrm{CPs},X)} (x),
\end{equation}
where $\hat\mcH^{(\textrm{CPs},X)} (x)$ is the same as in Eq.~\eqref{eq:CPQM_UnitaryUnravelingHamiltonianDensity}\footnote{With the sum starting from $n+1$, if one wants to be pedantic.} and
\begin{equation}\label{eq:FixedCPHamiltonianDensity}
\hat\mcH_n (x) := \sum_{j=1}^{n} g(x-x_j) \hQ_I (x) \hX_j.
\end{equation}
To take the PSL limit as in Sec.~\ref{Sec:EvolutionPSL_Limit}, we replace the unselected CP background by its Gaussian limit, temporarily retaining $\Upsilon$ for the selected qubits as a bookkeeping parameter. Thus,
\begin{equation}\label{eq:FlashHamiltonianDensityPSL}
\hat\mcW_n (x)
=
\sqrt{\Upsilon} \hat\mcH_n (x) + \sqrt{\lambda}\xi(x)\hQ_I (x).
\end{equation}
Using $\Omega_{\rm CP}\ket{\mathbf{1}_n}=0$, and defining
\begin{equation}\label{eq:FlashKrausOperatorFiniteUpsilon}
\hK_{n,\Upsilon}^{(\xi)}
:=
\Upsilon^{-n/2}
\bra{\mathbf{1}_n}
[\hU_n^{(\xi)}]_{\Sigma_0}^{{+\infty}}
\ket{\mathbf{0}_n},
\end{equation}
where $[\hU_n^{(\xi)}]_{\Sigma_0}^{{+\infty}}$ is the TS evolution operator generated on quantum matter and the $n$ selected CP-qubits by $\hat\mcW_n(x)$ for a given realization of the Gaussian noise $\xi(x)$, we can write
\begin{equation}\label{eq:FixedPointFlashProbability}
\mbP_{\rm F} (x_1,\dots,x_n|\tl{\rho}_{\Sigma_0})
\!=\!
\Upsilon^n \mbE_\xi\! \Tr_S\! \prtg{\hK_{n,\Upsilon}^{(\xi)}\! \prt{\tl{\rho}_{\Sigma_0}-\Omega_S}\!\hK_{n,\Upsilon}^{(\xi)\dg}}.
\end{equation}

To obtain the intensity $\mu_F^{(n)}(x_1,\dots,x_n|\tl{\rho}_{\Sigma_0})$, we multiply $\mbP_{\rm F} (x_1,\dots,x_n|\tl{\rho}_{\Sigma_0})$ by the $n$-th order intensity $\mu_C^n$ of the Poisson sprinkling. In Ref.~\cite{Piccione2026NonMarkovianPSL}, the corresponding PSL limit is derived for the non-relativistic model. We adapt that calculation to the present TS formulation in Appendix~\ref{APPSec:FlashIntensityDetailedCalculation}. The result is
\begin{multline}\label{eq:FlashIntensityPSL}
\mu_F^{(n)}(x_1,\dots,x_n|\tl{\rho}_{\Sigma_0})
=\\
\lambda^n \mbE_\xi \Tr_S \prtg{\hK_n^{(\xi)} \prt{\tl{\rho}_{\Sigma_0}-\Omega_S}\hK_n^{(\xi)\dg}},
\end{multline}
where $\hK_n^{(\xi)}:=\lim_{\Upsilon\to 0}\hK_{n,\Upsilon}^{(\xi)}$ is given by
\begin{multline}\label{eq:FlashKrausOperatorPSL}
\hK_n^{(\xi)}
=
\sum_{m=n}^{\infty}
\frac{(-i)^m \lambda^{(m-n)/2}}{m!}
\int_{\Sigma_0}^{{+\infty}}\! \dd[4]{y_1} \cdots \dd[4]{y_m}
\\
\times
\hat\mcM_{n,m}^{(\xi)} (y_1,\dots,y_m),
\end{multline}
with\footnote{The operator $\hat\mcM_{n,m}^{(\xi)}$ depends on the realization of $\xi$ and on the fixed flash locations $x_1,\dots,x_n$, but contains no factor of $\lambda$ after the $\Upsilon\to 0$ limit. Its combinatorial structure is made explicit in Appendix~\ref{APPSec:FlashIntensityDetailedCalculation}.}
\begin{multline}\label{eq:FlashKernelPSL}
\hat\mcM_{n,m}^{(\xi)} (y_1,\dots,y_m)
:=
\lim_{\Upsilon\to 0}
\frac{1}{\Upsilon^{n/2}\lambda^{(m-n)/2}}
\\
\times
\bra{\mathbf{1}_n}\mcT_{\hat\mcW_n}
\prtg{\prod_{j=1}^{m}\hat\mcW_n(y_j)}\ket{\mathbf{0}_n},
\qquad m\geq n.
\end{multline}

\section{Applications to a real scalar field\label{Sec:KleinGordonApplication}}

In this section, we apply our proposal (in the PSL limit) to the simplest QFT, i.e., the QFT of a free real massive scalar field with Hamiltonian density
\begin{equation}\label{eq:KleinGordonHamiltonianDensity}
\hat \mcH (x) = \frac{1}{2}\hat \pi^2 (x) + \frac{1}{2}\prt{\nabla \hat \varphi (x)}^2 + \frac{1}{2} m^2 \hat\varphi^2 (x),
\end{equation}
where $\hat\varphi (x)$ is the field operator, $\hat \pi (x)$ is the conjugate momentum field satisfying $[\hat\varphi (t,\bx),\hat \pi (t,\by)] = i \delta^{(3)}(\bx -\by)$, and $m>0$ denotes the mass of the bosons. Moreover, following the example of Ref.~\cite{Gundhi2026RelativisticCollapseStatisticalDynamics}, we choose \mbox{$\hQ_I(x)=\ :\!\hat\varphi^2(x)\!:/2$}.

Before starting our analysis, let us make a remark.
Even though our modified dynamics and our interpretation of it give exact meaning only to the flash probabilities, in this section we will compute expectation values for various operators, such as those for total energy and particle number. The expectation is that these calculations still give information about properties of the systems even though one cannot properly associate well-defined probability distributions with the projectors composing any given Hermitian operator.

\subsection{Bogoliubov transformation, particle creation and momentum transfer\label{SubSec:BogoliubovTransformation}}

We start by choosing a Lorentz frame and we fix the initial state $\tl{\rho}_0$ associated with the constant-time hyperplane with $t=0$. Thus, for every realization of the noise $\xi$ in Eq.~\eqref{eq:StochasticHamiltonianUnraveling_PSL}, the unitary operator evolving the system is
\begin{equation}
\hU_\xi (t) = \mcT_{\hQ} \exp{-i\sqrt{\lambda}\int_0^t \dd[4]{x} \xi(x) \hQ_I (x)}.
\end{equation}
Since the generator of the unitary transformation is quadratic in the free field operators, the evolution induces a so-called Bogoliubov transformation~\cite[Sec.~II.A.2]{Review_Weedbrook2012GaussianQuantumInformation} on the creation ($\ha_\bk^\dg$) and annihilation operators ($\ha_\bk$). In Appendix~\ref{APPSec:EnergyParticleNumberIncreaseBogoliubov}, we perform all calculations inside a cubic box with periodic boundary conditions. Hereafter, we report the results in the continuum limit. The annihilation and creation operators obey
\begin{equations}\label{eq:BogoliubovTrasformations}
\hU_\xi^\dg (t) \ha_\bp \hU_\xi (t)
&=
\int \dd[3]{\bk} \prtq{\alpha_\xi (\bp,\bk;t)\ha_\bk + \beta_\xi (\bp,\bk;t)\ha^\dg_\bk},
\\
\hU_\xi (t) \ha_\bp \hU_\xi^\dg (t)
&=
\int \dd[3]{\bk} \prtq{\alpha_\xi^* (\bk,\bp;t)\ha_\bk - \beta_\xi (\bk,\bp;t)\ha^\dg_\bk},
\end{equations}
where the operators $\ha_\bk$ and $\ha^\dg_{\bp}$ satisfy the commutation relation $[\ha_\bk,\ha_\bp^\dg] = \delta^{(3)} (\bk-\bp)$.
For each realization of $\xi (x)$, $\alpha_\xi (\bp,\bk;t)$ and $\beta_\xi (\bp,\bk;t)$ are complex-valued distributions equivalent to the Bogoliubov coefficients in the discrete setting. They satisfy the Bogoliubov relations reported in Appendix~\ref{APPSec:EnergyParticleNumberIncreaseBogoliubov}. In particular, at $t=0$ one has that $\alpha_\xi (\bp,\bk;0) = \delta^{(3)} (\bp-\bk)$ and $\beta_\xi (\bp,\bk;0)=0$, and, at first order in $\sqrt{\lambda}$, they are given by
\begin{equations}\label{eq:FirstOrderBogoliubovCoefficients}
\alpha_\xi (\bp,\bk;t) 
&\simeq
\delta^{(3)}(\bp-\bk)
-\frac{i\sqrt{\lambda}}{2(2\pi)^3} \int_0^t \dd[4]{x}
\xi (x) \frac{e^{i(p-k)\cdot x}}{\sqrt{\omega_\bp \omega_\bk}},
\\
\beta_\xi (\bp,\bk;t)
&\simeq
-\frac{i\sqrt{\lambda}}{2(2\pi)^3} \int_0^t \dd[4]{x}
\xi (x) \frac{e^{i(p+k)\cdot x}}{\sqrt{\omega_\bp \omega_\bk}},
\end{equations}
where we point out that $p = (\omega_\bp,\bp)$, $\omega_\bp = \sqrt{m^2 + \bp^2}$, and similarly for $k$ and $\omega_\bk$.

Given an initial state $\tl{\rho}_0$ and a realization $\xi (x)$ of the stochastic process, the state at an infinitesimal time $\delta t>0$ is given by
\begin{equation}
\tl{\rho}_{\delta t}
\simeq
\tl{\rho}_0 -i \comm{\hK_0}{\tl{\rho}_0}
-i \comm{\hK_+ +\hK_-}{\tl{\rho}_0-\Omega_S},
\end{equation}
where
\begin{equations}
\hK_0 &=
\frac{\sqrt{\lambda}}{2(2\pi)^3}
\int \dd[3]{\bp}\dd[3]{\bk} \int_{0}^{\delta t} \dd[4]{x}\xi(x)\frac{e^{i(p-k)\cdot x}}{\sqrt{\omega_\bp\omega_\bk}}\ha_\bp^\dg \ha_\bk,
\\
\hK_+ &=
\frac{\sqrt{\lambda}}{4(2\pi)^3}
\int \dd[3]{\bp}\dd[3]{\bk} \int_{0}^{\delta t} \dd[4]{x}\xi(x)\frac{e^{i(p+k)\cdot x}}{\sqrt{\omega_\bp\omega_\bk}}\ha_\bp^\dg \ha_\bk^\dg,
\end{equations}
and $\hK_- = \hK_+^\dg$. Indeed, one has that $[\hK_0,\Omega_S]=0$. The term $\hK_0$ creates and destroys excitations and is therefore responsible for momentum changes of particles. On the other hand, $\hK_\pm$ destroy and create pairs of particles. Moreover, the infinitesimal evolution applied to an annihilation operator gives
\begin{equation}
\hU_\xi^\dg (\delta t) \ha_\bp \hU_\xi (\delta t)
\simeq
\ha_\bp 
+ i\comm{\hK_0}{\ha_\bp}
+ i\comm{\hK_+}{\ha_\bp},
\end{equation}
where $[\hK_0, \ha_\bp]$ is a new annihilation operator while $[\hK_+,\ha_\bp]$ is a creation operator. Thus, by making a comparison with Eq.~\eqref{eq:BogoliubovTrasformations}, one can see that $\alpha_\xi (\bp,\bk;\delta t)$ does not change the particle number but is responsible for changes in particle momenta, while $\beta_\xi (\bp,\bk;\delta t)$ is associated with particle creation and annihilation. If one can stop the analysis at the first non-vanishing order in $\sqrt{\lambda}$, where Eq.~\eqref{eq:FirstOrderBogoliubovCoefficients} holds, this continues to be true at all times.

\subsection{Energy and particle number variation rates\label{SubSec:KleinGordonRates}}

We now consider two important global operators. The first one is the total energy of system $S$, which we write as $\hH_S = \int \dd[3]{\bk} \omega_\bk \ha^\dg_\bk \ha_\bk$. The second one is the particle-number operator $\hN = \int \dd[3]{\bk} \ha^\dg_\bk \ha_\bk$. For both operators, we can find their ``Heisenberg'' evolution by employing the Bogoliubov transformations of Sec.~\ref{SubSec:BogoliubovTransformation}. Choosing an initial state $\tl{\rho}_0$ such that $[\tl{\rho}_0,\hat N]=0$ and exploiting the statistical spatial-translation invariance of the dynamics, one can formally compute the change in the average particle number ($\Delta N_t$) and energy ($\Delta E_t$) of system $S$ by first performing the calculations inside a box with periodic boundary conditions and then taking the continuum limit (see detailed calculations in Appendix~\ref{APPSec:EnergyParticleNumberIncreaseBogoliubov}). The result for the variation in expected particle number is
\begin{equation}
\Delta N_t
=
2 \int \dd[3]{\bk} r_N (\bk;t)n_\bk,
\end{equation}
where $n_\bk := \Tr \{\ha^\dg_\bk \ha_\bk \tl{\rho}_0\}$ is the initial average particle density in mode $\bk$ and (formally)\footnote{Eqs.~\eqref{eq:ParticleNumberWavevectorCoefficient} and~\eqref{eq:EnergyIncreaseWavevectorCoefficient} should be seen as the continuum limit of calculations first performed inside a box. Especially in Eq.~\eqref{eq:EnergyIncreaseWavevectorCoefficient}, the problematic nature of the continuum limit is visible in the $\alpha$ coefficient at $t=0$. Indeed, Eq.~\eqref{eq:BogoliubovTrasformations} requires that $\alpha_\xi (\bp,\bk;0) = \delta^{(3)} (\bp-\bk)$, but this means that we would have $\abs{\delta^{(3)} (\bp-\bk)}^2$ in Eq.~\eqref{eq:EnergyIncreaseWavevectorCoefficient}.}
\begin{equation}\label{eq:ParticleNumberWavevectorCoefficient}
r_N (\bk;t)
=
\mbE_\xi \prtq{\int \dd[3]{\bp} \abs{\beta_\xi (\bp,\bk;t)}^2},
\end{equation}
which implies that $r_N (\bk;t) \geq 0$. Thus, $\Delta N_t \geq 0$ provided that $n_\bk \geq 0$ for every $\bk$.\footnote{This is automatically true for standard quantum states but, strictly speaking, it may not be true for elements of the set of physical states. Nevertheless, physical intuition dictates that this must be the case.} Without the DVS prescription, there would be an additional, state-independent contribution proportional to $\mbE_\xi \int \dd[3]{\bp}\dd[3]{\bk} \abs{\beta_\xi (\bp,\bk;t)}^2$ (i.e., formally proportional to $\int \dd[3]{\bk} r_N (\bk;t)$) which would correspond to particle creation from the vacuum. 
Such a term formally corresponds to the standard $\abs{\beta}^2$ vacuum contribution familiar from Bogoliubov particle production in curved spacetime~\cite{Book_Birrell1982QuantumFieldsCurvedSpace}, but (thanks to the DVS prescription) we have eliminated it here.
Regarding the energy variation, one gets instead
\begin{equation}\label{eq:EnergyIncreaseBogoliubovFinal}
\Delta E_t
=
\int \dd[3]{\bk} r_E (\bk;t) n_\bk,
\end{equation}
where (formally)
\begin{multline}\label{eq:EnergyIncreaseWavevectorCoefficient}
r_E (\bk;t)
=
\int \dd[3]{\bp} \mbE_\xi \Big[\\
(\omega_\bp-\omega_\bk)
\abs{\alpha_\xi(\bp,\bk;t)}^2
+
(\omega_\bp+\omega_\bk)
\abs{\beta_\xi(\bp,\bk;t)}^2
\Big].
\end{multline}
The contribution proportional to $\abs{\beta_\xi}^2$ is manifestly non-negative. On the other hand, since $\omega_\bp-\omega_\bk$ can be either positive or negative, the contribution proportional to $\abs{\alpha_\xi}^2$ does not seem to have, in general, a definite sign. A comparison between $r_E$ and $r_N$ suggests that the contribution proportional to $\abs{\beta_\xi}^2$ is connected with particle creation while the contribution proportional to $\abs{\alpha_\xi}^2$ is connected with particle diffusion. This would also physically explain why the former is non-negative while the latter is not manifestly so. However, $r_E$ is independent of the initial state of the field. Thus, we can consider an initial state such that $n_\bk \propto \delta^{(3)}(\bk)$, which would lead to $\Delta E_t \propto r_E (\mathbf 0;t)$ with
\begin{multline}\label{eq:EnergyIncreaseWavevectorCoefficient_ZeroMode}
r_E (\mathbf 0;t)
=
\int \dd[3]{\bp} \mbE_\xi \Big[\\
(\omega_\bp-m)
\abs{\alpha_\xi(\bp,\mathbf 0;t)}^2
+
(\omega_\bp+m)
\abs{\beta_\xi(\bp,\mathbf 0;t)}^2
\Big].
\end{multline}
Since $\omega_p \geq m$, the contribution proportional to $\abs{\alpha_\xi}^2$ is non-negative. Statistical Poincaré invariance then seems to imply that this is true even when $\bk \neq \mathbf 0$. The conclusion, however, does not follow rigorously because, while it is true that one can always select a Lorentz frame in which, for each mode, $\bk \to \mathbf 0$, this change of frame would also change the initial simultaneity slice $t=0$. This move is not allowed because the dynamics we are dealing with are instead defined by choosing a specific constant-time hyperplane as the initial Cauchy surface. That said, physically, this argument suggests that the true dynamics (the one obtained from Eq.~\eqref{eq:TrueCPQMDynamicsWithCorrelations} without neglecting the initial correlations) comes with energy-increasing diffusion of the relativistic particles, consistent with the arguments in Ref.~\cite{Donadi2023DiffusiveCollapse} for generic collapse models.

To get some more quantitative results, we can use the expansion to first order in $\sqrt{\lambda}$ of Eq.~\eqref{eq:FirstOrderBogoliubovCoefficients}.
One then finds that (see Appendix~\ref{APPSubSec:BogoliubovLongTime}),\footnote{We point out that one cannot simply substitute Eq.~\eqref{eq:FirstOrderBogoliubovCoefficients} into Eq.~\eqref{eq:ParticleNumberWavevectorCoefficient} because, as pointed out before, Eq.~\eqref{eq:ParticleNumberWavevectorCoefficient} is an ill-defined formal expression.} in the long-time limit,
\begin{equations}\label{eq:IncreaseFactorsLongTime}
r_N (\bk;t)
&\simeq
\frac{\lambda t}{4(2\pi)^3}
\int \dd[3]{\bp} \frac{\tl{G}(p+k)}{\omega_\bp \omega_\bk},
\\
r_E(\bk;t)
&\simeq
\frac{\lambda t}{4(2\pi)^3}
\int
\frac{d^3\mathbf{p}}{\omega_{\mathbf{p}}\omega_{\mathbf{k}}}
\Big[\\
&\quad
\left(\omega_{\mathbf{p}}-\omega_{\mathbf{k}}\right)
\tl{G}(p-k)
+
\left(\omega_{\mathbf{p}}+\omega_{\mathbf{k}}\right)
\tl{G}(p+k)
\Big].
\end{equations}
where $\tl{G}(q)=\int \dd[4]{r} e^{i q\cdot r} G(r)$ is the Fourier transform of $G(r)$. It also holds that $\tl{G}(q)=\abs{\tl{g}(q)}^2$, where $\tl{g}(q)$ is the Fourier transform of $g(x)$ (see Appendix~\ref{APPSubSec:CorrelationFunctionProperties}).
We remark that the results in the long-time limit have more physical relevance than those at finite times because the map $\tl{\Phi}$ is already derived under the assumption that it correctly describes the dynamics in this long-time regime.

By exploiting the Lorentz invariance of both $\tl{G}$ and $\dd[3]{\bp}/\omega_\bp$, we can further simplify the expressions for $r_N$ and $r_E$ (see Appendix~\ref{APPSubSec:InvariantRateReduction}):
\begin{equations}\label{eq:IncreaseFactorsLongTimeSimplified}
r_N (\bk;t)
&\simeq
\frac{\lambda t}{32 \pi^2 m^2 \omega_\bk}
\! \int_{4 m^2}^{\infty}\!\! \dd{s} \sqrt{s(s-4m^2)}\tl{G}_T (s),
\\
r_E (\bk;t)
&\simeq
\frac{\lambda t}{64\pi^2 m^4}\int_{4m^2}^{\infty}\!\! \dd{s} s \sqrt{s(s-4m^2)}\tl{G}_T (s)
\\
&\quad +
\frac{\lambda t}{64\pi^2 m^4}
\!\int_0^\infty\!\!\!\!\!\! \dd{s}
s\sqrt{s\! \left(s \!+\! 4m^2\right)}\tl{G}_S(s).
\end{equations}
In the above equations, $\tl{G}_T (q^2) := \tl{G}(q)$ for $q^2 \geq 0$ and $\tl{G}_S (-q^2) := \tl{G}(q)$ for $q^2 \leq 0$. In other words, $\tl{G}_T$ encodes the timelike contributions of $\tl{G}$ and $\tl{G}_S$ its spacelike contributions.
In particular, the $\tl{G}_T$ contribution comes from $\beta_\xi$ and is thus responsible for pair creation while the $\tl{G}_S$ contribution comes from $\alpha_\xi$ and is thus responsible for particle diffusion (see Sec.~\ref{SubSec:BogoliubovTransformation}).
Eq.~\eqref{eq:IncreaseFactorsLongTimeSimplified} also shows that, at order $\lambda$, $r_E (\bk;t) = r_E (\mathbf 0; t) \geq 0$, thus providing a perturbative long-time result consistent with the argument below Eq.~\eqref{eq:EnergyIncreaseWavevectorCoefficient_ZeroMode}. 
Both $r_N$ and $r_E$ have the correct dependence on $\omega_\bk$ required by Lorentz invariance. Indeed, considering $n_\bk \propto \delta^{(3)}(\bk-\bp_0)$, one has that
\begin{equation}
\Delta N_t \propto \frac{t}{\omega_{\bp_0}},
\qquad
\Delta E_t \propto t.
\end{equation}
In fact, the number of particles created between $t=0$ and $t>0$ is a scalar quantity that can be used as a clock. In our case, this relation is linear in the coordinate time $t$. Thus, if the initial particles are moving with velocity $\bv_0 =\bp_0/\omega_{\bp_0}$, the ``clock ticking'' should slow down by the factor $\gamma_{\bv_0}^{-1} = m/\omega_{\bp_0}$, where $\gamma_{\bv_0}$ is the Lorentz factor. Similarly, if $\Delta E_t$ increases linearly when particles are at rest, it increases linearly in the proper time $\tau_{\bp_0}$ of the particles. However, since $\Delta E_t$ is the time component of a four-vector, the two Lorentz factors coming from $t \to \tau_{\bp_0}$ and $\omega_{\mathbf 0} \to \omega_{\bp_0}$ cancel out.

With respect to Ref.~\cite{Gundhi2026RelativisticCollapseStatisticalDynamics}, our freedom in choosing a $G(r)$ such that the long-time number and energy rate integrals converge at order $\lambda$ is restricted by the conditions on $g(x)$ stated in Eq.~\eqref{eq:CondotionsOnSmearingFunction}. It is, however, possible to find such functions.
This analysis is carried out in Appendix~\ref{APPSubSec:RateConvergence}, in which we find an example:
\begin{multline}\label{eq:ConvergentSmearingProfile}
g (x)
=
\Theta(-x^0)\Theta(x^2)
\frac{x^2/r_C^2}{512 \pi}
\Big[\\
96-24(x^2/r_C^2) + (x^2/r_C^2)^2\Big] e^{-x^2/4 r_C^2},
\end{multline}
where $r_C$ is the quantity usually called ``collapse radius'' in the literature on spontaneous collapse models~\cite{Review_Bassi2003Dynamical,Review_Bassi2013Models,Review_Bassi2023CollapseModels,Review_Carlesso2025CollapseModelsShortReview}.

\subsection{First-order intensities\label{SubSec:KleinGordon_FirstOrderIntensities}}

We now specialize Eq.~\eqref{eq:FlashIntensityPSL} to the first-order intensity, i.e., to the one-point intensity of the flash process. We compute it at the first non-vanishing order in $\lambda$. The detailed calculation performed in a cubic box with periodic boundary conditions can be found in Appendix~\ref{APPSec:FirstOrderIntensityDetailedCalculation}.

Let $x$ be sufficiently far in the future of the initial Cauchy surface $\Sigma_0$ and define
\begin{equation}\label{eq:LateTimeOneFlashOperator}
\hC_x
:=
\int_{\Sigma_0}^{+\infty}\!\!\!\dd[4]{y}
 g(y-x)\hQ_I(y)
\simeq
\int\dd[4]{y}
 g(y-x)\hQ_I(y),
\end{equation}
where the approximation stems from considering CPs far in the future of $\Sigma_0$.
Specializing Eq.~\eqref{eq:FlashKrausOperatorPSL} to $n=1$ gives $\hK_1^{(\xi)}=-i\hC_x+O(\sqrt{\lambda})$. The term of order $\sqrt{\lambda}$ is linear in $\xi$ and therefore has zero stochastic average. Consequently,
\begin{multline}\label{eq:FirstOrderIntensityLeadingLambda}
\mu_F^{(1)}(x|\tl{\rho}_{\Sigma_0})
=
\lambda\Tr_S\prtg{
\hC_x\prt{\tl{\rho}_{\Sigma_0}-\Omega_S}\hC_x
}
+O(\lambda^2)
\\
=
\lambda\prtq{
\Tr_S\prtg{\hC_x^2\tl{\rho}_{\Sigma_0}}
-
\bra{\Omega}\hC_x^2\ket{\Omega}
}
+O(\lambda^2).
\end{multline}
The subtraction in Eq.~\eqref{eq:FirstOrderIntensityLeadingLambda} is essential for vacuum preservation and makes $\mu_F^{(1)}(x|\Omega_S)$ vanish, but makes general positivity non-trivial. As shown in Appendix~\ref{APPSubSec:FirstOrderIntensity_PositivityDiagonalStates}, for every standard positive state that is diagonal in particle number, we have, at leading order in $\lambda$, that $\mu_F^{(1)}(x|\tl{\rho}_{\Sigma_0})\geq 0$. This result does not extend to arbitrary coherent superpositions of different particle-number sectors. In particular, generic coherences between the vacuum and the non-vacuum even sector can make the leading expression in Eq.~\eqref{eq:FirstOrderIntensityLeadingLambda} negative (see examples in Appendix~\ref{APPSubSec:FirstOrderIntensity_Counterexample}). Hence, positivity of the flash intensities supplies a genuine restriction on the physical state space, consistently with the restricted-positivity construction of Sec.~\ref{SubSec:RestrictedPositivity}. Moreover, this leading one-point result does not establish positivity at higher orders in $\lambda$ or positivity of higher-order intensities; these remain part of the open restricted-positivity problem.

Some nice results can be obtained by assuming that the initial state is a momentum eigenstate (see Appendix~\ref{APPSubSec:FirstOrderIntensity_MomentumEigenstates}). In this case, the intensity is spatially uniform and the total one-flash rate is
\begin{equation}\label{eq:FirstOrderTotalFlashRateMomentumEigenstate}
\Gamma_F (\bk_0)
:=
\int \dd[3]{\bx} \mu_{F}^{(1)}(t,\bx|\bk_0)
=
\Gamma_{\alpha} (\bk_0) + \Gamma_{\beta} (\bk_0),
\end{equation}
where
\begin{equations}\label{eq:FirstOrderFlashRatesLorentzInvariant}
\Gamma_\alpha(\bk_0)
&=
\frac{\lambda}{32\pi^2m^2\omega_{\bk_0}}
\int_0^\infty\dd{s} 
\sqrt{s(s+4m^2)}\tl{G}_S(s),
\\
\Gamma_\beta(\bk_0)
&=
\frac{\lambda}{32\pi^2m^2\omega_{\bk_0}}
\int_{4m^2}^\infty\dd{s} 
\sqrt{s(s-4m^2)}\tl{G}_T(s).
\end{equations}
In particular, this implies that $\Gamma_F(\bk_0) = (m/\omega_{\bk_0})\Gamma_F(\mathbf 0)$, which is the expected time-dilation law for the number of flashes generated by a single momentum eigenstate. 
Moreover, comparing Eq.~\eqref{eq:FirstOrderFlashRatesLorentzInvariant} with Eq.~\eqref{eq:IncreaseFactorsLongTimeSimplified}, one finds, at leading order in $\lambda$,
\begin{equation}
\lim_{t\to\infty}\frac{r_N(\bk_0;t)}{t}
=
\Gamma_\beta(\bk_0).
\end{equation}
This relation establishes a quantitative connection between flash statistics and the computed change in particle number, supporting the physical interpretation of standard quantum observables within the framework.

\subsection{Nonrelativistic limit\label{SubSec:KleinGordon_NonRelativisticLimit}}

The nonrelativistic limit of the present model requires more than considering an initial state with a fixed number of particles, each with momentum much smaller than $m$, the particles' mass. The collapse interaction can create additional particles or, after many individually soft kicks, heat the particles out of the low-energy sector. Given a timescale $t$, we want both these effects to be negligible given the number of particles, i.e., we want that $\Delta N_t/N \ll 1$ and $\Delta E_t/N \ll m$. Given that, in the long-time limit [see Eq.~\eqref{eq:IncreaseFactorsLongTimeSimplified}],
\begin{equations}
\frac{\Delta N_t}{N} 
&\simeq 
\frac{2 m r_N (\mathbf 0;t)}{N} \int \dd[3]{\bk} \frac{n_\bk}{\omega_\bk}
\leq
2 r_N (\mathbf 0;t),
\\
\frac{\Delta E_t}{N}
&\simeq
r_E (\mathbf 0;t),
\end{equations}
our requirements are satisfied if
\begin{equation}\label{eq:NonRelativisticLimitRequirements}
r_N (\mathbf 0;t) \ll 1,
\qquad
r_E (\mathbf 0;t) \ll m.
\end{equation}
We also demand the occurrence of hard kicks to be negligible, which we formalize by demanding that the spacelike spectrum $\tl{G}_S(s)$ is negligible for $s\gtrsim m^2$.

Assuming the previously stated conditions are satisfied, we now derive the nonrelativistic limit of the real massive scalar field with collapse operator \mbox{$\hQ_I(x)=\ :\!\hat\varphi^2(x)\!:/2$}.
We start by separating the field operator $\hat\varphi(x)$ in positive and negative frequency terms, i.e., $\hat\varphi(x) = \hat\varphi_{+} (x) + \hat\varphi_{-} (x)$ where
\begin{equation}
\hat\varphi_{+} (x) :=\int \frac{\dd[3]{\bk}}{\sqrt{2\omega_\bk (2\pi)^3}}e^{-i k \cdot x}\ha_\bk,
\end{equation}
and $\hat\varphi_{-} (x)=[\hat\varphi_{+} (x)]^\dg$.
By assuming that only modes with $\abs{\bk} \ll m$ are non-negligibly populated 
we can approximate the normalization factor $1/\sqrt{2\omega_\bk}$ by $1/\sqrt{2m}$, while retaining $\omega_\bk$ in the time-dependent phases.
Then, we define the operator
\begin{equation}
\hat\psi_I(t,\bx)
=
\frac{1}{(2\pi)^{3/2}}
\int \dd[3]{\bk} 
e^{-i(\omega_\bk-m)t+i\bk\cdot\bx}\ha_\bk,
\end{equation}
and obtain the following leading-order expression within the low-momentum sector:
\begin{equation}\label{eq:NonRelativisticKleinGordonField}
\hat\varphi (t,\bx)
\simeq
\frac{1}{\sqrt{2m}}
\Big[
e^{-imt}\hat\psi_I(t,\bx)
+
e^{imt}\hat\psi_I^\dg(t,\bx)
\Big].
\end{equation}
Consequently, the collapse density becomes
\begin{equation}\label{eq:NonRelativisticQuadraticCollapseDensity}
\hQ_I(t,\bx)
\simeq
\frac{1}{2m^2}\hmu_I(t,\bx)
+
\frac{1}{4m}
\Big[
e^{-2imt}\hat\psi_I^2(t,\bx)+\hc
\Big],
\end{equation}
where $\hmu_I:=m\hat\psi_I^\dg\hat\psi_I$.
The first term preserves particle number, while the remaining terms create or annihilate pairs. Under the assumptions above, these oscillating terms can be neglected within the rotating-wave approximation.
The operator $\hmu$ is not local in the relativistic QFT sense because its definition uses the positive-frequency annihilation and creation operators.
This is why it can annihilate the vacuum without contradicting the Reeh--Schlieder theorem.
The Schr\"odinger-picture interaction with a CP at $x_C$ then reduces to
\begin{equation}\label{eq:NonRelativisticCollapsePointInteraction}
\sqrt{\Upsilon}\hL_{x_C}(t)
\rightarrow
\frac{\sqrt{\Upsilon}}{2m^2}
\int\dd[3]{\by} 
g\prt{t-x_C^0,\by-\bx_C}
\hmu(\by).
\end{equation}
This has exactly the structure of the nonrelativistic non-Markovian CPQM model of Ref.~\cite{Piccione2026NonMarkovianPSL}, with the mass-density operator as the collapse operator. Moreover, we stress that the effective mass-density interaction given by the above equation annihilates the vacuum; the DVS modification vanishes.

There is a distinction between the low-energy limit at finite $c$ described above and a strict $c\to\infty$ limit.
The latter leads to a Markovian collapse model when the invariant profile parameters, including $r_C$, are held fixed.
One way to see this is to restore $x^0=ct$, where $c$ is the speed of light.
Then, a Lorentz-invariant profile $g(x)$ characterized by an invariant length $r_C$ has a coordinate-time correlation scale $\tau_C\sim r_C/c$, which tends to zero as $c\to\infty$.
Alternatively, consider the spacelike contribution $\tl{G}_S(-q^2)$, which is the only contribution retained in the low-energy limit.
Restoring $c$, this can be written as $\tl{G}_S(\bq^2-\nu^2/c^2)$, where $\nu$ is the coordinate-time frequency.
Hence, $\tl{G}_S(\bq^2-\nu^2/c^2)\to\tl{G}_S(\bq^2)$ as $c\to\infty$; the limiting spectrum is independent of $\nu$ and therefore corresponds, up to an overall normalization absorbed into the effective coupling, to white noise in time.
As shown in Appendix~\ref{APPSubSec:MarkovianSmearing}, the smearing profile $g_{r_C} (\bx)$ of the Markovian limit can be obtained as
\begin{equation}\label{eq:MarkovianSmearingReduction}
g_{r_C} (\bx) = \mcN\int_{-\infty}^{+\infty}\!\!\!\!\! \dd{t} c\, g(c t,\bx),
\end{equation}
where $\mcN$ is a normalization constant such that $\int \dd[3]{\bx} g_{r_C} (\bx) = 1$, and $\mcN^{-1}$ can be absorbed by the coupling $\Upsilon$. Using the profile of Eq.~\eqref{eq:ConvergentSmearingProfile}, one gets
\begin{equation}\label{eq:MarkovianSmearingProfileExample}
g_{r_C}(\bx)
=
\frac{9}{32\pi r_C^3}
U\prt{\frac{5}{2},-\frac{1}{2},\frac{\bx^2}{4r_C^2}},
\end{equation}
where $U(a,b,z)$ is the confluent hypergeometric function of the second kind\footnote{See \url{https://mathworld.wolfram.com/ConfluentHypergeometricFunctionoftheSecondKind.html}.}.

\begin{figure}[t]
\centering
\includegraphics[width=\linewidth]{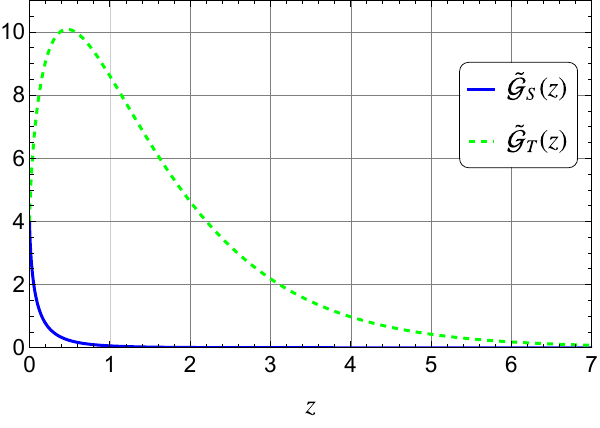}
\caption{Plot of the functions $\tl{\mcG}_T(z)$ and $\tl{\mcG}_S (z)$. For $z \gg 1$ they both decay as $\tl{\mcG}_{T/S}(z) \sim 144 z^{-6}$.}
\label{fig:PlotSpectralCorrelation}
\end{figure}

For the $g(x)$ profile given in Eq.~\eqref{eq:ConvergentSmearingProfile}, the explicit formulas for $\tl G_T(s)$ and $\tl G_S(s)$ follow from Eq.~\eqref{APPeq:ExplicitSpectralFormSmearingFunction} in Appendix~\ref{APPSubSec:RateConvergence}, using $\tl G_T(s)=\abs{\tl g_T(s)}^2$ and $\tl G_S(s)=\abs{\tl g_S(s)}^2$.
We now restore $\hbar$ and $c$. Then, we introduce\footnote{We recall that $\lambdabar:= \hbar/m c$ is the reduced Compton wavelength of a particle with mass $m$.}
\begin{equation}\label{eq:VariableSubstitutionMarkovianLimit}
z=r_C^2s,
\quad
a=\frac{m c r_C}{\hbar}=\frac{r_C}{\lambdabar},
\quad
\tl G_{T,S}(s)=r_C^8\tl{\mcG}_{T,S}(z),
\end{equation}
where\footnote{Notice that $\tl{\mcG}_T$ and $\tl{\mcG}_S$ are dimensionless and independent of $r_C$. See \url{https://reference.wolfram.com/language/ref/ExpIntegralEi.html} for the definition of $\operatorname{Ei}(z)$.}
\begin{equations}
\tl{\mcG}_T(z)
&=
\left[z^2-5z+2-ze^{-z}(z^2-6z+6)\operatorname{Ei}(z)\right]^2
\\
&\quad
+\pi^2z^2e^{-2z}(z^2-6z+6)^2,
\\
\tl{\mcG}_S(z)
&=
\left[z^2+5z+2+ze^z(z^2+6z+6)\operatorname{Ei}(-z)\right]^2,
\end{equations}
which are plotted in Fig.~\ref{fig:PlotSpectralCorrelation}. Using
\begin{equation}
\dd{s}=\frac{\dd{z}}{r_C^2},
\qquad
\sqrt{s\prt{s\pm\frac{4m^2c^2}{\hbar^2}}}
=
\frac{1}{r_C^2}\sqrt{z(z\pm4a^2)},
\end{equation}
the SI form of Eq.~\eqref{eq:IncreaseFactorsLongTimeSimplified}, evaluated at $\bk=\mathbf0$, gives
\begin{equations}
r_N(\mathbf0;t)
&\simeq
\frac{\lambda t\hbar^3r_C^4}{32\pi^2m^3c^2}
\int_{4a^2}^\infty\dd{z}
\sqrt{z(z-4a^2)} \tl{\mcG}_T(z),
\\
r_E(\mathbf0;t)
&\simeq
\frac{\lambda t\hbar^5r_C^2}{64\pi^2m^4c^2}
\int_{4a^2}^\infty\dd{z}
z\sqrt{z(z-4a^2)} \tl{\mcG}_T(z)
\\
&\quad+
\frac{\lambda t\hbar^5r_C^2}{64\pi^2m^4c^2}
\int_0^\infty\dd{z}
z\sqrt{z(z+4a^2)} \tl{\mcG}_S(z).
\end{equations}
The integrals are dimensionless, while $r_N$ is dimensionless and $r_E$ has units of energy. Setting $r_C=10^{-7} \mathrm{m}$ (a typical value in the collapse model literature~\cite{Review_Bassi2013Models}) and using the electron mass $m_e$ gives $a_e \simeq 2.59\times10^5$, while using the proton mass $m_0$ gives $a_0 \simeq4.75\times10^8$. In both cases $a \gg 1$. 
Since $a \gg 1$, the timelike integrals only sample $z\geq4a^2\gg1$,
where $\tl{\mcG}_T(z) \simeq 144z^{-6}$.
In the spacelike integral, the dominant contribution comes from
$z\ll a^2$, allowing us to approximate
$\sqrt{z(z+4a^2)} \simeq 2 a \sqrt{z}$ while retaining the full
function $\tl{\mcG}_S(z)$.
Then, by numerical integration we get that
\begin{equations}
r_N(\mathbf0;t)
&\simeq
1.8\times 10^{-4} \frac{\lambda t}{r_C^4} \frac{\hbar^{11}}{m^{11} c^{10}},
\\
r_E(\mathbf0;t)
&\simeq
5.4\times 10^{-4} \frac{\lambda t}{r_C^4}\frac{\hbar^{11}}{m^{10}c^{8}}
+
5.2\times 10^{-4} \lambda t r_C^3 \frac{\hbar^4}{m^3 c}.
\end{equations}

For the standard, mass proportional, Markovian CSL model, the energy increase per particle is given by (see Appendix~\ref{APPSubSec:HeatingRateCSL})
\begin{equation}\label{eq:HeatingCoefficientCSL}
r_{\rm CSL} (t) \simeq 0.33\times \frac{\lambda_{\rm CSL} \hbar^2 m}{m_0^2 r_C^2}t,
\end{equation}
where $\lambda_{\rm CSL}$ is the collapse rate, i.e., the decoherence rate of one particle of reference mass $m_0$ in spatial superposition with distance $d\gg r_C$.
To make a better comparison, we computed the above rate for the CSL model not by employing a Gaussian smearing but the Markovian one corresponding to our $g(x)$ [see Eq.~\eqref{eq:MarkovianSmearingReduction}].
Thus, by equating the second term of $r_E (\mathbf 0;t)$ (the one coming from $\tl{\mcG}_S$ and assumed to be much higher than the other one) we obtain that
\begin{equation}
\lambda \simeq 6.29 \times 10^{2} \frac{m^4 c}{m_0^2 \hbar^2} \frac{\lambda_{\rm CSL}}{r_C^5}.
\end{equation}
By substituting this back into $r_N$ and $r_E$ we finally obtain
\begin{equations}
r_N(\mathbf0;t)
&\simeq
0.11 \frac{\hbar^9}{m^7 m_0^2 c^9} \frac{\lambda_{\rm CSL}}{r_C^9}t,
\\
r_E(\mathbf0;t)
&\simeq
0.34 \times \frac{\hbar^9}{m^6 m_0^2 c^7}\frac{\lambda_{\rm CSL}}{r_C^9}t
+
0.33\times \frac{m \hbar^2}{m_0^2} \frac{\lambda_{\rm CSL}}{r_C^2}t.
\end{equations}
To get a sense of the above quantities, we now set $t= 3\times 10^7 \rm{s}$ (roughly a year), $\lambda_{\rm CSL}=10^{-16} \rm{Hz}$ and $r_C = 10^{-7} \rm{m}$ (the so-called GRW values~\cite{Review_Bassi2013Models,Adler2019NeutronStarSpontaneousHeating,XENONCollaboration2026CollapseConstraints}) and $m=m_e$, the electron mass. We get
\begin{equations}
r_N (\mathbf0;t) &\simeq 1.9 \times 10^{-65},
\\
r_E (\mathbf 0;t) &\simeq \prt{3.0 \times 10^{-59} + 2.2\times 10^{-21}} \rm{eV}.
\end{equations}
Indeed, considering that $m_e c^2 \simeq 5.1 \times 10^{5} \rm{eV}$, both conditions in Eq.~\eqref{eq:NonRelativisticLimitRequirements} are satisfied by a large margin. 
Moreover, the first term of $r_E$, the one linked to particle creation, is vastly smaller than the second one.
Relativistic spacelike momentum transfers are also suppressed: for $a \gg 1$, transfers capable of producing relativistic electron momenta probe $z \gtrsim a^2$, where $\tl{\mcG}_S(z) \sim 144z^{-6}$.
Of course, quantitative conclusions about electron behavior must be taken cautiously given that the Hamiltonian $\hH_S$ we used is not that of the Dirac field.

We have shown that a sensible nonrelativistic limit can be obtained under the condition (among others) that $a\gg 1$, i.e., $r_C \gg \lambdabar$, where $\lambdabar$ is the reduced Compton wavelength of the particles.
This can be given an intuitive physical interpretation as follows.
Since a collapse model localizes particles on the scale $r_C$, it makes sense that particle production is very suppressed when the localization scale is much larger than the Compton wavelength.

\section{Discussion\label{Sec:Discussion}}

\subsection{On the loss of general positivity\label{SubSec:Discussion_Positivity}}

In this work, we have proposed a framework for constructing relativistic spontaneous collapse models. To solve the vacuum stability problem~\cite{Pearle1999Relativistic,Pearle2015RelativisticCSL,Myrvold2017RelativisticMarkovian} we introduced the Dynamical Vacuum Subtraction (DVS) prescription.
The price for this is that the standard space of positive, unit-trace operators is no longer the appropriate physical state space.
Instead, given an initial Cauchy surface and a disposition of CPs in its future, physical states are defined operationally as those leading, through the modified dynamics, to well-defined, normalized, and non-negative probabilities for every finite flash/no-flash history.

A full characterization of this state space constitutes a challenging mathematical question. The answer may depend on the interaction operator $\hat Q_I(x)$, the smearing function $g(x)$, and the underlying QFT.
The analysis of Schlegel et al.~\cite{Schlegel2026ClassicalQuantumRepresentations} suggests that this difficulty may partly reflect the chosen representation.
Under the Wigner--Weyl correspondence, classical and quantum states can be represented in the same mathematical spaces but are selected by different positivity requirements: classical phase-space distributions are positive although their Hilbert-space representatives need not be, whereas quantum density operators are positive although their Wigner functions need not be.
Thus, positivity is an admissibility condition tied to the chosen representation, rather than an intrinsic property of its mathematical carrier space.
The unit-trace Hilbert-space operators used here may therefore not provide the most natural representation of our physical states; an alternative representation might turn the requirement of non-negative flash/no-flash probabilities into a simpler positivity condition.
Moreover, the resulting state space is not expected to be merely a subset of the standard one\footnote{See the qubit example in Sec.~\ref{SubSec:RestrictedPositivity}.}: some positive operators fail restricted positivity, while some operators that are not positive in the usual sense may nevertheless assign consistent probabilities to all allowed events.

The Klein--Gordon analysis provides preliminary evidence that this state space is nontrivial.
In the long-time PSL regime, the leading one-point flash intensity is non-negative for every standard positive state satisfying $[\hat N,\tl{\sigma}_0]=0$ [Sec.~\ref{SubSec:KleinGordon_FirstOrderIntensities}].
Together with finite and non-negative particle-creation and energy-increase rates for suitable smearing functions, their expected relativistic scaling with the initial particle momenta [Sec.~\ref{SubSec:KleinGordonRates}], and the proper nonrelativistic limit [Sec.~\ref{SubSec:KleinGordon_NonRelativisticLimit}], this provides encouraging evidence for the viability of the framework.
At the same time, suitable coherences between the vacuum and the non-vacuum even-particle sector yield negative leading intensities, thereby excluding otherwise standard positive states.

Let us also comment on another fact: lack of positivity is not unique to the present framework.
Quadratic gravity adds curvature-squared terms to the Einstein--Hilbert action, making the theory perturbatively renormalizable but introducing a massive spin-two ghost whose one-particle states have negative norm and thereby threaten unitarity~\cite{Review_Salvio2018QuadraticGravity,Review_Piva2023HigherDerivativeQuantumGravity,Stelle1977RenormalizationHigherDerivativeGravity}.
Rather than ending the research programme, this has motivated attempts to modify the quantization prescription or the definition of the physical sector~\cite{Anselmi2018QuantumGravityFakeonsMicrocausality,Donoghue2019UnitarityStabilityUnstableGhosts,Bateman2026EscapeOstrogradskyGhostParity}.
Related examples include Lee--Wick theories with negative-metric states~\cite{Review_Wise2010LeeWickStandardModel} and Gupta--Bleuler quantization, which employs an indefinite state space before selecting the physical sector~\cite{Review_Gottschalk2006IndefiniteMetric}.
These constructions differ mathematically from DVS and involve different notions of non-positivity, but illustrate the limited point relevant here: failure of a conventional positivity requirement is not, by itself, a no-go theorem.
The decisive question is whether one can identify a consistent physical state space on which all physical events are assigned non-negative probabilities.

Finally, the unmodified long-time statistical map $\Phi$ derived in Sec.~\ref{Sec:EvolutionPSL_Limit} formally coincides with the colored-noise evolution analyzed in Ref.~\cite{Gundhi2026RelativisticCollapseStatisticalDynamics}.
The DVS-corrected map $\tl{\Phi}$ differs by the vacuum-subtraction term in Eq.~\eqref{eq:ModifiedLongTimeStatisticalEvolution}, whereas Ref.~\cite{Gundhi2026RelativisticCollapseStatisticalDynamics} addresses the vacuum-instability problem by prescribing that observables be normal ordered at all times.
This prescription reproduces some expectation values obtained with DVS [Sec.~\ref{SubSec:ExpectationValuesEvolution_DVS}], but operates at a different level: the construction of Ref.~\cite{Gundhi2026RelativisticCollapseStatisticalDynamics} is statistical and does not provide a microscopic collapse rule assigning conditional physical states once the collapse outcomes are specified.
Moreover, normal ordering removes the vacuum contribution from the expectation values considered there without making the vacuum invariant under the underlying stochastic dynamics, since generally \mbox{$:\!\hat Q_I(x)\!:\ket{\Omega}\neq c\ket{\Omega}$} for any $c\in\mathbb{C}$.
Nor is normal ordering a positivity-preserving operation.\footnote{For a harmonic oscillator in a squeezed state with $\langle\hX^2\rangle_r=e^{-2r}/2$ and $\hX=(\ha+\ha^\dg)/\sqrt{2}$, one has $:\!\hX^2\!: =\hX^2-1/2$ and hence $\langle:\!\hX^2\!:\rangle_r=-(1-e^{-2r})/2<0$, although $\hX^2$ is positive.}
These considerations motivated us to address vacuum stability directly at the level of the microscopic dynamics through DVS.

\subsection{On wavefunction collapse\label{SubSec:Discussion_Collapse}}

As emphasized in Ref.~\cite{Mucino2026DemystifyingRelativisticCollapse}, the possibility of assigning a quantum state to each Cauchy surface is conceptually relevant, as in the Aharonov--Albert formalism for relativistic collapse.
The present framework, however, requires assigning a distinguished role to the initial Cauchy surface $\Sigma_0$: probabilities are defined with respect to this surface [see Eq.~\eqref{eq:GeneralConditionalProbability_DVS}], as is the state-update rule [see Eq.~\eqref{eq:ModifiedCollapseRule_DVS}].
This still allows one to associate a state $\tl{\sigma}_\Sigma^{\bF}$ with every Cauchy surface $\Sigma\succ\Sigma_0$ and every realized partial flash history between $\Sigma_0$ and $\Sigma$, but makes the quantum-state dynamics explicitly history-dependent, and therefore non-Markovian at this level.
We do not regard this temporal nonlocality as intrinsically incompatible with a relativistic collapse theory: there is no evident reason why all information relevant for future stochastic predictions should be encoded in a single operator on the current Cauchy surface.
One should therefore distinguish the quantum state associated with a surface from the complete predictive state, which may also contain information about the realized flash history.
Importantly, just like the DVS modification, the modified collapse prescription preserves foliation independence: the special role of $\Sigma_0$ does not select any preferred foliation between initial and final surfaces.
Moreover, the prescription reduces exactly to the standard one whenever the DVS correction vanishes, i.e., when the standard dynamics already preserves the vacuum.

A useful comparison is provided by Tumulka's relativistic GRW flash models~\cite{Tumulka2006relativistic,Tumulka2009rGRWf,Tumulka2021relativistic}.
The original models describe $N$ non-interacting, distinguishable relativistic particles, while Ref.~\cite{Tumulka2021relativistic} extends the construction to interacting distinguishable particles.
Already in the non-interacting model, the quantum state on a Cauchy surface is not sufficient to determine the subsequent flash process: for each particle, one must also specify its most recent flash, which acts as the seed for the next one.
Nevertheless, the model has a renewal structure: once a new flash for the $j$-th particle occurs, the quantum state is updated and the previous seed of that particle can be discarded.
In the interacting extension this simple renewal property is lost: the conditional distribution of future flashes retains additional dependence on the preceding flash history through a history-dependent operator.
Thus, the need to enlarge the data associated with a Cauchy surface beyond the quantum state alone is already present in relativistic GRW models, and explicit history dependence is not peculiar to our construction.

Finally, by analogy with Tumulka’s suggestion that the initial seed flashes lie at the Big Bang~\cite{Tumulka2006relativistic}, the distinguished role of $\Sigma_0$ may acquire a more natural interpretation in a future extension to curved spacetime.
In a cosmological spacetime with an initial boundary or singularity, one could choose $\Sigma_0$ as close as possible to that boundary or singularity (heuristically, to the Big Bang) and regard the subsequent stochastic dynamics as conditioned on cosmological initial data.
Its special role would then resemble the specification of initial conditions for the Universe rather than the introduction of a preferred foliation.
Moreover, in this specific case, it would make perfect sense to assume that the CP-qubits are initialized in the state $\ket{0}$. Whether the dependence on $\Sigma_0$ is ultimately fundamental, or can instead be absorbed into a richer notion of state associated with arbitrary Cauchy surfaces, remains an interesting open question.

\subsection{On substantive Poincaré invariance\label{SubSec:Discussion_SubstantiveRelativity}}

Whether a quantum theory is relativistic in a substantive sense requires consideration beyond the covariance of its equations; representative discussions can be found in Refs.~\cite{Book_Tumulka2022Foundations,Struyve2026SubstantiveLorentzInvariance,Mucino2026DemystifyingRelativisticCollapse} and references therein.

Our proposed framework possesses several properties relevant to this discussion.
Flashes provide a local ontology (local beables in Bell's terminology~\cite[p.~205]{Bell1987JumpsV2}) of spacetime events, while the dynamics and their associated probabilities are independent of the interpolating foliation.
The modified Aharonov--Albert prescription of Sec.~\ref{SubSec:CollapseUpdateRules_DVS} assigns states to arbitrary Cauchy surfaces to the future of $\Sigma_0$.
Moreover, Sec.~\ref{SubSec:NoSignalingExternalPotentials} establishes microscopic parameter independence under local external potentials and the corresponding absence of superluminal signaling.
The distinguished role of $\Sigma_0$, discussed in Sec.~\ref{SubSec:Discussion_Collapse}, nevertheless remains a separate issue from foliation independence.
A comprehensive assessment under the further criteria proposed in the literature lies beyond the scope of the present work.

\subsection{On the application of CPQM to alternative relativistic theories\label{SubSec:Discussion_AlternativeRelativisticTheories}}

Relativistic quantum theories need not be QFTs.
Multi-time wave functions and unitary hypersurface evolutions provide mathematically well-defined fixed-particle-number models, although the implementation of interactions among particles is strongly constrained~\cite{Petrat2014MultiTimeNoInteractionPotentials,Lienert2015RelativisticNParticleModel,Lienert2018DirectInteractionLightCones}.
On the one hand, these theories have a narrower phenomenological scope than standard QFT.
On the other hand, they may have no vacuum sector at all, or, when a vacuum is present, they need not possess the local vacuum structure responsible for the Reeh--Schlieder obstruction~\cite{Reeh1961ReehSchliederTheorem,Fewster2020AQFTIntroduction} mentioned at the beginning of Sec.~\ref{Sec:ModifyingQM}.
Thus, pursuing the CPQM approach for constructing a relativistic collapse model may lead to a theory like that of Sec.~\ref{Sec:CPQM_ModelBuilding} without the need to introduce DVS.
It may also provide a route toward relativistic collapse models for interacting indistinguishable particles, which are not covered by the interacting rGRWf construction of Ref.~\cite{Tumulka2021relativistic}.

\subsection{On external potentials and curved spacetime\label{SubSec:Discussion_ExternalPotentialsCurvedSpacetime}}

In the Tomonaga-Schwinger formalism, interaction-picture observables are generically surface-dependent even in standard QFT.
In Sec.~\ref{Sec:KleinGordonApplication}, the particle-number operator could be treated as time-independent because we considered constant-time hyperplanes and $[\hN,\hH_S]=0$.
If an external potential is included in the standard QFT evolution $\hV_0^t$, this simplification need not hold because, in general, $\hN_I(t) = (\hV_0^t)^\dg \hN \hV_0^t \neq \hN$. Thus, $\Tr \{\Omega_S\hN_I(t) \}$ may vary while the reference state $\Omega_S$ remains fixed in the interaction picture.
Particle production induced by the external potential is then encoded in the standard evolution of the observable, whereas DVS removes only the additional vacuum excitation caused by CPQM. 
However, since DVS leaves the reference vacuum invariant, no flashes would occur when starting from it, even if the external potential produces particles according to the evolved particle-number observable.
The implications of this absence of flashes for the interpretation of particle production require further investigation.

A similar strategy may apply on globally hyperbolic curved spacetimes after choosing a physically admissible initial reference state, such as a Hadamard state, and factoring out the corresponding standard QFT evolution~\cite{Book_Birrell1982QuantumFieldsCurvedSpace,Juarez2018CollapseModelsHadamardCondition}. An investigation of how to extend the framework presented in this paper to curved background spacetime is left for future work.

\subsection{On the gravitization of quantum mechanics\label{SubSec:Discussion_GravitizationQM}}

Gravity need not necessarily be quantized; one may instead try to modify quantum theory so that it can consistently interact with classical gravitational degrees of freedom~\cite{Penrose2014Gravitization,Kafri2014LOCCGravity,Tilloy2016CSLGravity,Diosi2017GKLSforLOCC,Oppenheim2023PostQuantum,Galley2023ConsistentQuantumClassicalGravity}.
In the nonrelativistic case, this can often be done through a measurement-and-feedback mechanism in which spontaneous collapse models play the role of the measurement part~\cite{Tilloy2024HybridDynamics}.
This can also be done when the collapse model has jump dynamics~\cite{Tilloy2018GRWGravity}, as in the Newtonian GPSL model~\cite{Piccione2025NewtonianPSL,Piccione2026EnergyIncreaseGPSL}.
Moreover, constructing a relativistic collapse model is sometimes considered a necessary step toward developing a full theory of classical gravity and quantum matter~\cite{Diosi2024HybridRelativisticDynamicsDifficulties,Tilloy2024HybridDynamics}.

From this perspective, the present framework seems quite promising.
However, some of its features do not make it immediately suitable for this kind of construction.
In fact, when constructing a nonrelativistic flash-based model of classical--quantum Newtonian gravity, one naturally assumes that the one-flash intensity scales linearly with the smeared mass density.
However, for collapse operators of the type considered here, the nonrelativistic collapse operator becomes proportional to a density (e.g., the mass density), while the corresponding leading one-flash intensity is quadratic in that operator rather than linear.
Therefore, the present proposal would likely require further modifications if meant to be used for constructing a theory of relativistic quantum matter coupled to classical gravity.

\section{Conclusions\label{Sec:Conclusions}}

We have developed a framework for relativistic spontaneous collapse in which Poisson-distributed Collapse Points carry internal qubits.
Each such qubit is initialized in the state $\ket{0}$ and coupled to a local QFT operator in the causal past of its collapse point.
It spontaneously collapses to either $\ket{0}$ or $\ket{1}$ at its Collapse Point, when it is no longer interacting, and the outcome $\ket{1}$ defines a \emph{flash} at that point.
The construction is statistically Poincaré invariant and foliation-independent.
To prevent particle production from the vacuum, we introduced the Dynamical Vacuum Subtraction (DVS) prescription, which preserves linearity, trace preservation, and the composition law while making the matter--CP vacuum an exact fixed point.
The price is that the standard space of positive density operators must be replaced by an operationally defined space of physical states, required to assign non-negative probabilities to every finite flash/no-flash history.
A modified, history-dependent Aharonov--Albert prescription allows the assignment of states to arbitrary Cauchy surfaces to the future of a fixed initial surface.
We also established microscopic parameter independence under local external potentials and the corresponding absence of superluminal signaling.
In the nonrelativistic regime, DVS is expected to become trivial so that the standard positive state space and the ordinary nonrelativistic collapse dynamics are recovered.

In the PSL limit, we derived, in the long-time regime, the effective non-Markovian evolution of quantum matter and formal expressions for all joint flash intensities.
For a free massive scalar field with a specific choice of qubit-matter interaction, we obtained, at leading order and for suitable smearing functions, finite particle-number and energy-increase rates with the expected relativistic momentum dependence.
We also found that the leading one-point flash intensity is non-negative for standard positive states diagonal in particle number, while suitable coherences between different particle-number sectors can violate this condition.
Finally, we identified conditions under which standard nonrelativistic collapse dynamics are recovered in this example.

The main open problem is the full characterization of the physical state space. 
Nevertheless, the present results show that vacuum stability, statistical Poincaré invariance, foliation-independent collapse, statistical linear dynamics, no-signaling, finite leading-order long-time excitation rates, and recovery of standard nonrelativistic dynamics can be accommodated in a single framework.
Further directions include extensions to interacting QFTs and curved spacetime, applications to alternative relativistic particle theories, and the possible use of flashes as sources of classical gravity.

\section*{Acknowledgements}

N. P. thanks Angelo Bassi for numerous and extensive discussions about this work.
N. P. thanks Anirudh Gundhi, Lajos Diósi, Roderich Tumulka, Valter Moretti, Oliviero Angeli, and Sandro Donadi for useful discussions about this work.
N. P. acknowledges the use of ChatGPT for language polishing, LaTeX assistance, bibliographic searches, and assistance with calculations, particularly in Sec.~\ref{Sec:KleinGordonApplication}.
All AI-assisted material was critically reviewed by the author, who takes full responsibility for the content of the manuscript.
N. P. acknowledges support from INFN and the University of Trieste.
N. P. also acknowledges support from the European Union's Horizon Europe research and innovation programme under the Marie Skłodowska-Curie Grant Agreement No.~101150889 (CPQM).

\clearpage
\bibliography{AuxiliaryFiles/Bibliography}

\clearpage
\onecolumngrid
\appendix

\section{Mathematical properties of the modified unitary evolution\label{APPSec:ModifiedUnitaryDynamics_MathProperties}}

The TS equation corresponding to the evolution given by $\tl{\mcU}_{\Sigma_0}^{\Sigma_1}$ can be found by first writing Eq.~\eqref{eq:TomonagaSchwingerDifferentialEquation_UnitaryVersion} in the super-operator form
\begin{equation}
\frac{\delta}{\delta \Sigma(x)} \mcU_{\Sigma_0}^{\Sigma}
=
\mcL_x \mcU_{\Sigma_0}^{\Sigma}
= -i\comm{\mcH(x)}{\mcU_{\Sigma_0}^{\Sigma}}.
\end{equation}
Therefore, applying this to a generic operator $\hA$, we get that
\begin{equation}
\frac{\delta}{\delta \Sigma(x)} \tl{\mcU}_{\Sigma_0}^{\Sigma}\hA
=
- i \comm{\hat\mcH(x)}{\mcU_{\Sigma_0}^{\Sigma}\hA - \Tr{\hA}\mcU_{\Sigma_0}^{\Sigma}\Omega}.
\end{equation}
When $\hA$ is a state, one gets that
\begin{equation}\label{eq:ModifiedInteractionPictureDynamics_UnitaryEvolution}
\frac{\delta}{\delta \Sigma(x)} \tl{\mcU}_{\Sigma_0}^{\Sigma}\sigma_{\Sigma_0}
=
- i \comm{\hat\mcH(x)}{\mcU_{\Sigma_0}^{\Sigma}\prt{\sigma_{\Sigma_0} - \Omega}},
\end{equation}
and by substituting the physical evolution we can rewrite the above as
\begin{equation}\label{APPeq:ModifiedTomonagaSchwingerDynamics_DensityOperator}
\frac{\delta}{\delta \Sigma(x)} \tl{\sigma}_{\Sigma}
=
- i \comm{\hat\mcH(x)}{\tl{\sigma}_{\Sigma}-\Omega}.
\end{equation}
By integrating over constant-time slices, we also get a modified Liouville-von Neumann equation [cf. Eq.~\eqref{eq:StandardInteractionPictureDynamics}]:
\begin{equation}\label{APPeq:ModifiedInteractionPictureDynamics_DensityOperator}
\dv{t} \tl{\sigma}_{t}
=
- i \comm{\hH_I (t)}{\tl{\sigma}_{t}-\Omega},
\end{equation}
where we implicitly set $t_0=0$ to lighten the notation. Eq.~\eqref{APPeq:ModifiedTomonagaSchwingerDynamics_DensityOperator} can be rewritten as follows:
\begin{equation}
\frac{\delta}{\delta \Sigma(x)} \tl{\sigma}_{\Sigma}
=
\tl{\mcL}_x (\tl{\sigma}_{\Sigma}),
\qq{where}
\tl{\mcL}_x
:= \mcL_x - \Tr{\bullet}\mcL_x\Omega
= \mcL_x \prt{\bullet-\Tr{\bullet}\Omega}.
\end{equation}
The super-operator $\tl{\mcL}_x$ inherits the commutation properties for spacelike separation from the super-operator $\mcL_x$ (i.e., $[\mcL_x,\mcL_y ]\hA=0$):\footnote{We assume here that $\hA$ is trace-class. We also recall that the vacuum $\Omega$ is trace class as it is a pure state density operator.}
\begin{multline}
(x-y)^2<0
\implies
\comm{\tl{\mcL}_x}{\tl{\mcL}_y}\hA
=
\tl{\mcL}_x\tl{\mcL}_y\hA - \tl{\mcL}_y\tl{\mcL}_x\hA
=
\mcL_x\tl{\mcL}_y\hA - \mcL_y\tl{\mcL}_x \hA
=\\
\mcL_x\prt{\mcL_y\hA - \Tr{\hA}\mcL_y\Omega} - \mcL_y\prt{\mcL_x\hA - \Tr{\hA}\mcL_x\Omega}
=
\comm{\mcL_x}{\mcL_y}\hA -\Tr{\hA}\comm{\mcL_x}{\mcL_y}\Omega
=0.
\end{multline}
The super-operator $\tl{\mcL}_x$, however, is not local in the standard QFT sense as it does not commute with every standard super-operator. For example, it need not commute with the standard $\mcL_y$:
\begin{equation}
(x-y)^2<0
\implies
\comm{\tl{\mcL}_x}{\mcL_y}\hA
=
\tl{\mcL}_x\mcL_y\hA - \mcL_y\tl{\mcL}_x \hA
=
\mcL_x\mcL_y\hA - \mcL_y\prt{\mcL_x \hA - \Tr{\hA}\mcL_x\Omega}
=
\Tr{\hA}\mcL_y\mcL_x\Omega.
\end{equation}
There is, however, a sense in which $\tl{\mcL}_x$ is still local, as we will see after introducing the \enquote{Heisenberg} point of view.

Now that we have defined this new way of evolving the quantum state, we should obtain something akin to the Heisenberg picture in standard quantum mechanics. This is done by defining the adjoint map $[\tl{\mcU}_{\Sigma_0}^{\Sigma_1}]^\dg$ by requiring that
\begin{equation}
\Tr{\sigma_{\Sigma_0}[\tl{\mcU}_{\Sigma_0}^{\Sigma_1}]^\dg \hA_\Sigma}
=
\Tr{\hA_\Sigma \tl{\mcU}_{\Sigma_0}^{\Sigma_1}\sigma_{\Sigma_0}},
\end{equation}
for any operator $\hA_\Sigma$, which we allow to be hypersurface dependent, as is the case, for example, for operators that do not commute with $\hH_S$. This gives
\begin{equation}
[\tl{\mcU}_{\Sigma_0}^{\Sigma_1}]^\dg \hA_{\Sigma_0}
=
[\mcU_{\Sigma_0}^{\Sigma_1}]^\dg \hA_{\Sigma_0}
+
\Tr{\Omega\prt{1-[\mcU_{\Sigma_0}^{\Sigma_1}]^\dg}\hA_{\Sigma_0}}\Id,
\end{equation}
where $\Id$ is the identity operator and we recall that [see Eq.~\eqref{eq:FormalSolutionTS}] $\mcU_{\Sigma_0}^{\Sigma_1} \hA = \hU_{\Sigma_0}^{\Sigma_1}\hA \hU_{\Sigma_1}^{\Sigma_0}$ while $[\mcU_{\Sigma_0}^{\Sigma_1}]^\dg \hA = \hU_{\Sigma_1}^{\Sigma_0}\hA \hU_{\Sigma_0}^{\Sigma_1}$. The adjoint map is unital, i.e., $[\tl{\mcU}_{\Sigma_0}^{\Sigma_1}]^\dg \Id = \Id$. However, it does not map effects to effects, which is the \enquote{Heisenberg} counterpart of not preserving positivity. An effect $\hE$ is a self-adjoint ($\hE^\dg=\hE$) and positive ($\forall\ \ket{\psi}\in\mcH,\ \ev{\hE}{\psi}\geq 0$) operator such that $\Id-\hE$ is also self-adjoint and positive. The Heisenberg dynamics induced by $[\tl{\mcU}_{\Sigma_0}^{\Sigma_1}]^\dg$ does not preserve this property:
\begin{equation}
\ev{[\tl{\mcU}_{\Sigma_0}^{\Sigma_1}]^\dg \hE_{\Sigma_0}}{\psi}
=
\ev{[\mcU_{\Sigma_0}^{\Sigma_1}]^\dg \hE_{\Sigma_0}}{\psi}
+
\Tr{\hE_{\Sigma_0}\Omega}-\Tr{\Omega[\mcU_{\Sigma_0}^{\Sigma_1}]^\dg \hE_{\Sigma_0}}.
\end{equation}

We can also derive the TS equation for operators. First, we notice that
\begin{equation}
\frac{\delta}{\delta \Sigma(x)} \prt{[\hU_{\Sigma_0}^{\Sigma}]^\dg \hA_{\Sigma} \hU_{\Sigma_0}^{\Sigma}}
=
i [\hU_{\Sigma_0}^{\Sigma}]^\dg \prt{\comm{\hat\mcH(x)}{\hA_\Sigma} -i\frac{\delta \hA_\Sigma}{\delta \Sigma(x)}}\hU_{\Sigma_0}^{\Sigma}
=
i [\mcU_{\Sigma_0}^{\Sigma}]^\dg \prt{\comm{\hat\mcH(x)}{\hA_\Sigma} -i\frac{\delta \hA_\Sigma}{\delta \Sigma(x)} }.
\end{equation}
Then, we obtain
\begin{multline}
\frac{\delta}{\delta \Sigma(x)} \prt{[\tl{\mcU}_{\Sigma_0}^{\Sigma}]^\dg \hA_\Sigma}
=
i [\mcU_{\Sigma_0}^{\Sigma}]^\dg \prt{\comm{\hat\mcH(x)}{\hA_\Sigma}}
+
[\mcU_{\Sigma_0}^{\Sigma}]^\dg \prt{\frac{\delta \hA_\Sigma}{\delta \Sigma(x)}}
+\\
- i\Tr{\Omega[\mcU_{\Sigma_0}^{\Sigma}]^\dg \prt{\comm{\hat\mcH(x)}{\hA_\Sigma}}}\Id
+\Tr{\Omega\prt{1-[\mcU_{\Sigma_0}^{\Sigma}]^\dg}\prt{\frac{\delta \hA_\Sigma}{\delta \Sigma(x)}}}\Id,
\end{multline}
which can be rewritten as
\begin{equation}
\frac{\delta}{\delta \Sigma(x)} \prt{[\tl{\mcU}_{\Sigma_0}^{\Sigma}]^\dg \hA_\Sigma}
=
-i\Tr{\Omega\comm{\hat\mcH(x)}{\hA_\Sigma}}\Id
+i[\tl{\mcU}_{\Sigma_0}^{\Sigma}]^\dg\prt{\comm{\hat\mcH(x)}{\hA_\Sigma}}
+[\tl{\mcU}_{\Sigma_0}^{\Sigma}]^\dg \prt{\frac{\delta \hA_\Sigma}{\delta \Sigma(x)}}
\end{equation}
In analogy to Eq.~\eqref{eq:ModifiedInteractionPictureDynamics_UnitaryEvolution}, one can focus on constant-time hyperplanes and integrate over them so as to obtain
\begin{equation}\label{APPeq:ModifiedHeisenbergInteractionPictureDynamics}
\dv{t}\prt{\tl{\mcU}_{t}^\dg \hA_t}
=
-i\Tr{\Omega\comm{\hH_I (t)}{\hA_t}}\Id
+i \tl{\mcU}_{t}^\dg \prt{\comm{\hH_I (t)}{\hA_t}}
+\tl{\mcU}_{t}^\dg \prt{\pdv{\hA_t}{t}}
\end{equation}
If we now define $\tl{A}_t:= \tl{\mcU}_t^\dg \hA_t$ and $\tl{H}_t := \tl{\mcU}_t^\dg \hH_I (t)$, we get that
\begin{equation}
\dv{t} \tl{A}_t 
= 
i \prt{\comm{\tl{H}_t}{\tl{A}_t}-\Tr{\Omega\comm{\tl{H}_t}{\tl{A}_t}}\Id}
+\tl{\mcU}_{t}^\dg \prt{\pdv{\hA_t}{t}},
\end{equation}
which is the equivalent of the dynamics in the Heisenberg picture.
We can now see in which sense $\tl{\mcL}_x$ is local. Consider its adjoint $\tl{\mcL}_x^\dg$, which is the local generator of $\tl{\mcU}^\dg$. We have
\begin{equation}
\tl{\mcL}_x^{\dg}:= i\comm{\mcH(x)}{\bullet} -i\Tr{\Omega\comm{\mcH(x)}{\bullet}}\Id,
\qquad
\comm{\mcH(x)}{\hB(y)}=0
\implies
\tl{\mcL}_x^{\dg} \hB(y) = 0.
\end{equation}
Thus, if $\hB(y)$ is a local operator at $y$ in the standard sense, it does not evolve under $\tl{\mcL}_x^\dg$ when $x$ and $y$ are spacelike separated.

Let us define the super-operator $\Delta_\Omega \hA$ as follows:
\begin{equation}
\Delta_\Omega \hA
:=\hA - \Tr{\hA}\Omega,
\qquad
\tl{\mcU}_{\Sigma_0}^{\Sigma_1}\hA
=
\mcU_{\Sigma_0}^{\Sigma_1}\Delta_\Omega \hA + \Tr{\hA}\Omega,
\qquad
\Delta_\Omega \tl{\mcU}_{\Sigma_0}^{\Sigma_1}
=
\mcU_{\Sigma_0}^{\Sigma_1}\Delta_\Omega.
\end{equation}
In the above set of equations, we see that $\Delta_\Omega \hA$, $\tl{\mcU}$, and $\mcU$ satisfy an intertwining relation. Moreover, the evolution can be written as
\begin{equation}
\tl{\sigma}_{\Sigma_1}-\Omega
=
\mcU_{\Sigma_0}^{\Sigma_1}\prt{\tl{\sigma}_{\Sigma_0}-\Omega}
\implies
\norm{\tl{\sigma}_{\Sigma_1}-\Omega}
=
\norm{\tl{\sigma}_{\Sigma_0}-\Omega},
\end{equation}
where the equality of the norms holds for any unitarily invariant norm, i.e., the vacuum is not attractive in the dynamics generated by $\tl{\mcU}_{\Sigma_0}^{\Sigma}$. Furthermore, a generic unitarily invariant norm between states is preserved:
\begin{equation}
\norm{\tl{\mcU}_{\Sigma_0}^{\Sigma}\prt{\tl{\sigma}_{\Sigma_0}-\tl{\sigma}'_{\Sigma_0}}}
=
\norm{\mcU_{\Sigma_0}^{\Sigma}\prt{\tl{\sigma}_{\Sigma_0}-\tl{\sigma}'_{\Sigma_0}}}
=
\norm{\tl{\sigma}_{\Sigma_0}-\tl{\sigma}'_{\Sigma_0}}.
\end{equation}

\clearpage
\section{Calculations for the collapse rule modification\label{APPSec:CollapseRuleModification}}

We now prove Eq.~\eqref{eq:CompleteHistoryComposition_DVS}, used in Sec.~\ref{SubSec:CollapseUpdateRules_DVS}.
We assume that the realized flash history between $\Sigma_0$ and $\Sigma_1$ contains at least one flash.
Since the vacuum has all CP-qubits in $\ket{0}$, this implies $\mcF_{\Sigma_0}^{\Sigma_1}\Omega=0$.
Moreover, the complete-history projector satisfies $\mcF_{\Sigma_0}^{\Sigma_2} = \mcF_{\Sigma_1}^{\Sigma_2}\mcF_{\Sigma_0}^{\Sigma_1}$, $\mcF_{\Sigma_0}^{\Sigma_2}\mcF_{\Sigma_0}^{\Sigma_1} = \mcF_{\Sigma_0}^{\Sigma_2}$,
and, as discussed in Sec.~\ref{SubSec:CollapseUpdateRules_DVS}, $\comm{\mcU_{\Sigma_1}^{\Sigma_2}}{\mcF_{\Sigma_0}^{\Sigma_1}}=0$. 
It follows that $\mcF_{\Sigma_0}^{\Sigma_2}\Omega=0$ and
\begin{equation}\label{APPeq:CompleteHistoryVacuumCancellation}
\mcF_{\Sigma_0}^{\Sigma_2}\mcU_{\Sigma_1}^{\Sigma_2}\Omega
=
\mcF_{\Sigma_1}^{\Sigma_2}\mcF_{\Sigma_0}^{\Sigma_1}\mcU_{\Sigma_1}^{\Sigma_2}\Omega
=
\mcF_{\Sigma_1}^{\Sigma_2}\mcU_{\Sigma_1}^{\Sigma_2}\mcF_{\Sigma_0}^{\Sigma_1}\Omega
=
0.
\end{equation}

We can now apply the left-hand side of Eq.~\eqref{eq:CompleteHistoryComposition_DVS} to a generic trace-class operator $\hA$.
Using the definition of $\tl{\mcU}$, we first obtain
\begin{equation}
\mcF_{\Sigma_0}^{\Sigma_2}\tl{\mcU}_{\Sigma_1}^{\Sigma_2}\mcF_{\Sigma_0}^{\Sigma_1}\tl{\mcU}_{\Sigma_0}^{\Sigma_1}\hA
=
\mcF_{\Sigma_0}^{\Sigma_2}\mcU_{\Sigma_1}^{\Sigma_2}\mcF_{\Sigma_0}^{\Sigma_1}\tl{\mcU}_{\Sigma_0}^{\Sigma_1}\hA
+
\Tr{\mcF_{\Sigma_0}^{\Sigma_1}\tl{\mcU}_{\Sigma_0}^{\Sigma_1}\hA}
\mcF_{\Sigma_0}^{\Sigma_2}\prt{\Omega-\mcU_{\Sigma_1}^{\Sigma_2}\Omega}.
\end{equation}
The second term vanishes because of
Eq.~\eqref{APPeq:CompleteHistoryVacuumCancellation} and
$\mcF_{\Sigma_0}^{\Sigma_2}\Omega=0$.
Using
$\comm{\mcU_{\Sigma_1}^{\Sigma_2}}{\mcF_{\Sigma_0}^{\Sigma_1}}=0$
and
$\mcF_{\Sigma_0}^{\Sigma_2}\mcF_{\Sigma_0}^{\Sigma_1}=\mcF_{\Sigma_0}^{\Sigma_2}$,
we therefore find
\begin{equation}
\mcF_{\Sigma_0}^{\Sigma_2}\tl{\mcU}_{\Sigma_1}^{\Sigma_2}\mcF_{\Sigma_0}^{\Sigma_1}\tl{\mcU}_{\Sigma_0}^{\Sigma_1}\hA
=
\mcF_{\Sigma_0}^{\Sigma_2}\mcU_{\Sigma_1}^{\Sigma_2}\tl{\mcU}_{\Sigma_0}^{\Sigma_1}\hA.
\end{equation}
Applying once more the definition of $\tl{\mcU}$ gives
\begin{equation}
\mcF_{\Sigma_0}^{\Sigma_2}\mcU_{\Sigma_1}^{\Sigma_2}\tl{\mcU}_{\Sigma_0}^{\Sigma_1}\hA
=
\mcF_{\Sigma_0}^{\Sigma_2}\mcU_{\Sigma_1}^{\Sigma_2}\mcU_{\Sigma_0}^{\Sigma_1}\hA
+
\Tr{\hA}
\mcF_{\Sigma_0}^{\Sigma_2}\mcU_{\Sigma_1}^{\Sigma_2}
\prt{\Omega-\mcU_{\Sigma_0}^{\Sigma_1}\Omega}
=
\mcF_{\Sigma_0}^{\Sigma_2}\mcU_{\Sigma_0}^{\Sigma_2}\hA
-
\Tr{\hA}
\mcF_{\Sigma_0}^{\Sigma_2}\mcU_{\Sigma_0}^{\Sigma_2}\Omega,
\end{equation}
where we used $\mcU_{\Sigma_1}^{\Sigma_2}\mcU_{\Sigma_0}^{\Sigma_1}=\mcU_{\Sigma_0}^{\Sigma_2}$ and Eq.~\eqref{APPeq:CompleteHistoryVacuumCancellation}.
On the other hand,
\begin{equation}
\mcF_{\Sigma_0}^{\Sigma_2}\tl{\mcU}_{\Sigma_0}^{\Sigma_2}\hA
=
\mcF_{\Sigma_0}^{\Sigma_2}\mcU_{\Sigma_0}^{\Sigma_2}\hA
+
\Tr{\hA} 
\mcF_{\Sigma_0}^{\Sigma_2}\prt{\Omega-\mcU_{\Sigma_0}^{\Sigma_2}\Omega}
=
\mcF_{\Sigma_0}^{\Sigma_2}\mcU_{\Sigma_0}^{\Sigma_2}\hA
-
\Tr{\hA} 
\mcF_{\Sigma_0}^{\Sigma_2}\mcU_{\Sigma_0}^{\Sigma_2}\Omega,
\end{equation}
since $\mcF_{\Sigma_0}^{\Sigma_2}\Omega=0$. Hence,
\begin{equation}
\mcF_{\Sigma_0}^{\Sigma_2}\tl{\mcU}_{\Sigma_1}^{\Sigma_2}\mcF_{\Sigma_0}^{\Sigma_1}\tl{\mcU}_{\Sigma_0}^{\Sigma_1}
=
\mcF_{\Sigma_0}^{\Sigma_2}\tl{\mcU}_{\Sigma_0}^{\Sigma_2},
\end{equation}
which proves Eq.~\eqref{eq:CompleteHistoryComposition_DVS}.

\clearpage
\section{Derivation of the formal CPTP map in the long-time limit\label{APPSec:ProofReducedDynamicsFormulaBetweenCurvedSurfaces}}

In this Appendix, we derive Eq.~\eqref{eq:GeneralFormalMapTomonagaSchwinger} directly from the unitary unraveling in the PSL limit. 

For every fixed realization of the noise $W$ in Eq.~\eqref{eq:StochasticHamiltonianUnraveling_PSL}, the integrability argument of Sec.~\ref{Sec:CPQM_ModelBuilding} applies because $\xi(x)$ is a number and the super-operators $\mcQ_I(x)$ and $\mcQ_I(y)$ commute at spacelike separation. Inserting Eq.~\eqref{eq:StochasticHamiltonianUnraveling_PSL} into Eq.~\eqref{eq:FormalSolutionTS} and averaging over $W$ gives\footnote{We point out that Eq.~\eqref{APPeq:AverageFixedNoiseMap} shows that the map of Eq.~\eqref{eq:GeneralFormalMapTomonagaSchwinger} is formally CPTP because the map inside the stochastic average is unitary for every $W$.}
\begin{equation}\label{APPeq:AverageFixedNoiseMap}
\Phi_{\Sigma_0}^{\Sigma_1}=\mbE_W\prtq{\mcT_{\mcQ}\exp{-i\sqrt{\lambda}\int_{\Sigma_0}^{\Sigma_1}\dd[4]{x}\xi(x)\mcQ_I(x)}}.
\end{equation}
Expanding Eq.~\eqref{APPeq:AverageFixedNoiseMap} over the full spacetime region between the two Cauchy surfaces yields
\begin{equation}\label{APPeq:DysonExpansionCurvedSurfaces}
\Phi_{\Sigma_0}^{\Sigma_1}=\sum_{n=0}^{\infty}\frac{\prt{-i\sqrt{\lambda}}^n}{n!}\int_{\Sigma_0}^{\Sigma_1}\prod_{j=1}^{n}\dd[4]{x_j}\mbE_W\prtq{\prod_{j=1}^{n}\xi(x_j)}\mcT_{\mcQ}\prtg{\prod_{j=1}^{n}\mcQ_I(x_j)}.
\end{equation}
In the above equation, Isserlis' theorem~\cite{Munthekaas2025ShortProofIsserlis} tells us that odd-order terms vanish while, for $n=2k$ (with $k=1,2,\dots$), we have that
\begin{equation}\label{APPeq:WickPairingsCurvedSurfaces}
\mbE_W\prtq{\prod_{j=1}^{2k}\xi(x_j)}= \sum_{P\in\mathfrak P_{2k}}\prod_{\{a,b\}\in P}G(x_a-x_b),
\qquad
\abs{\mathfrak P_{2k}} = \frac{(2k)!}{2^k k!},
\end{equation}
where $P$ denotes a partition of $n$ into disjoint pairs and $\mathfrak P_{2k}$ is the set of such partitions. Thus, we get that
\begin{equation}\label{APPeq:DysonExpansionCurvedSurfacesIsserlisApplied}
\Phi_{\Sigma_0}^{\Sigma_1}=\sum_{k=0}^{\infty}\frac{(-1)^k \lambda^k}{(2k)!} \sum_{P\in\mathfrak P_{2k}}\int_{\Sigma_0}^{\Sigma_1}\prod_{j=1}^{2k}\dd[4]{x_j} \prtq{\prod_{\{a,b\}\in P}G(x_a-x_b)} \mcT_{\mcQ}\prtg{\prod_{j=1}^{2k}\mcQ_I(x_j)}.
\end{equation}
Because of the time-ordering and the fact that $G(x-y)=G(y-x)$, every partition gives the same contribution, so that
\begin{equations}\label{APPeq:FormalMapCurvedSurfacesDerived}
\Phi_{\Sigma_0}^{\Sigma_1}
&=\sum_{k=0}^{\infty}\frac{(-1)^k \lambda^k}{2^k k!} \int_{\Sigma_0}^{\Sigma_1}\prod_{j=1}^{2k}\dd[4]{x_j} \prtq{G(x_1-x_2)\cdots G(x_{2k-1}-x_{2k})} \mcT_{\mcQ}\prtg{\prod_{j=1}^{2k}\mcQ_I(x_j)},
\\
&=
\sum_{k=0}^{\infty} \frac{1}{k!}
\mcT_{\mcQ}\prtg{\prtq{-\frac{\lambda}{2}\int_{\Sigma_0}^{\Sigma_1}\dd[4]{y}\dd[4]{z}G(y-z)\mcQ_I(y)\mcQ_I(z)}^k},
\\
&=
\mcT_{\mcQ}\exp{-\frac{\lambda}{2}\int_{\Sigma_0}^{\Sigma_1}\dd[4]{y}\dd[4]{z}G(y-z)\mcQ_I(y)\mcQ_I(z)}.
\end{equations}
This proves Eq.~\eqref{eq:GeneralFormalMapTomonagaSchwinger}.

\clearpage
\section{Flash intensities completely characterize the stochastic flash process\label{APPSec:PointProcessMathematics}}

Adapting Appendix~A of Ref.~\cite{Piccione2026NonMarkovianPSL}, in this Appendix we summarize how, and in what sense, the intensities $\mu_F^{(n)}$ of Sec.~\ref{Sec:FlashIntensities} completely characterize the stochastic flash process. Even though the formulas in Sec.~\ref{Sec:FlashIntensities} are (approximately) valid only when all flash locations are far in the future of $\Sigma_0$, we will here consider $\mu_F^{(n)}(x_1,\dots,x_n|\tl{\rho}_{\Sigma_0})$ to be valid for all $x$ in spacetime. Since the formulas are intended to predict the flash statistics in the future of $\Sigma_0$, we assume that, whenever one of the $x_j$ is in the past of $\Sigma_0$, we have that $\mu_F^{(n)}(x_1,\dots,x_n|\tl{\rho}_{\Sigma_0})=0$. All the equations presented in this Appendix are taken from chapter 5 of Ref.~\cite{Book_Daley2003PointProcessesVol1}. 
We assume that the mathematical conditions underlying these formulas, including absolute convergence of the local Janossy expansions, are satisfied. The general theory of stochastic point processes is presented in Refs.~\cite{Book_Daley2003PointProcessesVol1,Book_Daley2008PointProcessesVol2}.

Given a bounded spacetime region $\mcA$, one can compute another set of quantities called the local Janossy densities \cite[Def.~5.4.IV, p.~137; Eq.~(5.4.14), p.~138]{Book_Daley2003PointProcessesVol1}:
\begin{equation}
j_\mcA^{(n)}(x_1,\dots,x_n|\tl{\rho}_{\Sigma_0})
=
\sum_{k=0}^{\infty} \frac{(-1)^k}{k!} \int_{\mcA^k} \mu_F^{(n+k)} (x_1,\dots,x_n,z_1,\ldots,z_k|\tl{\rho}_{\Sigma_0}) \dd[4]{z_1}\cdots \dd[4]{z_k}.
\end{equation}
Here, the subscript $\mcA^k$ in the integral indicates that each variable $z_1,\dots,z_k$ should be integrated over the region $\mcA$. The quantity $j_\mcA^{(n)}(x_1,\dots,x_n|\tl{\rho}_{\Sigma_0})\dd[4]{x_1}\cdots\dd[4]{x_n}$ gives the probability that there are exactly $n$ flashes in $\mcA$ at locations $\dd[4]{x_1},\dots,\dd[4]{x_n}$.

We can then introduce the family of stochastic variables $N_\mcA$, which count how many flashes occur in the bounded spacetime region $\mcA$. The probability $\mbP_{\tl{\rho}_{\Sigma_0}} (N_\mcA =n)$ of obtaining a certain number $n$ of flashes inside the region $\mcA$ is then given by~\cite[Eq.~(5.3.12), p.~130]{Book_Daley2003PointProcessesVol1}\footnote{The quantity $\sum_{r=0}^{\infty} (r!)^{-1} J_{n+r} \prt{\mcA^{(n)}\times (\mcA^{c})^{(r)}}$ in Eq.~(5.3.12) of Ref.~\cite{Book_Daley2003PointProcessesVol1} corresponds to $\int_{\mcA^n} j_\mcA^{(n)}(x_1,\dots,x_n|\tl{\rho}_{\Sigma_0})\dd[4]{x_1}\cdots\dd[4]{x_n}$ according to \cite[Def.~5.4. IV, p.~137]{Book_Daley2003PointProcessesVol1}.}
\begin{equation}
\mbP_{\tl{\rho}_{\Sigma_0}} (N_\mcA=n) 
= 
\frac{1}{n!}\int_{\mcA^n} j_\mcA^{(n)}(x_1,\dots,x_n|\tl{\rho}_{\Sigma_0})\dd[4]{x_1}\cdots\dd[4]{x_n}.
\end{equation}
More generally, for a finite family of disjoint bounded regions
\(\mcA_1,\ldots,\mcA_q\), with \(\mcC:= \mcA_1\cup\cdots\cup \mcA_q\), one has~\cite[Eq.~(5.3.13), p.~130]{Book_Daley2003PointProcessesVol1}\footnote{The quantity $\sum_{r=0}^{\infty} (r!)^{-1} J_{n+r} \prt{\mcA_1^{(n_1)}\times \cdots \mcA_q^{(n_q)} \times (\mcC^{c})^{(r)}}$ in Eq.~(5.3.13) of Ref.~\cite{Book_Daley2003PointProcessesVol1} corresponds to $\int_{\mcA_1^{n_1}}\!\!\cdots\!\int_{\mcA_q^{n_q}}
j_\mcC^{(n_1+\cdots+n_q)}
(x_1^{(1)},\ldots,x_{n_q}^{(q)}|\tl{\rho}_{\Sigma_0})
\prod_{\alpha=1}^{q}
\prod_{j=1}^{n_\alpha}\dd[4]{x_j^{(\alpha)}}$ according to \cite[Def.~5.4. IV, p.~137]{Book_Daley2003PointProcessesVol1}.}
\begin{equation}
\mbP_{\tl{\rho}_{\Sigma_0}}
\left(
N_{\mcA_1}=n_1,\ldots,N_{\mcA_q}=n_q
\right)
=
\frac{1}{n_1!\cdots n_q!}
\int_{\mcA_1^{n_1}}\!\!\cdots\!\int_{\mcA_q^{n_q}}
j_\mcC^{(n_1+\cdots+n_q)}
(x_1^{(1)},\ldots,x_{n_q}^{(q)}|\tl{\rho}_{\Sigma_0})
\prod_{\alpha=1}^{q}
\prod_{j=1}^{n_\alpha}\dd[4]{x_j^{(\alpha)}}.
\end{equation}
The collection of all such finite-dimensional count distributions completely
characterizes the point process.\footnote{See also pages 26 and 27 of Ref.~\cite{Book_Daley2008PointProcessesVol2}, in particular Corollary 9.2.IV.}
Hence, under the assumptions stated above,
the intensities $\mu_F^{(n)}$ determine all flash counting probabilities in bounded spacetime regions.

\clearpage
\section{Details on the derivation of the flash intensities\label{APPSec:FlashIntensityDetailedCalculation}}

In this Appendix, we derive Eqs.~\eqref{eq:FlashIntensityPSL}, \eqref{eq:FlashKrausOperatorPSL}, and~\eqref{eq:FlashKernelPSL}. This Appendix constitutes the TS counterpart of the calculation presented in Ref.~\cite{Piccione2026NonMarkovianPSL}.

For a fixed realization of $\xi$, the relevant TS evolution operator is
\begin{equation}
[\hU_n^{(\xi)}]_{\Sigma_0}^{{+\infty}}
=
\mcT_{\hat\mcW_n}
\exp{-i\int_{\Sigma_0}^{{+\infty}}\dd[4]{y}\hat\mcW_n(y)}.
\end{equation}
Its Dyson expansion gives
\begin{equation}\label{APPeq:FlashKrausDysonExpansion}
\hK_{n,\Upsilon}^{(\xi)}
:=
\Upsilon^{-n/2}
\bra{\mathbf{1}_n}
[\hU_n^{(\xi)}]_{\Sigma_0}^{{+\infty}}
\ket{\mathbf{0}_n}
=
\Upsilon^{-n/2}
\sum_{m=0}^{\infty}\frac{(-i)^m}{m!}
\int_{\Sigma_0}^{{+\infty}}\dd[4]{y_1}\cdots\dd[4]{y_m}
\bra{\mathbf{1}_n}\mcT_{\hat\mcW_n}
\prtg{\hat\mcW_n(y_1)\cdots\hat\mcW_n(y_m)}
\ket{\mathbf{0}_n}.
\end{equation}
To isolate the dependence on $\Upsilon$ and $\lambda$, define the matter operators
\begin{equation}
\hA_k (y) := g(y-x_k)\hQ_I(y),
\qquad
\hB^{(\xi)} (y) := \xi(y)\hQ_I (y).
\end{equation}
Equation~\eqref{eq:FlashHamiltonianDensityPSL} can then be written as
\begin{equation}\label{APPeq:FlashHamiltonianDensityDecomposition}
\hat\mcW_n(y)
=
\sqrt{\Upsilon}\sum_{k=1}^{n}\hA_k(y)\hX_k
+
\sqrt{\lambda}\hB^{(\xi)}(y).
\end{equation}

Consider a monomial obtained by expanding the product of the $m$ factors in Eq.~\eqref{APPeq:FlashKrausDysonExpansion}, and let $r_k$ be the number of factors proportional to $\hX_k$. Since the $\hX_k$ act on different CP-qubits and commute with all matter operators, its CP-qubit matrix element contains
\begin{equation}
\bra{\mathbf{1}_n}\prod_{k=1}^{n}\hX_k^{r_k}\ket{\mathbf{0}_n}
=
\prod_{k=1}^{n}\bra{1}\hX^{r_k}\ket{0}.
\end{equation}
This is nonzero if and only if every $r_k$ is odd. We may therefore write $r_k=2\ell_k+1$, with $\ell_k\geq 0$. If $R:=\sum_k r_k=n+2\sum_k\ell_k$ is the total number of insertions coming from the first term of Eq.~\eqref{APPeq:FlashHamiltonianDensityDecomposition}, then, after including the prefactor $\Upsilon^{-n/2}$ in Eq.~\eqref{APPeq:FlashKrausDysonExpansion}, the corresponding monomial scales as $\Upsilon^{-n/2}\Upsilon^{R/2}\lambda^{(m-R)/2}$.
In the PSL limit ($\Upsilon \to 0$) with $\lambda>0$ fixed, only the terms with $R=n$ can survive, implying that $l_k$ must be vanishing for every $k$.
Hence, every surviving term contains exactly one factor $\hA_k\hX_k$ for each selected CP-qubit and $m-n$ factors $\hB^{(\xi)}$. In particular, the matrix element vanishes for $m<n$.

For $m\geq n$, let $\mathrm{Inj}(n,m)$ denote the set of injective maps
$\sigma:\{1,\dots,n\}\to\{1,\dots,m\}$. There are $m!/(m-n)!$ of such maps.
The value $\sigma(k)$ specifies the Dyson slot ($1,\dots,m$) occupied by the unique insertion associated with the $k$-th selected CP-qubit ($1,\dots,n$). For each $\sigma\in\mathrm{Inj}(n,m)$, define
\begin{equation}\label{APPeq:FlashSlotOperator}
\hat\mcO_j^{(\sigma,\xi)}(y_j)
:=
\begin{cases}
 g\prt{y_j-x_{\sigma^{-1}(j)}}\hQ_I(y_j),
 & j\in\mathrm{Im}(\sigma),
\\
 \xi(y_j)\hQ_I(y_j),
 & j\notin\mathrm{Im}(\sigma),
\end{cases}
\end{equation}
where $\mathrm{Im}(f)$ denotes the image of the function $f$.
The TS ordering is inherited from the ordered product in Eq.~\eqref{APPeq:FlashKrausDysonExpansion}. Therefore, by introducing
\begin{equation}\label{APPeq:FlashKernelExplicit}
\hat\mcM_{n,m}^{(\xi)}(y_1,\dots,y_m)
:=
\mcT\prtg{
\sum_{\sigma\in\mathrm{Inj}(n,m)}
\prod_{j=1}^{m}\hat\mcO_j^{(\sigma,\xi)}(y_j)
},
\qquad m\geq n,
\end{equation}
one can equivalently write
\begin{equation}
\bra{\mathbf{1}_n}\mcT_{\hat\mcW_n}
\prtg{\hat\mcW_n(y_1)\cdots\hat\mcW_n(y_m)}
\ket{\mathbf{0}_n}
=
\Upsilon^{n/2}\lambda^{(m-n)/2}
\hat\mcM_{n,m}^{(\xi)}(y_1,\dots,y_m)
+
\mathrm{o}\!\prt{\Upsilon^{n/2}},
\end{equation}
for fixed $m$ and $\lambda>0$. Dividing by $\Upsilon^{n/2}\lambda^{(m-n)/2}$ and taking $\Upsilon\to 0$ proves Eq.~\eqref{eq:FlashKernelPSL}. Notice that $\hat\mcM_{n,m}^{(\xi)}$ contains $m!/(m-n)!$ terms, one for each injection, and contains no factor of either $\Upsilon$ or $\lambda$.

At the formal Dyson-series level used in the main text, substituting Eq.~\eqref{eq:FlashKernelPSL} into Eq.~\eqref{APPeq:FlashKrausDysonExpansion} and taking the PSL limit term by term gives
\begin{equation}
\hK_n^{(\xi)}
:=
\lim_{\Upsilon\to 0}\hK_{n,\Upsilon}^{(\xi)}
=
\sum_{m=n}^{\infty}
\frac{(-i)^m\lambda^{(m-n)/2}}{m!}
\int_{\Sigma_0}^{{+\infty}}\dd[4]{y_1}\cdots\dd[4]{y_m}
\hat\mcM_{n,m}^{(\xi)}(y_1,\dots,y_m),
\end{equation}
which is Eq.~\eqref{eq:FlashKrausOperatorPSL}. When expanding $\hat\mcM_{n,m}^{(\xi)}$, the factor $1/m!$ combines with the $m!/(m-n)!$ injective assignments in Eq.~\eqref{APPeq:FlashKernelExplicit}, leaving the factorial suppression $1/(m-n)!$.

It remains to derive the intensity. Before taking $\Upsilon\to0$, the Gaussian-background expression for the $n$-point intensity is the conditional flash probability~\eqref{eq:FixedPointFlashProbability} multiplied by the $n$-point intensity $\mu_C^n$ of the homogeneous Poisson sprinkling. Hence,
\begin{equation}
\mu_{F,\Upsilon,\mu_C}^{(n)}(x_1,\dots,x_n|\tl{\rho}_{\Sigma_0})
=
(\mu_C\Upsilon)^n
\mbE_\xi\Tr_S\prtg{
\hK_{n,\Upsilon}^{(\xi)}
\prt{\tl{\rho}_{\Sigma_0}-\Omega_S}
\hK_{n,\Upsilon}^{(\xi)\dg}
}.
\end{equation}
At the same formal level, using $\mu_C\Upsilon\to\lambda$ and $\hK_{n,\Upsilon}^{(\xi)}\to\hK_n^{(\xi)}$, we obtain
\begin{equation}
\mu_F^{(n)}(x_1,\dots,x_n|\tl{\rho}_{\Sigma_0})
=
\lambda^n
\mbE_\xi\Tr_S\prtg{
\hK_n^{(\xi)}
\prt{\tl{\rho}_{\Sigma_0}-\Omega_S}
\hK_n^{(\xi)\dg}
},
\end{equation}
which is Eq.~\eqref{eq:FlashIntensityPSL}.

\clearpage
\section{Energy and particle number variation rates: Calculations\label{APPSec:EnergyParticleNumberIncreaseBogoliubov}}

In this appendix, we perform the calculations needed to obtain the results in Sec.~\ref{SubSec:KleinGordonRates} of the main text. In order to do so, similarly to Ref.~\cite{Gundhi2026RelativisticCollapseStatisticalDynamics}, we consider a cubic box $\mcB$ of side $L$ and volume $V=L^3$ with periodic boundary conditions, i.e., we identify opposite faces so that $\hat\varphi(t,\bx+L\be_j)=\hat\varphi(t,\bx)$ for $j=1,2,3$. This implies that the allowed momenta are discrete and given by $\bk=2\pi\bn/L$, with $\bn\in\mathbb Z^3$. The corresponding plane waves satisfy $\int_{\mcB}\dd[3]{\bx}e^{i(\bk-\bq)\cdot\bx}=V\delta_{\bk,\bq}$, so that the continuum Dirac delta is replaced by a Kronecker delta. To preserve spatial translation invariance even at finite $L$, we require that the stochastic field obeys the same periodicity, $\xi(t,\bx+L\be_j)=\xi(t,\bx)$, and that the same holds for its correlation function. The continuum results are obtained in the limit $L\to\infty$, where $\sum_{\bk} \to V(2\pi)^{-3}\int\dd[3]{\bk}$ and $\delta_{\bk,\bq}\to (2\pi)^3 V^{-1}\delta^{(3)}(\bk-\bq)$.

In standard box quantization, the creation and annihilation operators satisfy~\cite[Sec.~2.2]{Book_Birrell1982QuantumFieldsCurvedSpace}
\begin{equation}
\comm{\ha_\bk}{\ha_\bp^\dg}=\delta_{\bk,\bp},
\end{equation}
and the interaction-picture Klein-Gordon field reads
\begin{equation}\label{APPeq:FieldOperatorBox}
\hat\varphi_I(t,\bx)
=
\frac{1}{\sqrt V}
\sum_\bk
\frac{1}{\sqrt{2\omega_\bk}}
\prtq{
 e^{-i\omega_\bk t+i\bk\cdot\bx}\ha_\bk
 +
 e^{i\omega_\bk t-i\bk\cdot\bx}\ha_\bk^\dg
},
\qquad
\omega_\bk:=\sqrt{m^2+\bk^2}.
\end{equation}
We also introduce the spatial Fourier coefficients of the real stochastic field,
\begin{equation}\label{APPeq:BoxFourierCoeffientsNoise}
\xi_\bk(t)
:=
\frac{1}{V}
\int_{\mcB}\dd[3]{\bx}
 e^{-i\bk\cdot\bx}\xi(t,\bx),
\qquad
\xi(t,\bx)=\sum_\bk e^{i\bk\cdot\bx}\xi_\bk(t),
\end{equation}
so that $\xi_{-\bk}(t)=\xi_\bk^*(t)$.

For a fixed realization of the noise, the unitary operator implementing the stochastic interaction is
\begin{equation}
\hU_\xi(t)
=
\mcT_{\hQ}
\exp{-i\sqrt\lambda
\int_0^t\dd{s}
\int_{\mcB}\dd[3]{\bx}
\xi(s,\bx)\hQ_I(s,\bx)},
\qquad
\hQ_I(x)=\frac{1}{2} :\!\hat\varphi_I^2(x)\!:.
\end{equation}
This is the finite-volume version of the unitary unraveling introduced in the main text.
Since the generator is quadratic in the creation and annihilation operators, the evolved operators are linear combinations of creation and annihilation operators~\cite[Sec.~II.A.2]{Review_Weedbrook2012GaussianQuantumInformation}. Therefore, the unitary evolution induces a bosonic Bogoliubov transformation~\cite[Sec.~3.2]{Book_Birrell1982QuantumFieldsCurvedSpace},
\begin{equation}\label{APPeq:BogoliubovTrasformations}
\hU_\xi^\dg(t)\ha_\bp\hU_\xi(t)
=
\sum_\bk
\prtq{
\alpha_\xi(\bp,\bk;t)\ha_\bk
+
\beta_\xi(\bp,\bk;t)\ha_\bk^\dg
},
\end{equation}
where $\alpha_\xi(\bp,\bk;t)$ and $\beta_\xi(\bp,\bk;t)$ are complex numbers.

Let us now derive the finite-volume evolution equations for these coefficients. We start by defining $\hH_{\xi,I}(t):= \sqrt{\lambda}\int_{\mcB}\dd[3]{\bx}\xi(t,\bx)\hQ_I(t,\bx)$. Then, from the box expansion of the field, we obtain
\begin{equation}
\comm{\ha_\bp}{\hat\varphi_I(t,\bx)}
=
\frac{e^{i\omega_\bp t-i\bp\cdot\bx}}
{\sqrt{2V\omega_\bp}},
\qquad
\comm{\ha_\bp}{\hQ_I(t,\bx)}
=
\comm{\ha_\bp}{\hat\varphi_I(t,\bx)}
\hat\varphi_I(t,\bx),
\end{equation}
from which it follows that
\begin{equation}
\comm{\ha_\bp}{\hH_{\xi,I}(t)}
=
\frac{\sqrt\lambda}{2V}
\sum_\bq
\frac{1}{\sqrt{\omega_\bp\omega_\bq}}
\int_{\mcB}\dd[3]{\bx}\xi(t,\bx)
\Big[
 e^{i(\omega_\bp-\omega_\bq)t}
 e^{-i(\bp-\bq)\cdot\bx}\ha_\bq
+
 e^{i(\omega_\bp+\omega_\bq)t}
 e^{-i(\bp+\bq)\cdot\bx}\ha_\bq^\dg
\Big].
\end{equation}
Using the fact that
\begin{equations}
\partial_t\prt{\hU_\xi^\dg (t) \ha_\bp \hU_\xi (t)}
&=
-i\hU_\xi^\dg(t)
\comm{\ha_\bp}{\hH_{\xi,I}(t)}
\hU_\xi(t),
\\
\partial_t\prt{\hU_\xi^\dg (t) \ha_\bp \hU_\xi (t)}
&=
\sum_\bk
\prtq{
\partial_t\alpha_\xi(\bp,\bk;t)\ha_\bk
+
\partial_t\beta_\xi(\bp,\bk;t)\ha_\bk^\dg
},
\end{equations}
we can compare the two right-hand sides above and equate the coefficients of $\ha_\bk$ and $\ha_\bk^\dg$, thus obtaining
\begin{equations}\label{APPeq:BoxBogoliubovEquations}
\partial_t\alpha_\xi(\bp,\bk;t)
&=
-\frac{i\sqrt\lambda}{2V}
\sum_\bq
\frac{1}{\sqrt{\omega_\bp\omega_\bq}}
\int_{\mcB}\dd[3]{\bx}\xi(t,\bx)
\Big[
 e^{i(\omega_\bp-\omega_\bq)t}
 e^{-i(\bp-\bq)\cdot\bx}
 \alpha_\xi(\bq,\bk;t)
+
 e^{i(\omega_\bp+\omega_\bq)t}
 e^{-i(\bp+\bq)\cdot\bx}
 \beta_\xi^*(\bq,\bk;t)
\Big],
\\
\partial_t\beta_\xi(\bp,\bk;t)
&=
-\frac{i\sqrt\lambda}{2V}
\sum_\bq
\frac{1}{\sqrt{\omega_\bp\omega_\bq}}
\int_{\mcB}\dd[3]{\bx}\xi(t,\bx)
\Big[
 e^{i(\omega_\bp-\omega_\bq)t}
 e^{-i(\bp-\bq)\cdot\bx}
 \beta_\xi(\bq,\bk;t)
+
 e^{i(\omega_\bp+\omega_\bq)t}
 e^{-i(\bp+\bq)\cdot\bx}
 \alpha_\xi^*(\bq,\bk;t)
\Big].
\end{equations}
The initial conditions for the dynamics of the Bogoliubov coefficients are
\begin{equation}
\alpha_\xi(\bp,\bk;0)=\delta_{\bp,\bk},
\qquad
\beta_\xi(\bp,\bk;0)=0.
\end{equation}
Moreover, using Eq.~\eqref{APPeq:BoxFourierCoeffientsNoise}, we can write Eq.~\eqref{APPeq:BoxBogoliubovEquations} in terms of the Fourier coefficients of the noise:
\begin{equations}\label{APPeq:BoxBogoliubovEquationsFourierModes}
\partial_t\alpha_\xi(\bp,\bk;t)
&=
-\frac{i\sqrt\lambda}{2}
\sum_\bq
\frac{
 e^{i(\omega_\bp-\omega_\bq)t}
 \xi_{\bp-\bq}(t)\alpha_\xi(\bq,\bk;t)
+
 e^{i(\omega_\bp+\omega_\bq)t}
 \xi_{\bp+\bq}(t)\beta_\xi^*(\bq,\bk;t)
}
{\sqrt{\omega_\bp\omega_\bq}},
\\
\partial_t\beta_\xi(\bp,\bk;t)
&=
-\frac{i\sqrt\lambda}{2}
\sum_\bq
\frac{
 e^{i(\omega_\bp-\omega_\bq)t}
 \xi_{\bp-\bq}(t)\beta_\xi(\bq,\bk;t)
+
 e^{i(\omega_\bp+\omega_\bq)t}
 \xi_{\bp+\bq}(t)\alpha_\xi^*(\bq,\bk;t)
}
{\sqrt{\omega_\bp\omega_\bq}}.
\end{equations}

The Bogoliubov coefficients are not completely arbitrary. Preservation of the canonical commutation relations gives the standard bosonic Bogoliubov identities~\cite[Sec.~3.2]{Book_Birrell1982QuantumFieldsCurvedSpace},
\begin{equations}
\sum_\bq
\prtq{
\alpha_\xi(\bp,\bq;t)\alpha_\xi^*(\bk,\bq;t)
-
\beta_\xi(\bp,\bq;t)\beta_\xi^*(\bk,\bq;t)
}
&=
\delta_{\bp,\bk},
\\
\sum_\bq
\prtq{
\alpha_\xi(\bp,\bq;t)\beta_\xi(\bk,\bq;t)
-
\beta_\xi(\bp,\bq;t)\alpha_\xi(\bk,\bq;t)
}
&=0.
\end{equations}
The inverse Bogoliubov transformation reads~\cite[Eq.~(69), Appendix~A.1]{Papageorgiou2019ParticleLocalizability}
\begin{equation}
\hU_\xi(t)\ha_\bk\hU_\xi^\dg(t)
=
\sum_\bp\Big[
\alpha_\xi^*(\bp,\bk;t)\ha_\bp
-\beta_\xi(\bp,\bk;t)\ha_\bp^\dg
\Big].
\end{equation}
Preservation of the canonical commutation relations under this inverse transformation implies
\begin{equation}\label{APPeq:InverseBogoliubovIdentity}
\sum_\bp\Big[
\alpha_\xi(\bp,\bk;t)\alpha_\xi^*(\bp,\bq;t)
-\beta_\xi(\bp,\bq;t)\beta_\xi^*(\bp,\bk;t)
\Big]
=
\delta_{\bq,\bk}.
\end{equation}
Rearranging this identity gives
\begin{equation}\label{APPeq:BogoliubovRelation}
\sum_\bp
\prtq{
\alpha_\xi(\bp,\bk;t)\alpha_\xi^*(\bp,\bq;t)
+
\beta_\xi(\bp,\bq;t)\beta_\xi^*(\bp,\bk;t)
}
=
\delta_{\bq,\bk}
+
2\sum_\bp
\beta_\xi(\bp,\bq;t)\beta_\xi^*(\bp,\bk;t).
\end{equation}

We now use the Bogoliubov transformation to characterize the evolution of the particle-number expectation value.
The DVS prescription\footnote{We recall that $\Tr \{ \hN\Omega_S \}=0$.} gives
\begin{equation}
\ev{\hN}_t
=
\Tr{\hN\tl{\Phi}_0^t\tl{\rho}_0}
=
\Tr{\mbE_\xi\prtq{\hU_\xi^\dg(t)\hN\hU_\xi(t)}
\prt{\tl{\rho}_0-\Omega_S}}.
\end{equation}
Using Eq.~\eqref{APPeq:BogoliubovTrasformations}, we obtain
\begin{multline}
\hU_\xi^\dg(t)\hN\hU_\xi(t)
=
\sum_\bp
\prtq{\hU_\xi^\dg(t)\ha_\bp^\dg\hU_\xi(t)}
\prtq{\hU_\xi^\dg(t)\ha_\bp\hU_\xi(t)}
=\\
=
\sum_{\bp,\bq,\bk}
\Big[
\alpha_\xi(\bp,\bk;t)\alpha_\xi^*(\bp,\bq;t)
\ha_\bq^\dg\ha_\bk
+
\beta_\xi(\bp,\bk;t)\alpha_\xi^*(\bp,\bq;t)
\ha_\bq^\dg\ha_\bk^\dg
+\\
+
\alpha_\xi(\bp,\bk;t)\beta_\xi^*(\bp,\bq;t)
\ha_\bq\ha_\bk
+
\beta_\xi(\bp,\bk;t)\beta_\xi^*(\bp,\bq;t)
\ha_\bq\ha_\bk^\dg
\Big].
\end{multline}
Let us now focus on the two contributions containing one creation and one annihilation operator. Normal-ordering the last term and using Eq.~\eqref{APPeq:BogoliubovRelation}, we get
\begin{multline}
\sum_{\bp,\bq,\bk}
\Big[
\alpha_\xi(\bp,\bk;t)\alpha_\xi^*(\bp,\bq;t)
\ha_\bq^\dg\ha_\bk
+
\beta_\xi(\bp,\bk;t)\beta_\xi^*(\bp,\bq;t)
\ha_\bq\ha_\bk^\dg
\Big]
=\\
=
\sum_{\bq,\bk}\ha_\bq^\dg\ha_\bk
\sum_\bp
\Big[
\alpha_\xi(\bp,\bk;t)\alpha_\xi^*(\bp,\bq;t)
+
\beta_\xi(\bp,\bq;t)\beta_\xi^*(\bp,\bk;t)
\Big]
+
\sum_{\bp,\bk}
\abs{\beta_\xi(\bp,\bk;t)}^2\Id
=\\
=
\hN
+
2\sum_{\bq,\bk}\ha_\bq^\dg\ha_\bk
\sum_\bp
\beta_\xi(\bp,\bq;t)\beta_\xi^*(\bp,\bk;t)
+
\sum_{\bp,\bk}
\abs{\beta_\xi(\bp,\bk;t)}^2\Id.
\end{multline}
The last term is the particle number generated from the vacuum by the unmodified unitary evolution, as in the standard Bogoliubov description of particle creation~\cite[Sec.~3.2]{Book_Birrell1982QuantumFieldsCurvedSpace}. It cancels when traced against $\tl{\rho}_0-\Omega_S$ because $\Tr{\tl{\rho}_0-\Omega_S}=0$.
Moreover, when $[\tl{\rho}_0,\hN]=0$, the terms proportional to $\ha^\dg\ha^\dg$ and $\ha\ha$ have vanishing expectation values. Therefore, we obtain
\begin{equation}\label{APPeq:FirstFormParticleNumberIncrease}
\Delta N_t^{(L)}
:=
\ev{\hN}_t-\ev{\hN}_0
=
2\sum_{\bq,\bk}
R_N^{(L)}(\bq,\bk;t)
\Tr{\ha_\bq^\dg\ha_\bk\tl{\rho}_0},
\end{equation}
where
\begin{equation}
R_N^{(L)}(\bq,\bk;t)
:=
\mbE_\xi
\sum_\bp
\beta_\xi(\bp,\bq;t)\beta_\xi^*(\bp,\bk;t).
\end{equation}

We now exploit the statistical translation invariance of the dynamics. Let us define $\xi_\ba(t,\bx):=\xi(t,\bx-\ba)$.
The stochastic correlation depends only on spacetime differences, which means that $\xi_\ba(t,\bx)$ is statistically distributed as $\xi (t,\bx)$. Let $\hT(\ba)$ be the unitary translation operator such that
\begin{equation}
\hT(\ba)\ha_\bp\hT^\dg(\ba)
=e^{i\bp\cdot\ba}\ha_\bp,
\qquad
\hU_{\xi_\ba}(t) :=\hT(\ba)\hU_\xi(t)\hT^\dg(\ba).
\end{equation}
Applying this relation to Eq.~\eqref{APPeq:BogoliubovTrasformations}, we obtain
\begin{equation}
\alpha_{\xi_\ba}(\bp,\bk;t)
=e^{-i(\bp-\bk)\cdot\ba}\alpha_\xi(\bp,\bk;t),
\qquad
\beta_{\xi_\ba}(\bp,\bk;t)
=e^{-i(\bp+\bk)\cdot\ba}\beta_\xi(\bp,\bk;t).
\end{equation}
Since $\xi_\ba$ and $\xi$ share the same probability law,
\begin{equation}
R_N^{(L)}(\bq,\bk;t)
=e^{i(\bk-\bq)\cdot\ba}R_N^{(L)}(\bq,\bk;t)
\end{equation}
for every translation $\ba$ on the box. Therefore,
\begin{equation}
R_N^{(L)}(\bq,\bk;t)
=
\delta_{\bq,\bk}r_N^{(L)}(\bk;t),
\end{equation}
with
\begin{equation}\label{APPeq:ParticleNumberWavevectorCoefficient}
 r_N^{(L)}(\bk;t)
=
\mbE_\xi
\sum_\bp
\abs{\beta_\xi(\bp,\bk;t)}^2.
\end{equation}
Thus, at finite volume, the non-negativity of the dynamical coefficient $r_N^{(L)}$ is manifest. Consequently,
\begin{equation}\label{APPeq:ParticleNumberIncreaseBox}
\Delta N_t^{(L)}
=
2\sum_\bk r_N^{(L)}(\bk;t)n_\bk^{(L)},
\qquad
n_\bk^{(L)}
:=
\Tr{\ha_\bk^\dg\ha_\bk\tl{\rho}_0}.
\end{equation}
For an ordinary positive state, $n_\bk^{(L)}\geq0$ automatically. For the generalized states considered in the main text, positivity of all $n_\bk^{(L)}$ is an additional condition under which Eq.~\eqref{APPeq:ParticleNumberIncreaseBox} implies $\Delta N_t^{(L)}\geq0$.

We now repeat the same calculation for the normal-ordered free Hamiltonian $\hH_S=\sum_\bp\omega_\bp\ha_\bp^\dg\ha_\bp$.
Using Eq.~\eqref{APPeq:BogoliubovTrasformations}, we obtain
\begin{multline}
\hU_\xi^\dg(t)\hH_S\hU_\xi(t)
=
\sum_{\bp,\bq,\bk}\omega_\bp
\Big[
\alpha_\xi(\bp,\bk;t)\alpha_\xi^*(\bp,\bq;t)
\ha_\bq^\dg\ha_\bk
+
\beta_\xi(\bp,\bk;t)\alpha_\xi^*(\bp,\bq;t)
\ha_\bq^\dg\ha_\bk^\dg
+\\
\alpha_\xi(\bp,\bk;t)\beta_\xi^*(\bp,\bq;t)
\ha_\bq\ha_\bk
+
\beta_\xi(\bp,\bk;t)\beta_\xi^*(\bp,\bq;t)
\ha_\bq\ha_\bk^\dg
\Big].
\end{multline}
Normal ordering the last term gives (among others) a contribution that is proportional to the identity and thus canceled by the DVS prescription. As in the particle-number calculation, the terms proportional to $\ha^\dg\ha^\dg$ and $\ha\ha$ have vanishing expectation values when $[\tl{\rho}_0,\hN]=0$. We can therefore define
\begin{equation}
S_E^{(L)}(\bq,\bk;t)
:=
\mbE_\xi
\sum_\bp\omega_\bp
\prtq{
\alpha_\xi(\bp,\bk;t)\alpha_\xi^*(\bp,\bq;t)
+
\beta_\xi(\bp,\bq;t)\beta_\xi^*(\bp,\bk;t)
}.
\end{equation}
The same translation argument used for $R_N^{(L)}$ shows that
\begin{equation}
S_E^{(L)}(\bq,\bk;t)
=
\delta_{\bq,\bk}s_E^{(L)}(\bk;t),
\implies
\ev{\hH_S}_t
=
\sum_\bk s_E^{(L)}(\bk;t)n_\bk^{(L)}.
\end{equation}
Since $\ev{\hH_S}_0=\sum_\bk\omega_\bk n_\bk^{(L)}$, we obtain
\begin{equation}\label{APPeq:EnergyIncreaseBogoliubovDiagonal}
\Delta E_t^{(L)}
:=
\ev{\hH_S}_t-\ev{\hH_S}_0
=
\sum_\bk r_E^{(L)}(\bk;t)n_\bk^{(L)},
\qquad
r_E^{(L)}(\bk;t)
:=
s_E^{(L)}(\bk;t)-\omega_\bk.
\end{equation}
Let us now make $r_E^{(L)}(\bk;t)$ more explicit. At finite volume, the diagonal coefficient is the ordinary sum
\begin{equation}
s_E^{(L)}(\bk;t)
=
\mbE_\xi
\sum_\bp\omega_\bp
\prtq{
\abs{\alpha_\xi(\bp,\bk;t)}^2
+
\abs{\beta_\xi(\bp,\bk;t)}^2
}.
\end{equation}
Setting $\bq=\bk$ in Eq.~\eqref{APPeq:InverseBogoliubovIdentity} gives
\begin{equation}
\sum_\bp
\prtq{
\abs{\alpha_\xi(\bp,\bk;t)}^2
-
\abs{\beta_\xi(\bp,\bk;t)}^2
}
=1.
\end{equation}
Multiplying this identity by $\omega_\bk$ and subtracting it from $s_E^{(L)}(\bk;t)$ yields
\begin{equation}\label{APPeq:EnergyIncreaseBogoliubovFinal}
 r_E^{(L)}(\bk;t)
=
\mbE_\xi
\sum_\bp
\Big[
(\omega_\bp-\omega_\bk)
\abs{\alpha_\xi(\bp,\bk;t)}^2
+
(\omega_\bp+\omega_\bk)
\abs{\beta_\xi(\bp,\bk;t)}^2
\Big].
\end{equation}
The contribution proportional to $\abs{\beta_\xi}^2$ is manifestly non-negative. The contribution proportional to $\abs{\alpha_\xi}^2$ does not have a definite sign in general because $\omega_\bp-\omega_\bk$ can be either positive or negative.

As mentioned at the beginning of this appendix, the continuum results are obtained in the limit $L\to\infty$, where $\sum_{\bk}\to (2\pi)^{-3} V \int\dd[3]{\bk}$ and $\delta_{\bk,\bq}\to (2\pi)^3 V^{-1}\delta^{(3)}(\bk-\bq)$. We then introduce the shorthand notation $\Delta_L := (2\pi)^3/V$ so that the box-continuum correspondences now read $\Delta_L\sum_{\bk}f(\bk) \to \int\dd[3]{\bk} f(\bk)$ and $\delta_{\bk,\bq} \to \Delta_L\delta^{(3)}(\bk-\bq)$.
To distinguish box and continuum quantities in the following equations, we temporarily restore a superscript $(L)$ to indicate the box quantities. 
As in quantum optics, when going from a discrete to a continuum description of waveguide optical modes~\cite{Ciccarello2017Collision,Maffei2022Closed,Piccione2024EnergyExchangesTiltedQubitLight}, the mode operators and the Bogoliubov coefficients must be rescaled as
\begin{equation}
\ha_\bk^{(L)}
=\sqrt{\Delta_L}\ha_\bk,
\qquad
\alpha_\xi^{(L)}(\bp,\bk;t)
=\Delta_L\alpha_\xi(\bp,\bk;t),
\qquad
\beta_\xi^{(L)}(\bp,\bk;t)
=\Delta_L\beta_\xi(\bp,\bk;t),
\end{equation}
from which it follows that $ n_\bk^{(L)} =\Delta_L n_\bk$.
Indeed, with these definitions, the initial condition $\alpha_\xi^{(L)}(\bp,\bk;0)=\delta_{\bp,\bk}$ becomes $\alpha_\xi(\bp,\bk;0)=\delta^{(3)}(\bp-\bk)$. Dividing Eq.~\eqref{APPeq:BoxBogoliubovEquations} by $\Delta_L$, together with $\int_{\mcB_L}\dd[3]{\bx}\to\int_{\mathbb R^3}\dd[3]{\bx}$, gives the equations in the continuum limit
\begin{equations}
\partial_t\alpha_\xi(\bp,\bk;t)
&=
-\frac{i\sqrt\lambda}{2(2\pi)^3}
\int
\frac{\dd[3]{\bq}\dd[3]{\bx}}
{\sqrt{\omega_\bp\omega_\bq}}
\xi(t,\bx)
\Big[
 e^{i(\omega_\bp-\omega_\bq)t}
 e^{-i(\bp-\bq)\cdot\bx}
 \alpha_\xi(\bq,\bk;t)
+
 e^{i(\omega_\bp+\omega_\bq)t}
 e^{-i(\bp+\bq)\cdot\bx}
 \beta_\xi^*(\bq,\bk;t)
\Big],
\\
\partial_t\beta_\xi(\bp,\bk;t)
&=
-\frac{i\sqrt\lambda}{2(2\pi)^3}
\int
\frac{\dd[3]{\bq}\dd[3]{\bx}}
{\sqrt{\omega_\bp\omega_\bq}}
\xi(t,\bx)
\Big[
 e^{i(\omega_\bp-\omega_\bq)t}
 e^{-i(\bp-\bq)\cdot\bx}
 \beta_\xi(\bq,\bk;t)
+
 e^{i(\omega_\bp+\omega_\bq)t}
 e^{-i(\bp+\bq)\cdot\bx}
 \alpha_\xi^*(\bq,\bk;t)
\Big].
\end{equations}
We also have that
\begin{equation}
R_N^{(L)}(\bq,\bk;t)
=\Delta_L R_N(\bq,\bk;t),
\qquad
S_E^{(L)}(\bq,\bk;t)
=\Delta_L S_E(\bq,\bk;t).
\end{equation}
Hence, if the diagonal coefficients have limits $r_N^{(L)}(\bk;t)\to r_N(\bk;t)$ and $s_E^{(L)}(\bk;t)\to s_E(\bk;t)$, the continuum kernels are defined by
\begin{equation}\label{APPeq:DiagonalKernelsSaferContinuumDefinition}
R_N(\bq,\bk;t)
=
\delta^{(3)}(\bq-\bk)r_N(\bk;t),
\qquad
S_E(\bq,\bk;t)
=
\delta^{(3)}(\bq-\bk)s_E(\bk;t),
\end{equation}
where $r_N(\bk;t)$ and $s_E(\bk;t)$ are implicitly defined.
Using $r_E(\bk;t):=s_E(\bk;t)-\omega_\bk$, Eqs.~\eqref{APPeq:ParticleNumberIncreaseBox} and \eqref{APPeq:EnergyIncreaseBogoliubovDiagonal} then become
\begin{equation}
\Delta N_t
=
2\int\dd[3]{\bk}r_N(\bk;t)n_\bk,
\qquad
\Delta E_t
=
\int\dd[3]{\bk}r_E(\bk;t)n_\bk.
\end{equation}
The continuum expressions in Eqs.~\eqref{eq:ParticleNumberWavevectorCoefficient} and~\eqref{eq:EnergyIncreaseWavevectorCoefficient} are formal reformulations of Eqs.~\eqref{APPeq:ParticleNumberWavevectorCoefficient} and~\eqref{APPeq:EnergyIncreaseBogoliubovFinal}. However, in the continuum, $\alpha_\xi$ and $\beta_\xi$ can be distributional kernels and may contain singular contributions like Dirac delta distributions. In such a case, products such as $\abs{\alpha_\xi}^2$ and $\abs{\beta_\xi}^2$ are not well-defined. The safer continuum definitions of $r_N$ and $s_E=r_E+\omega_\bk$ are instead the coefficients of the diagonal Dirac delta in the two-momentum kernels $R_N$ and $S_E$, as written in Eq.~\eqref{APPeq:DiagonalKernelsSaferContinuumDefinition}.

\clearpage
\section{Noise correlations, long-time rates, and a convergent interaction profile\label{APPSec:EnergyIncrease}\label{APPSec:NoiseCorrelationMathProperties}}

In this Appendix, we derive some properties of the noise correlation function, the long-time rate formulas used in Sec.~\ref{SubSec:KleinGordonRates}, and construct the interaction profile of Eq.~\eqref{eq:ConvergentSmearingProfile}.

\subsection{Properties of the noise correlation function\label{APPSubSec:CorrelationFunctionProperties}}

For a real smearing function, the stationary covariance of Eq.~\eqref{eq:NoiseCorrelationFunction} is formally
\begin{equation}
G(x-y)=\int\dd[4]{r}g(x-r)g(y-r),\qquad G(r)=G(-r).
\end{equation}
A simultaneous translation of $x$ and $y$ is absorbed in the integration variable. A proper orthochronous Lorentz transformation preserves the measure and the scalar function $g$, including its past-light-cone support. Consequently,
\begin{equation}
G(\Lambda(x-y))=G(x-y),\qquad \tl G(\Lambda q)=\tl G(q),
\end{equation}
where $\tl G(q)=\int\dd[4]{r}e^{iq\cdot r}G(r)$. With the Fourier convention used in the main text, the autocorrelation identity reads
\begin{equation}\label{APPeq:AutoCorrelationFunctionMomentumSpace}
\tl G(q)=\int\dd[4]{r}e^{iq\cdot r}G(r)=\tl g(q)\tl g(-q)=|\tl g(q)|^2\geq0.
\end{equation}
Here $\tl g(-q)=\tl g(q)^*$ follows from the reality of $g$. Thus, although the timelike Fourier transform of a past-supported $g$ can distinguish the sign of $q^0$, its squared modulus presents only the two branches $\tl G_T$ and $\tl G_S$ used in the main text.

\subsection{Long-time limits of the Bogoliubov coefficients\label{APPSubSec:BogoliubovLongTime}}

Here we prove Eq.~\eqref{eq:IncreaseFactorsLongTime} of the main text. As in the previous Appendix, we perform our calculations in a cubic box $\mcB$ of side $L$ and volume $V=L^3$ with periodic boundary conditions. Using the Fourier components of the periodic noise [cf. Eq.~\eqref{APPeq:BoxFourierCoeffientsNoise}] we get that
\begin{equation}
\mbE_\xi\prtq{\xi_\bq(s)\xi_\bq^*(u)}
=
\frac{1}{V} \int_{\mcB}\dd[3]{\br}
G_L(s-u,\br)e^{-i\bq\cdot\br},
\qquad
G_L (s-u,\bx-\by) := \mbE_\xi \prtq{\xi (s,\bx) \xi (u,\by)},
\end{equation}
where $G_L$ is the box-equivalent of the correlation distribution $G$ of the main text. Expanding Eq.~\eqref{APPeq:BoxBogoliubovEquationsFourierModes} in $\sqrt{\lambda}$ gives
\begin{equations}
\alpha_\xi(\bp,\bk;t)
&=
\delta_{\bp,\bk}
-
\frac{i\sqrt\lambda}{2\sqrt{\omega_\bp\omega_\bk}}
\int_0^t\dd{s} 
 e^{i(\omega_\bp-\omega_\bk)s}\xi_{\bp-\bk}(s)
+O(\lambda),
\\
\beta_\xi(\bp,\bk;t)
&=
-
\frac{i\sqrt\lambda}{2\sqrt{\omega_\bp\omega_\bk}}
\int_0^t\dd{s} 
 e^{i(\omega_\bp+\omega_\bk)s}\xi_{\bp+\bk}(s)
+O(\lambda).
\end{equations}
Therefore, we get that
\begin{multline}
\mbE_\xi\abs{\beta_\xi(\bp,\bk;t)}^2
=
\frac{\lambda}{4 \omega_\bp \omega_\bk}
\int_0^t \dd{s}\dd{u} e^{i(\omega_\bp+\omega_\bk)(s-u)}
\mbE\prtq{\xi_{\bp+\bk}(s)\xi_{\bp+\bk}(u)^*}
+O(\lambda^2)
=\\
=
\frac{\lambda}{4 \omega_\bp \omega_\bk}
\int_0^t \dd{s}\dd{u} e^{i(\omega_\bp+\omega_\bk)(s-u)}
\int_{\mcB} \frac{\dd[3]{\br}}{V}
G_L(s-u,\br)e^{-i(\bp+\bk)\cdot\br}
+O(\lambda^2)
=\\
=
\frac{\lambda}{4V\omega_\bp\omega_\bk}
\int_{-t}^{t}\dd{\tau}(t-\abs{\tau})
\int_{\mcB}\dd[3]{\br}
G_L(\tau,\br)
e^{i(\omega_\bp+\omega_\bk)\tau-i(\bp+\bk)\cdot\br}
+O(\lambda^2),
\end{multline}
where, in the last equality, we set $\tau=s-u$ and integrate over
$\max(0,-\tau)\leq u\leq\min(t,t-\tau)$, obtaining the
factor $t-\abs{\tau}$. Similarly, for $\bp\neq\bk$, we get
\begin{multline}
\mbE_\xi\abs{\alpha_\xi(\bp,\bk;t)}^2
=
\frac{\lambda}{4\omega_\bp\omega_\bk}
\int_0^t\dd{s}\int_0^t\dd{u}
e^{i(\omega_\bp-\omega_\bk)(s-u)}
\mbE_\xi\prtq{
\xi_{\bp-\bk}(s)\xi_{\bp-\bk}^*(u)}
+O(\lambda^2)
=\\
=
\frac{\lambda}{4V\omega_\bp\omega_\bk}
\int_0^t\dd{s}\int_0^t\dd{u}
e^{i(\omega_\bp-\omega_\bk)(s-u)}
\int_{\mcB}\dd[3]{\br}
G_L(s-u,\br)e^{-i(\bp-\bk)\cdot\br}
+O(\lambda^2)
=\\
=
\frac{\lambda}{4V\omega_\bp\omega_\bk}
\int_{-t}^{t}\dd{\tau}(t-\abs{\tau})
\int_{\mcB}\dd[3]{\br}
G_L(\tau,\br)
e^{i(\omega_\bp-\omega_\bk)\tau-i(\bp-\bk)\cdot\br}
+O(\lambda^2).
\end{multline}
Later, as can be seen in Eq.~\eqref{eq:EnergyIncreaseWavevectorCoefficient} of the main text, the coefficient $\mbE_\xi\abs{\alpha_\xi(\bp,\bk;t)}^2$ will be multiplied by $(\omega_\bp-\omega_\bk)$. Therefore, it is not necessary to investigate what happens for $\bp=\bk$.

We can now compute the coefficient $r_N$ from Eq.~\eqref{APPeq:ParticleNumberWavevectorCoefficient}. Since the sum $\sum_\bp$ over momenta may not commute with the long-time limit we are going to investigate, we introduce a momentum cutoff that we then eliminate at the end. We get
\begin{equation}
r_{N,\epsilon}^{(L)}(\bk;t)
=
\mbE_\xi
\sum_\bp
\abs{\beta_\xi(\bp,\bk;t)}^2 e^{-\epsilon \bp^2}
\simeq
\frac{\lambda t}{4V\omega_\bk}
\int_{-t}^{t}\dd{\tau} \prt{1-\frac{\abs{\tau}}{t}}
\int_{\mcB}\dd[3]{\br}
G_L(\tau,\br)
\sum_\bp \frac{e^{i(\omega_\bp+\omega_\bk)\tau-i(\bp+\bk)\cdot\br}}{\omega_\bp}e^{-\epsilon \bp^2}.
\end{equation}
If one now assumes that $G_L (\tau,\br)$ decays to zero fast enough for $\tau \to \infty$, one can compute the long-time limit by means of the following equalities
\begin{equation}
\lim_{t\to\infty} \int_{-t}^{t}\dd{\tau} \int_{\mcB}\dd[3]{\br} 
\prt{1-\frac{\abs{\tau}}{t}} G_L (\tau,\br)e^{i \nu\tau-i \bq\cdot\br}
=
\lim_{t\to\infty} \int_{-t}^{t}\dd{\tau} \int_{\mcB}\dd[3]{\br}
G_L (\tau,\br)e^{i\nu\tau-i\bq\cdot\br}
=
\tl{G}_L (\nu,\bq),
\end{equation}
thus getting
\begin{equation}
r_{N,\epsilon}^{(L)}(\bk;t)
\simeq
\frac{\lambda t}{4V\omega_\bk} \sum_\bp \frac{\tl{G}_L (\omega_\bp+\omega_\bk,\bp+\bk)}{\omega_\bp}e^{-\epsilon \bp^2}.
\end{equation}
Taking then the continuum limit (and then $\epsilon \to 0$) gives
\begin{equation}
r_N (\bk;t)\simeq \frac{\lambda t}{4 (2\pi)^3}\int \dd[3]{\bp} \frac{\tl{G} (p+k)}{\omega_\bp \omega_\bk}.
\end{equation}

Similarly (we omit the regulator $e^{-\epsilon \bp^2}$ in the following equations), using Eq.~\eqref{APPeq:EnergyIncreaseBogoliubovFinal},
we obtain
\begin{multline}
r_E^{(L)}(\bk;t)
=
\mbE_\xi\sum_\bp
\Big[
(\omega_\bp-\omega_\bk)
\abs{\alpha_\xi(\bp,\bk;t)}^2
+
(\omega_\bp+\omega_\bk)
\abs{\beta_\xi(\bp,\bk;t)}^2
\Big]
\simeq\\
\simeq
\frac{\lambda t}{4V\omega_\bk}
\int_{-t}^{t}\dd{\tau}
\prt{1-\frac{\abs{\tau}}{t}}
\int_{\mcB}\dd[3]{\br}
G_L(\tau,\br)
\sum_\bp\frac{1}{\omega_\bp}
\Big[
(\omega_\bp-\omega_\bk)
e^{i(\omega_\bp-\omega_\bk)\tau-i(\bp-\bk)\cdot\br}
+
(\omega_\bp+\omega_\bk)
e^{i(\omega_\bp+\omega_\bk)\tau-i(\bp+\bk)\cdot\br}
\Big].
\end{multline}
As noted above, the factor $\omega_\bp-\omega_\bk$ eliminates
the diagonal contribution from $\alpha_\xi$.
Taking the long-time limit as in the calculation of $r_N^{(L)}$
gives
\begin{equation}
r_E^{(L)}(\bk;t)
\simeq
\frac{\lambda t}{4V\omega_\bk}
\sum_\bp\frac{1}{\omega_\bp}
\Big[
(\omega_\bp-\omega_\bk)
\tl G_L(\omega_\bp-\omega_\bk,\bp-\bk)
+
(\omega_\bp+\omega_\bk)
\tl G_L(\omega_\bp+\omega_\bk,\bp+\bk)
\Big].
\end{equation}
Finally, taking the continuum limit yields
\begin{equation}
r_E(\bk;t)
\simeq
\frac{\lambda t}{4(2\pi)^3}
\int\frac{\dd[3]{\bp}}{\omega_\bp\omega_\bk}
\Big[
(\omega_\bp-\omega_\bk)\tl G(p-k)
+
(\omega_\bp+\omega_\bk)\tl G(p+k)
\Big].
\end{equation}

\subsection{Invariant momentum integrals\label{APPSubSec:InvariantRateReduction}}

The measure $\dd[3]{\bp}/\omega_\bp$ and the scalar functions $\tl G(p\pm k)$ are both Lorentz invariant. Thus, within the integrals, one can make a change of coordinates so that $\bk=0$. Then, one has that
\begin{equation}
\int \frac{\dd[3]{\bp}}{\omega_\bp} \tl{G} (p+k)
=
\int \frac{\dd[3]{\bp}}{\omega_\bp} \tl{G} (p+(m,\mathbf 0)).
\end{equation}
Indeed, in the above equation, the argument of $\tl{G}$ is timelike. We therefore set $s=(p+(m,\mathbf 0))^2 = 2m(\omega_\bp + m)$ and define $\tl{G}_T (q^2) := \tl{G}(q)$ for $q^2 >0$. Setting also $P=\sqrt{\bp^2}$ we have that 
\begin{equation}
\frac{1}{2m} \dd s = \frac{P}{\omega_P}\dd P
\implies
\frac{P}{2m} \dd s = \frac{P^2}{\omega_P}\dd P
\implies
\frac{\sqrt{s(s-4m^2)}}{4 m^2} \dd s = \frac{P^2}{\omega_P}\dd P.
\end{equation}
Therefore, we get that
\begin{equation}
\int \frac{\dd[3]{\bp}}{\omega_\bp} \tl{G} (p+(m,\mathbf 0))
=
4\pi \int_0^\infty \frac{\dd P}{\omega_P} P^2 \tl{G}_T \prt{2m(\omega_P+m)}
=
\frac{\pi}{m^2}\int_{4 m^2}^\infty \dd{s} \sqrt{s(s-4m^2)}\tl{G}_T (s),
\end{equation}
which gives the first line of Eq.~\eqref{eq:IncreaseFactorsLongTimeSimplified}.

To obtain the second line of Eq.~\eqref{eq:IncreaseFactorsLongTimeSimplified}, we consider the four vector quantity $(P_{\pm} (\bk))^\mu := \int\dd[3]{\bp} (p \pm k )^\mu\tl G(p\pm k)/\omega_\bp$, whose spatial components must vanish when selecting coordinates such that $\bk=0$ because $\tl{G}$ is rotationally invariant. Thus, we have that
\begin{equation}
\frac{1}{\omega_\bk}
\int\frac{\dd[3]{\bp}}{\omega_\bp}
(\omega_\bp\pm\omega_\bk)\tl G(p\pm k)
=
\frac{(P_\pm(\bk))^0}{\omega_\bk}
=
\frac{(P_\pm(\mathbf 0))^0}{m}
=
\int\frac{\dd[3]{\bp}}{\omega_\bp}
\prt{\frac{\omega_\bp\pm m}{m}}
\tl G\bigl(p\pm(m,\mathbf 0)\bigr).
\end{equation}
The second contribution in Eq.~\eqref{eq:IncreaseFactorsLongTime} can then be computed similarly to the $r_N$ case by setting $s=(p+(m,\mathbf 0))^2 = 2m(\omega_\bp + m)$. The first one, with $\tl{G}(p-(m,\mathbf 0))$, can be computed setting $s=-(p-(m,\mathbf 0))^2 = 2m(\omega_\bp - m)$. This latter contribution is due to the spacelike part of $\tl{G}$ so that we define $\tl{G}_S (-q^2):= \tl{G}(q)$ for $q^2 <0$. For both contributions, one can see that $(\omega_\bp\pm m)/m$ becomes $s/(2 m^2)$.

\subsection{Convergence and construction of a smearing profile\label{APPSubSec:RateConvergence}}

We now investigate the convergence of the integrals in Eq.~\eqref{eq:IncreaseFactorsLongTimeSimplified}.
We consider spectra $\tl G_T$ and $\tl G_S$ that are locally bounded on their respective integration domains, including the lower endpoints. This ensures convergence over every finite integration interval.
At large $s$, the energy-rate weights grow as $s^2$. Since $\tl G_T$ and $\tl G_S$ are nonnegative, ultraviolet convergence therefore requires
\begin{equation}\label{APPeq:UltravioletConvergenceCriteria}
\int_{4m^2}^\infty\dd s s^2[\tl G_T(s)+\tl G_S(s)]<\infty.
\end{equation}
It also ensures ultraviolet convergence of the number and flash rates [see Eq.~\eqref{eq:FirstOrderFlashRatesLorentzInvariant}], whose weights have one fewer power of $s$.

One can build an interesting $\tl{g}(q)$ which satisfies the above criteria in the following way. First we consider the function
\begin{equation}
\Delta_-(x;\sigma)
=
\Theta(-x^0)\Theta(x^2)\frac{1}{8\pi}J_0\prt{\sqrt{\sigma x^2}}
=
\lim_{\varepsilon\to0^+}
\int\frac{\dd[4]{q}}{(2\pi)^4}e^{-iq\cdot x}
\frac{1}{(\sigma-q^2+i\varepsilon q^0)^2},
\implies
\prt{\Box+\sigma}^2\Delta_-(x;\sigma)=\delta^{(4)}(x),
\end{equation}
where $J_0(x)$ is the Bessel function of the first kind\footnote{See \url{https://reference.wolfram.com/language/ref/BesselJ.html}.}. The function $\Delta_-(x;\sigma)$ is the past-supported Green function of the Klein--Gordon operator squared.
Using the squared Klein--Gordon operator avoids a Dirac delta singularity on the light-cone surface.
The Fourier transform of $\Delta_-(x;\sigma)$ contains a pole for $q^2=\sigma$ and would lead to an ill-defined autocorrelation function. Therefore, we now smear it by defining
\begin{equation}
g(x)=\int_0^\infty\dd{\sigma}f_{r_C}(\sigma)\Delta_-(x;\sigma),
\qquad
f_{r_C}(\sigma):=\sigma^2r_C^6\prt{3-\sigma r_C^2}e^{-\sigma r_C^2},
\implies
\int_0^\infty\dd{\sigma}f_{r_C}(\sigma)=0.
\end{equation}
The integration can be done explicitly, giving
\begin{equation}
g(x)
=
\Theta(-x^0)\Theta(x^2)
\frac{x^2/r_C^2}{512\pi}
\prtq{96-24(x^2/r_C^2)+(x^2/r_C^2)^2}e^{-x^2/(4r_C^2)}.
\end{equation}
We also have that (in a distributional sense)
\begin{equation}
\tl{g}(q)
=
\lim_{\varepsilon\to0^+}\int_0^\infty\dd{\sigma}
\frac{f_{r_C}(\sigma)}{(\sigma-q^2+i\varepsilon q^0)^2}.
\end{equation}
We can separate the behavior for spacelike and timelike $q$.
For timelike momenta, we define $\tl{g}_T(q^2):=\tl{g}(q)$ for $q^0>0$.
For $q^0<0$, reality of $g$ gives $\tl{g}(q)=\tl{g}_T(q^2)^*$.
For spacelike momenta, we define $\tl{g}_S(-q^2):=\tl{g}(q)$.
Thus, the spectra entering the previous subsections and Eq.~\eqref{eq:IncreaseFactorsLongTimeSimplified} are $\tl{G}_T(s)=\abs{\tl{g}_T(s)}^2$ and $\tl{G}_S(s)=\abs{\tl{g}_S(s)}^2$.
We get that
\begin{equations}\label{APPeq:ExplicitSpectralFormSmearingFunction}
\tl{g}_T(s)
&=
r_C^4\left(-i r_C^2s e^{-r_C^2s}\left(r_C^2s\left(r_C^2s-6\right)+6\right)\left(\pi-i\text{Ei}\left(r_C^2s\right)\right)+r_C^2s\left(r_C^2s-5\right)+2\right),
\\
\tl{g}_S(s)
&=
r_C^4\left(r_C^2s e^{r_C^2s}\left(r_C^4s^2+6r_C^2s+6\right)\text{Ei}\left(-r_C^2s\right)+r_C^4s^2+5r_C^2s+2\right).
\end{equations}
The above expressions are continuous for $s>0$ and satisfy
\begin{equation}
\lim_{s\to0^+}\tl g_T(s)=\lim_{s\to0^+}\tl g_S(s)=2r_C^4.
\end{equation}
Thus, $\tl G_T$ and $\tl G_S$ are locally bounded on their respective integration domains, including the lower endpoints.

To obtain the large-$s$ behavior, we expand the denominator
of the spectral representation as follows:
\begin{equation}
\frac{1}{(\sigma-q^2)^2}
=
\frac{1}{(q^2)^2}
\frac{1}{(1-\sigma/q^2)^2}
=
\sum_{n=0}^\infty
\frac{(n+1)\sigma^n}{(q^2)^{n+2}},
\qquad \sigma<\abs{q^2}.
\end{equation}
Although this series does not converge over the entire
$\sigma$ integration domain, the exponential decay of
$f_{r_C}(\sigma)$ suppresses the contribution of large values of $\sigma$.
For $\abs{q^2}r_C^2\gg1$, expanding in the region
$\sigma\ll\abs{q^2}$ therefore gives an asymptotic expansion
in terms of the moments of $f_{r_C}$:
\begin{equation}
\tl g(q)
=
\frac{1}{(q^2)^2}
\prtq{\int_0^\infty\dd{\sigma}f_{r_C}(\sigma)}
+
\frac{2}{(q^2)^3}
\prtq{\int_0^\infty\dd{\sigma}\sigma f_{r_C}(\sigma)}
+
O\prt{(q^2)^{-4}}.
\end{equation}
For the timelike branch, the pole prescription also gives
$\operatorname{Im}\tl g_T(s)=-\pi f_{r_C}'(s)$.\footnote{The pole prescription refers to the term
$+i\varepsilon q^0$ and the limit $\varepsilon\to0^+$
in the spectral representation of $\tl g(q)$.
For $q^2=s>0$ and $q^0>0$, setting $\eta=\varepsilon q^0$
gives
\begin{equation*}
\operatorname{Im}\frac{1}{(\sigma-s+i\eta)^2}
=
\frac{\partial}{\partial\sigma}
\frac{\eta}{(\sigma-s)^2+\eta^2}.
\end{equation*}
Integrating by parts, with vanishing boundary terms
because $f_{r_C}(0)=0$ and $f_{r_C}$ decays exponentially,
we obtain
\begin{equation*}
\operatorname{Im}\tl g_T(s)
=
-\lim_{\eta\to0^+}\int_0^\infty\dd{\sigma} 
f_{r_C}'(\sigma)
\frac{\eta}{(\sigma-s)^2+\eta^2}
=-\pi f_{r_C}'(s),
\end{equation*}
where we used
$\eta/[(\sigma-s)^2+\eta^2]\to\pi\delta(\sigma-s)$
in the distributional sense.
For the chosen $f_{r_C}$, this imaginary part is
exponentially small at large $s$.}
This contribution is exponentially small at large $s$
and does not change the inverse-power asymptotics.

For our choice of $f_{r_C}$, the first two moments are
\begin{equation}
\int_0^\infty\dd{\sigma}f_{r_C}(\sigma)=0,
\qquad
\int_0^\infty\dd{\sigma}\sigma f_{r_C}(\sigma)
=
-\frac{6}{r_C^2}.
\end{equation}
The vanishing zeroth moment eliminates the
$(q^2)^{-2}$ term, giving
\begin{equation}
\tl g(q)
=
-\frac{12}{r_C^2(q^2)^3}
+
O\prt{(q^2)^{-4}}.
\end{equation}
Thus, setting $q^2=s$ for the timelike branch and $q^2=-s$
for the spacelike branch, we obtain
\begin{equation}
\tl g_T(s)\sim-12r_C^{-2}s^{-3},
\qquad
\tl g_S(s)\sim+12r_C^{-2}s^{-3}.
\end{equation}
Consequently, $\tl G_T(s),\tl G_S(s)=O(s^{-6})$,
so the integrand in
Eq.~\eqref{APPeq:UltravioletConvergenceCriteria}
decays as $s^{-4}$.
The ultraviolet convergence criterion is therefore satisfied.

\clearpage
\section{First-order flash intensity for the Klein--Gordon field\label{APPSec:FirstOrderIntensityDetailedCalculation}}

In this Appendix, we derive the results of Sec.~\ref{SubSec:KleinGordon_FirstOrderIntensities}, using again the cubic box with periodic boundary conditions of Appendix~\ref{APPSec:EnergyParticleNumberIncreaseBogoliubov}.
To this end, as in Appendix~\ref{APPSubSec:BogoliubovLongTime}, we work directly with box quantities and take the continuum limit at the end. We introduce a real box smearing function $g_L$ satisfying
\begin{equation}\label{APPeq:PeriodicizedCollapseProfile}
g_L(\tau,\br+L\be_j)=g_L(\tau,\br),
\qquad
g_L(\tau,\br)=0\quad\text{for }\tau>0.
\end{equation}
All spatial integrations below are restricted to $\mcB$. We leave the dependence of the field operators and states on $L$ implicit.

Let $x$ be spacetime position of a selected CP and $\Sigma_T$ a Cauchy surface to its future. Since the selected CP-qubit has decoupled by $\Sigma_T$, subsequent matter-only evolution cancels from the trace defining its flash intensity. For $n=1$, the first two terms of Eq.~\eqref{APPeq:FlashKernelExplicit} are
\begin{equation}
\hat\mcM_{1,1}^{(\xi)}(y)
=g_L(y-x)\hQ_I(y),
\qquad
\hat\mcM_{1,2}^{(\xi)}(y,z)
=\mcT\prtg{g_L(y-x)\hQ_I(y)\xi(z)\hQ_I(z)}
+\mcT\prtg{\xi(y)\hQ_I(y) g_L(z-x)\hQ_I(z)}.
\end{equation}
Relabeling $y$ and $z$ in the second contribution and inserting these expressions into Eq.~\eqref{eq:FlashKrausOperatorPSL} gives
\begin{equation}\label{APPeq:OneFlashKrausPerturbativeExpansion}
\hK_1^{(\xi)}
=-i\hC_x
-\sqrt{\lambda}\int_{\Sigma_0}^{\Sigma_T}\dd[4]{y}\dd[4]{z}
\mcT\prtg{g_L(y-x)\hQ_I(y)\xi(z)\hQ_I(z)}
+O(\lambda),
\end{equation}
where $\hC_x$ is the box counterpart of the operator in Eq.~\eqref{eq:LateTimeOneFlashOperator}, $\hC_x := \int_{\Sigma_0}^{\Sigma_T}\dd[4]{y} g_L(y-x)\hQ_I(y)$.
The term proportional to $\sqrt{\lambda}$ is linear in the centered noise and has zero stochastic average. Thus, the terms of order $\lambda^{3/2}$ in Eq.~\eqref{eq:FlashIntensityPSL} vanish, leaving
\begin{equation}\label{APPeq:OneFlashIntensityLeadingOrder}
\mu_F^{(1)}(x|\tl{\rho}_{\Sigma_0})
=\lambda\Tr_S\prtg{\hC_x(\tl{\rho}_{\Sigma_0}-\Omega_S)\hC_x}
+O(\lambda^2)
=\lambda\prtq{\Tr_S\prtg{\hC_x^2\tl{\rho}_{\Sigma_0}}
-\bra{\Omega}\hC_x^2\ket{\Omega}}
+O(\lambda^2),
\end{equation}
where we used the Hermiticity of $\hC_x$. The above equation constitutes the box counterpart of Eq.~\eqref{eq:FirstOrderIntensityLeadingLambda}.
For a CP sufficiently far in the future of $\Sigma_0$, we use the late-time approximation in Eq.~\eqref{eq:LateTimeOneFlashOperator} and extend the time integral from $-\infty$ to $+\infty$, while keeping the spatial integration over $\mcB$:
\begin{equation}\label{APPeq:FlashOperatorBox}
\hC_x
\simeq
\int_{-\infty}^{+\infty}\dd{y^0}
\int_{\mcB}\dd[3]{\by} 
g_L(y-x)\hQ_I(y).
\end{equation}

\subsection{Positivity of first-order intensities for states diagonal in particle number\label{APPSubSec:FirstOrderIntensity_PositivityDiagonalStates}}

In the box $\mcB$ of volume $V=L^3$, the allowed momenta are $\bq=2\pi\bn/L$, with $\bn\in\mathbb Z^3$.
For every allowed $\bq$, the Fourier transform of $g_L(\tau,\br)$ is given by $\tl g_L(q^0,\bq) = \int_{-\infty}^{+\infty}\dd{\tau} \int_{\mcB}\dd[3]{\br}  g_L(\tau,\br)e^{iq^0\tau-i\bq\cdot\br}$.
Its autocorrelation function $G_L$ is $G_L(\tau,\br) = \int_{-\infty}^{+\infty}\dd{v} \int_{\mcB}\dd[3]{\bz} g_L(v+\tau,\bz+\br)g_L(v,\bz)$ and, by assumption, corresponds to the one used in Appendix~\ref{APPSubSec:BogoliubovLongTime}.
As in continuous case, we have that $\tl G_L(q^0,\bq) = \abs{\tl g_L(q^0,\bq)}^2$.

In the box, the real Klein--Gordon field operator is given by Eq.~\eqref{APPeq:FieldOperatorBox}. Therefore, the collapse operator \mbox{$\hQ_I(y) =\ :\!\hat \varphi^2 (y)\!:/2$} reads
\begin{equation}\label{APPeq:QuadraticCollapseOperatorBox}
\hQ_I(y)=\frac{1}{4V}\sum_{\bp,\bk}
\frac{1}{\sqrt{\omega_\bp\omega_\bk}}
\Big[e^{-i(p+k)\cdot y}\ha_\bp\ha_\bk
+2e^{i(p-k)\cdot y}\ha_\bp^\dg\ha_\bk
+e^{i(p+k)\cdot y}\ha_\bp^\dg\ha_\bk^\dg\Big].
\end{equation}
Inserting this into Eq.~\eqref{APPeq:FlashOperatorBox} one can rewrite $\hC_x$ as
\begin{equation}\label{APPeq:OneFlashOperatorNumberDecomposition}
\hC_x=\hD_x+\hB_x+\hB_x^\dg,
\qq{where}
\hD_x=\sum_{\bp,\bk}d_x(\bp,\bk)\ha_\bp^\dg\ha_\bk,
\qquad
\hB_x^\dg=\frac12\sum_{\bp,\bk}\kappa_x(\bp,\bk)\ha_\bp^\dg\ha_\bk^\dg,
\end{equation}
with
\begin{equation}\label{APPeq:OneFlashBoxKernels}
d_x(\bp,\bk)
=\frac{e^{i(p-k)\cdot x}\tl g_L(p-k)}{2V\sqrt{\omega_\bp\omega_\bk}},
\qquad
\kappa_x(\bp,\bk)
=\frac{e^{i(p+k)\cdot x}\tl g_L(p+k)}{2V\sqrt{\omega_\bp\omega_\bk}}.
\end{equation}
Since $g_L$ is real, $\tl g_L(-q)=\tl g_L(q)^*$, so $d_x(\bp,\bk)^*=d_x(\bk,\bp)$ and $\hD_x$ is Hermitian. The kernel related to pair creation and annihilation instead obeys $\kappa_x(\bp,\bk)=\kappa_x(\bk,\bp)$.

For a normalized state $\ket{\Psi_N}$ with fixed particle number $N$, the three terms of $\hC_x\ket{\Psi_N}$ belong to mutually orthogonal sectors. Hence,
\begin{equation}\label{APPeq:FixedNumberCNorm}
\bra{\Psi_N}\hC_x^2\ket{\Psi_N}
=
\norm{\hD_x\Psi_N}^2+\norm{\hB_x\Psi_N}^2
+\norm{\hB_x^\dg\Psi_N}^2
=
\norm{\hD_x\Psi_N}^2
+2\norm{\hB_x\Psi_N}^2
+\ev{\comm{\hB_x}{\hB_x^\dg}}{\Psi_N}.
\end{equation}
The vacuum contribution is $\norm{\hB_x^\dg\Omega}^2$. To isolate it, we write
\begin{equation}
[\ha_\bp\ha_\bk,\ha_\bq^\dg\ha_\br^\dg]
=\delta_{\bk,\bq}\ha_\bp\ha_\br^\dg
+\delta_{\bk,\br}\ha_\bp\ha_\bq^\dg
+\delta_{\bp,\bq}\ha_\br^\dg\ha_\bk
+\delta_{\bp,\br}\ha_\bq^\dg\ha_\bk.
\end{equation}
Normal ordering the first two terms and using $\kappa_x(\bp,\bk)=\kappa_x(\bk,\bp)$ gives
\begin{equation}\label{APPeq:PairOperatorCommutator}
[\hB_x,\hB_x^\dg]
=\frac12\sum_{\bp,\bk}\abs{\kappa_x(\bp,\bk)}^2\Id
+\sum_{\bp,\bk,\bq}\kappa_x(\bp,\bq)\kappa_x(\bk,\bq)^*
\ha_\bp^\dg\ha_\bk.
\end{equation}
In particular,
\begin{equation}\label{APPeq:VacuumPairNormTrace}
\bra{\Omega}\hC_x^2\ket{\Omega}
=
\bra{\Omega}\comm{\hB_x}{\hB_x^\dg}\ket{\Omega}
=
\frac{1}{2}\sum_{\bp,\bk}\abs{\kappa_x(\bp,\bk)}^2.
\end{equation}
Combining these identities yields
\begin{equation}\label{APPeq:FixedNumberFirstOrderIntensityPositiveIdentity_V1}
\bra{\Psi_N}\hC_x^2\ket{\Psi_N}
-\bra{\Omega}\hC_x^2\ket{\Omega}
=
\norm{\hD_x\Psi_N}^2+2\norm{\hB_x\Psi_N}^2
+\sum_{\bp,\bk,\bq}\kappa_x(\bp,\bq)\kappa_x(\bk,\bq)^*
\ev{\ha_\bp^\dg\ha_\bk}{\Psi_N}.
\end{equation}
The last addendum can then be rewritten to make its positivity manifest:
\begin{equation}\label{APPeq:FixedNumberFirstOrderIntensityPositiveIdentity_V2}
\bra{\Psi_N}\hC_x^2\ket{\Psi_N}
-\bra{\Omega}\hC_x^2\ket{\Omega}
=\norm{\hD_x\Psi_N}^2+2\norm{\hB_x\Psi_N}^2
+\sum_\bq\norm{\sum_\bp\kappa_x(\bp,\bq)^*\ha_\bp\Psi_N}^2
\geq0.
\end{equation}
Thus, every term remaining after the vacuum subtraction is non-negative. By linearity, the same holds for a positive density operator supported in a fixed-number sector. More generally, a standard positive state with $[\tl\rho_{\Sigma_0},\hN]=0$ has the decomposition
\begin{equation}
\tl\rho_{\Sigma_0}=\sum_{N=0}^{\infty}p_N\rho_N,
\qquad p_N\geq0,\qquad\sum_Np_N=1,
\end{equation}
where $\rho_N$ is normalized and supported in the $N$-particle sector. Thus,
\begin{equation}
\Tr_S\prtg{\hC_x^2\tl\rho_{\Sigma_0}}
-\bra{\Omega}\hC_x^2\ket{\Omega}
=\sum_{N\geq1}p_N\prtq{\Tr_S\prtg{\hC_x^2\rho_N}
-\bra{\Omega}\hC_x^2\ket{\Omega}}
\geq0.
\end{equation}
Together with Eq.~\eqref{APPeq:OneFlashIntensityLeadingOrder}, this proves the leading-order positivity result stated in the main text, i.e., that $[\tl\rho_{\Sigma_0},\hN]=0$ implies $\mu_F^{(1)}(x|\tl{\rho}_{\Sigma_0})\geq 0$ at first order in $\lambda$.

\subsection{Counterexample to positivity of first-order intensities\label{APPSubSec:FirstOrderIntensity_Counterexample}}

To exhibit the effect of coherences between particle-number sectors, define the unnormalized state $\ket{\eta_x}=(\Id-\Omega_S)\hC_x^2\ket{\Omega}$, where $\Omega_S$ denotes here the quantum matter vacuum projector.
The decomposition in Eq.~\eqref{APPeq:OneFlashOperatorNumberDecomposition} gives
$\ket{\eta_x} =\hD_x\hB_x^\dg\ket{\Omega} +(\hB_x^\dg)^2\ket{\Omega}$,
which lies in the two- and four-particle sectors. Whenever $\ket{\eta_x}\neq 0$, let
\begin{equation}
\ket{\chi_x}=\frac{\ket{\eta_x}}{\norm{\eta_x}},
\qquad
\ket{\Psi_\epsilon}
=\frac{\ket{\Omega}-\epsilon\ket{\chi_x}}{\sqrt{1+\epsilon^2}},
\qquad\epsilon>0.
\end{equation}
Since $\bra{\Omega}\hC_x^2\ket{\chi_x}=\norm{\eta_x}$, direct expansion gives
\begin{equation}\label{APPeq:FirstOrderIntensityNegativeCoherenceExample}
\bra{\Psi_\epsilon}\hC_x^2\ket{\Psi_\epsilon}
-\bra{\Omega}\hC_x^2\ket{\Omega}
=\frac{-2\epsilon\norm{\eta_x}}{1+\epsilon^2}
+\frac{\epsilon^2}{1+\epsilon^2}
\prtq{\bra{\chi_x}\hC_x^2\ket{\chi_x}
-\bra{\Omega}\hC_x^2\ket{\Omega}}.
\end{equation}
The term linear in $\epsilon$ is negative and dominates for sufficiently small positive $\epsilon$. Thus, a standard positive pure state with coherence between the vacuum and the non-vacuum even sector can have a negative leading flash intensity.

\subsection{Momentum eigenstates analysis\label{APPSubSec:FirstOrderIntensity_MomentumEigenstates}}

Let $\ket{\bk_0}=\ha_{\bk_0}^\dg\ket{\Omega}$ be a normalized box momentum eigenstate. Since $\hB_x\ket{\bk_0}=0$, Eq.~\eqref{APPeq:FixedNumberFirstOrderIntensityPositiveIdentity_V1} reduces to
\begin{equation}
\bra{\bk_0}\hC_x^2\ket{\bk_0}
-\bra{\Omega}\hC_x^2\ket{\Omega}
=\sum_\bp\abs{d_x(\bp,\bk_0)}^2
+\sum_\bp\abs{\kappa_x(\bp,\bk_0)}^2.
\end{equation}
Therefore, we obtain the spatially uniform intensity
\begin{equation}\label{APPeq:OneParticlePlaneWaveIntensityBox}
\mu_{F,V}^{(1)}(x|\bk_0)
=\frac{\lambda}{4V^2\omega_{\bk_0}}\sum_\bp\frac{1}{\omega_\bp}
\prtq{\tl G_L(p-k_0)+\tl G_L(p+k_0)}+O(\lambda^2).
\end{equation}
Integrating over $\mcB$ gives, at leading order,
\begin{equation}\label{APPeq:OneParticleTotalFlashRateBox}
\Gamma_F^{(L)}(\bk_0)
=
\int_{\mcB}\dd[3]{\bx} \mu_{F,V}^{(1)}(x|\bk_0)
\simeq
\frac{\lambda}{4V\omega_{\bk_0}}
\sum_\bp\frac{\tl G_L(p-k_0)+\tl G_L(p+k_0)}{\omega_\bp}.
\end{equation}
As in Appendix~\ref{APPSubSec:BogoliubovLongTime}, we assume that the box spectra approach their continuum counterparts as $L\to\infty$. Using $V^{-1}\sum_\bp\to(2\pi)^{-3}\int\dd[3]{\bp}$ then gives the leading-order continuum rate
\begin{equation}
\Gamma_F(\bk_0)
=\frac{\lambda}{4(2\pi)^3\omega_{\bk_0}}
\int\frac{\dd[3]{\bp}}{\omega_\bp}
\prtq{\tl G(p-k_0)+\tl G(p+k_0)}
=\Gamma_\alpha(\bk_0)+\Gamma_\beta(\bk_0).
\end{equation}

The integration measure $\dd[3]{\bp}/\omega_\bp$ is Lorentz invariant. Apart from the explicit factor $1/\omega_{\bk_0}$, each integral is therefore a scalar and can be evaluated in the rest frame of $k_0$. Similar calculations to those of Appendix~\ref{APPSubSec:InvariantRateReduction} lead to
\begin{equations}\label{APPeq:FlashRateInvariantReduction}
\Gamma_\alpha(\bk_0)
&=
\frac{\lambda}{32\pi^2m^2\omega_{\bk_0}}
\int_0^{\infty}\dd{s}
\sqrt{s(s+4m^2)}\tl G_S(s),
\\
\Gamma_\beta(\bk_0)
&=
\frac{\lambda}{32\pi^2m^2\omega_{\bk_0}}
\int_{4m^2}^{\infty}\dd{s}
\sqrt{s(s-4m^2)}\tl G_T(s).
\end{equations}

\clearpage

\section{Calculations for nonrelativistic limit and comparison with Markovian CSL\label{APPSec:ComparisonMarkovianCSL}}

In this Appendix, we prove that the spatial smearing in Eq.~\eqref{eq:MarkovianSmearingReduction} is the correct one.
Then, we compute the associated CSL heating rate.
In this appendix, we restore $\hbar$ and $c$, with $x^0=ct$ and spatial wavevectors measured in inverse length.
The parameter $\lambda_{\rm CSL}$ denotes the large-separation decoherence rate for a particle of reference mass $m_0$.

\subsection{Markovian smearing\label{APPSubSec:MarkovianSmearing}}

For fixed $\bx$, the profile in
Eq.~\eqref{eq:ConvergentSmearingProfile} is integrable over $x^0$.
Thus, for a smooth test function $\chi(t)$,
\begin{equation}
\int_{-\infty}^{\infty}\dd{t}\,c\,g(ct,\bx)\chi(t)
=
\int_{-\infty}^{\infty}\dd{x^0}\,
g(x^0,\bx)\chi(x^0/c)
\underset{c\to\infty}{\longrightarrow}
\mcN^{-1} g_{r_C}(\bx)\chi(0).
\end{equation}
Consequently, $c\,g(ct,\bx)\to\delta(t) \mcN^{-1} g_{r_C}(\bx)$ in the distributional sense.

To compute $g_{r_C}$, set $s=ct$, $r=\abs{\bx}$, $b=r^2/(4r_C^2)$, and $u=(c^2 t^2-r^2)/(4r_C^2)$.
The support condition $s\leq-r$ gives
\begin{equation}
\dd{s}
=
-\frac{2r_C^2}{\sqrt{4r_C^2u+r^2}}\dd{u}
=
-\frac{r_C}{\sqrt{u+b}}\dd{u}.
\end{equation}
Therefore, [cf. Eq.~\eqref{eq:ConvergentSmearingProfile}]
\begin{equation}\label{APPeq:MarkovianMarginalIntegral}
g_{r_C}(r)
=
\mcN\int_{-\infty}^{-r} \dd{s} \frac{u}{8\pi}\prt{u^2-6u+6}e^{-u}
=
\mcN\frac{r_C}{8\pi}\int_{0}^{\infty} \dd{u} \frac{u}{\sqrt{u+b}}\prt{u^2-6u+6}e^{-u}
=
\mcN\frac{9r_C}{16\pi}
U\prt{\frac{5}{2},-\frac{1}{2},b},
\end{equation}
where $U$ is the confluent hypergeometric function of the second kind\footnote{See \url{https://mathworld.wolfram.com/ConfluentHypergeometricFunctionoftheSecondKind.html}.}.
Integration over $\bx$ gives $\mcN = r_C^{-4}/2$. Thus, the final form of $g_{r_C} (\bx)$ is
\begin{equation}
g_{r_C}(\bx)
=
\frac{9}{32\pi r_C^3}
U\prt{\frac{5}{2},-\frac{1}{2},\frac{\bx^2}{4r_C^2}}.
\end{equation}
Notice that the second moment of the above distribution is divergent, so $r_C^2$ does not represent its variance (unlike the fixed-variance convention adopted in Ref.~\cite{Piccione2026DiffusionMinimization}).
Computing the Fourier transform, we see that [see Eqs.~\eqref{APPeq:ExplicitSpectralFormSmearingFunction} and~\eqref{eq:VariableSubstitutionMarkovianLimit}]
\begin{equation}
\tl{g}_{r_C} (\bq)
=
\int \dd[3]{\bx} e^{-i\bq\cdot\bx}g_{r_C}(\bx)
=
\prtq{c\int \frac{\dd{t}\dd[3]{\bx}}{2 r_C^4} e^{i(\nu t-\bq\cdot\bx)} g(ct,\bx)}_{\nu=0}
\!\!=
\frac{\tl{g}(0,\bq)}{2 r_C^4}
=
\frac{\tl{g}_S (\bq^2)}{2 r_C^4}
\implies
\abs{\tl{g}_{r_C} (\bq)}^2
=
\frac{\tl\mcG_S (r_C^2 \bq^2)}{4}.
\end{equation}
Using Parseval's identity, we obtain
\begin{equations}\label{APPeq:MarkovianSmearingNorms}
\int\dd[3]{\bx} \abs{g_{r_C}(\bx)}^2
&=
\int \frac{\dd[3]{\bq}}{(2\pi)^3} \abs{\tl{g}_{r_C}(\bq)}^2
=
\int \frac{\dd[3]{\bq}}{(2\pi)^3} \frac{\tl\mcG_S (r_C^2 \bq^2)}{4}
=
\int_0^{\infty}  \frac{\dd{q}}{8 \pi^2} q^2 \tl\mcG_S (r_C^2 q^2)
=
\int_0^{\infty}\dd{z} \frac{\sqrt{z}\tl\mcG_S (z)}{16\pi^2 r_C^3} 
=
\frac{J_0}{16\pi^2r_C^3},
\\
\int\dd[3]{\bx}
\abs{\nabla g_{r_C}(\bx)}^2
&=
\int \frac{\dd[3]{\bq}}{(2\pi)^3} \bq^2 \abs{\tl{g}_{r_C}(\bq)}^2
=
\int_0^{\infty}  \frac{\dd{q}}{8 \pi^2} q^4 \tl\mcG_S (r_C^2 q^2)
=
\int_0^{\infty}\dd{z} \frac{z^{3/2}\tl\mcG_S (z)}{16\pi^2 r_C^5} 
=
\frac{J}{16\pi^2 r_C^5},
\end{equations}
where
\begin{equation}\label{APPeq:MarkovianSpectralMoments}
J_0
:=
\int_0^\infty\dd{z}\,
z^{1/2}\tl{\mcG}_S(z)
\simeq 0.251,
\qquad
J
:=
\int_0^\infty\dd{z}\,
z^{3/2}\tl{\mcG}_S(z)
\simeq 0.164.
\end{equation}

\subsection{CSL Heating\label{APPSubSec:HeatingRateCSL}}

For the comparison with CSL, we use the mass-proportional master equation~\cite{Review_Bassi2013Models}
\begin{equation}\label{APPeq:MarkovianCSLGenerator}
\mcL_{\rm CSL}[\rho]
=
-\frac{\gamma_{\rm CSL}}{2m_0^2}
\int\dd[3]{\bx}\,
[\hmu_f(\bx),[\hmu_f(\bx),\rho]],
\end{equation}
where $\hmu_f(\bx)= \int\dd[3]{\by}\,g_{r_C}(\bx-\by)\hmu(\by)$.
For one particle, $\hmu_f(\bx)=m g_{r_C}(\bx-\hat{\mathbf q})$.
We define the spatial autocorrelation
\begin{equation}
G_{r_C}(\bd)
:=
\int\dd[3]{\bx}\,
g_{r_C}(\bx)g_{r_C}(\bx+\bd).
\end{equation}
Then, the spatial decoherence rate at separation $\bd$ is
\begin{equation}
\Gamma_m(\bd)
=
\gamma_{\rm CSL}\frac{m^2}{m_0^2}
\prt{G_{r_C}(0)-G_{r_C}(\bd)}.
\end{equation}
Since $G_{r_C}(\bd)\to0$ at large separation, the
reference-mass collapse rate is
\begin{equation}\label{APPeq:CSLReferenceCollapseRate}
\lambda_{\rm CSL}
:=
\Gamma_{m_0}(\infty)
=
\gamma_{\rm CSL}\int\dd[3]{\bx} \abs{g_{r_C}(\bx)}^2
=
\gamma_{\rm CSL}
\frac{J_0}{16\pi^2r_C^3}.
\end{equation}
For a particle of mass $m$, the large-separation rate is
therefore $(m/m_0)^2\lambda_{\rm CSL}$.

The energy increase rate per particle is given by~\cite{Piccione2026DiffusionMinimization}
\begin{equation}\label{APPeq:CSLHeatingForMarginal}
\frac{\dd\langle H\rangle}{\dd{t}}
=
\frac{\gamma_{\rm CSL}\hbar^2m}{2m_0^2}
\int\dd[3]{\bx}\,
\abs{\nabla g_{r_C}(\bx)}^2
=
\frac{J}{2J_0}
\frac{\lambda_{\rm CSL}\hbar^2m}{m_0^2r_C^2}.
\end{equation}
Hence,
\begin{equation}
r_{\rm CSL}(t)
=
\frac{J}{2J_0}
\frac{\lambda_{\rm CSL}\hbar^2m}{m_0^2r_C^2}t,
\end{equation}
where $J/(2J_0)\simeq 0.327$, yielding the coefficient used in Eq.~\eqref{eq:HeatingCoefficientCSL} of the main text.

\end{document}

%% file: AuxiliaryFiles/preamble_packages.tex
\usepackage[T1]{fontenc}
\usepackage{graphicx}% Include figure files
\graphicspath{{./Figures/}}
\usepackage{dcolumn}% Align table columns on decimal point
\usepackage{bm}% bold math
\usepackage[dvipsnames,table,xcdraw]{xcolor}
\usepackage{hyperref}% add hypertext capabilities
\hypersetup{colorlinks,linkcolor={MidnightBlue},citecolor={MidnightBlue},urlcolor={MidnightBlue}}  
\usepackage{url} % for \url command in footnotes
\usepackage{tabularray}
\usepackage{array}
\usepackage{makecell}

\usepackage{wrapfig}

\usepackage{tensor}

\usepackage{physics}
\usepackage{amsfonts}
\usepackage{verbatim}
\usepackage{amsmath}
\usepackage{csquotes}
\usepackage{color}
\usepackage{soul}
\usepackage{amsthm}
\usepackage{bm}
\usepackage{graphicx}
\usepackage[normalem]{ulem}
\usepackage{mathtools}

\usepackage{yfonts}

\usepackage{verbatim}

%% file: AuxiliaryFiles/preamble_commands.tex
\newenvironment{equations}
{\begin{equation}\begin{aligned}}
		{\end{aligned}\end{equation}\ignorespacesafterend}

\newenvironment{equations*}
{\begin{equation*}\begin{aligned}}
		{\end{aligned}\end{equation*}\ignorespacesafterend}

\let\emptyset\varnothing

\newcommand{\prt}[1]{\left(#1\right)}

\newcommand{\CP}{\textrm{CP}}

\newcommand{\Schr}{Schr\"{o}dinger\ }

\newcommand{\dg}{\dagger}
\newcommand{\tl}{\tilde}

\newcommand{\prtq}[1]{\left[#1\right]}

\newcommand{\prtg}[1]{\left\{#1\right\}}

\newcommand{\prtqB}[1]{\Bigg[#1\Bigg]}

\newcommand{\sign}{\mathrm{sign}}
\newcommand{\mcA}{\mathcal{A}}
\newcommand{\mcB}{\mathcal{B}}
\newcommand{\mcC}{\mathcal{C}}

\newcommand{\mcF}{\mathcal{F}}
\newcommand{\mcG}{\mathcal{G}}
\newcommand{\mcH}{\mathcal{H}}

\newcommand{\mcL}{\mathcal{L}}
\newcommand{\mcM}{\mathcal{M}}
\newcommand{\mcN}{\mathcal{N}}
\newcommand{\mcO}{\mathcal{O}}

\newcommand{\mcQ}{\mathcal{Q}}
\newcommand{\mcR}{\mathcal{R}}

\newcommand{\mcT}{\mathcal{T}}
\newcommand{\mcU}{\mathcal{U}}

\newcommand{\mcW}{\mathcal{W}}

\newcommand{\mbE}{\mathbb{E}}

\newcommand{\mbP}{\mathbb{P}}

\newcommand{\Id}{\mathbb{I}}

\newcommand{\bF}{\mathbf{F}}

\newcommand{\hA}{\hat{A}}
\newcommand{\hB}{\hat{B}}
\newcommand{\hC}{\hat{C}}
\newcommand{\hD}{\hat{D}}
\newcommand{\hE}{\hat{E}}

\newcommand{\hH}{\hat{H}}

\newcommand{\hK}{\hat{K}}
\newcommand{\hL}{\hat{L}}

\newcommand{\hN}{\hat{N}}
\newcommand{\hO}{\hat{O}}
\newcommand{\hP}{\hat{P}}
\newcommand{\hQ}{\hat{Q}}
\newcommand{\hR}{\hat{R}}

\newcommand{\hT}{\hat{T}}
\newcommand{\hU}{\hat{U}}
\newcommand{\hV}{\hat{V}}

\newcommand{\hX}{\hat{X}}
\newcommand{\hY}{\hat{Y}}
\newcommand{\hZ}{\hat{Z}}

\newcommand{\ha}{\hat{a}}

\newcommand{\hc}{\textrm{h.c.}}

\newcommand{\hmu}{\hat{\mu}}

\newcommand{\bx}{\mathbf{x}}
\newcommand{\by}{\mathbf{y}}
\newcommand{\bz}{\mathbf{z}}

\newcommand{\bp}{\mathbf{p}}
\newcommand{\bq}{\mathbf{q}}
\newcommand{\bk}{\mathbf{k}}

\newcommand{\bd}{\mathbf{d}}

\newcommand{\br}{\mathbf{r}}

\newcommand{\bn}{\mathbf{n}}

\newcommand{\be}{\mathbf{e}}

\newcommand{\bv}{\mathbf{v}}

\newcommand{\ba}{\mathbf{a}}